\documentclass[aps,pre,reprint,twocolumn,superscriptaddress,showpacs]{revtex4-2}
\usepackage{amssymb,amsmath}
\usepackage[table]{xcolor}
\usepackage{graphicx}
\usepackage{mathtools}
\usepackage[export]{adjustbox}
\usepackage{overpic}
\usepackage[colorlinks=true,
 linkcolor=black,
 urlcolor=blue,
 citecolor=blue]{hyperref}
\usepackage{nicefrac}
\usepackage{multirow}
\usepackage{lipsum}
\usepackage[normalem]{ulem}
\usepackage{pbox}
\newcolumntype{C}[1]{>{\centering\let\newline\\\arraybackslash\hspace{0pt}}m{#1}}
\usepackage{verbatim}
\usepackage{enumitem}

\usepackage{framed,color}
\definecolor{shadecolor}{rgb}{0.85,0.80,0.80}

\definecolor{myorange}{RGB}{253, 184, 99}
\definecolor{mypurple}{RGB}{178, 171, 210}

\newcommand{\comments}[1]{}
\usepackage{multirow}

\usepackage{float}

\renewcommand{\vec}[1]{\mathbf{#1}}

\newcommand{\vx}{\vec{x}}
\newcommand{\vp}{\vec{p}}

\renewcommand{\vec}[1]{\mbox{\boldmath$#1$}}

\newcommand{\beq}{\begin{equation}}
\newcommand{\eeq}{\end{equation}}
\newcommand{\bal}{\begin{aligned}}
\newcommand{\eal}{\end{aligned}}

\newcommand{\be}{\begin{equation}}
\newcommand{\ee}{\end{equation}}
\newcommand{\bd}{\begin{displaymath}}
\newcommand{\ed}{\end{displaymath}}
\newcommand{\BE}{\begin{eqnarray}}
\newcommand{\EE}{\end{eqnarray}}

\newcommand{\erf}{{\rm erf}}

\newcommand{\id}{{\openone}}

\newcommand{\boldpsi}{{\mbox{\boldmath $\psi$}}}

\newcommand{\blambda}{{\mbox{\boldmath $\lambda$}}}

\newcommand{\jb}[1]{\textbf{\textcolor{green}{[JB: #1]}}}

\newcommand{\ess}{{{}_\mathcal{S}}}

\allowdisplaybreaks

\begin{document}
\title{Breakdown of random-matrix universality in persistent Lotka--Volterra communities}
\author{Joseph W. Baron}
\email{joseph-william.baron@phys.ens.fr}
\affiliation{Instituto de F{\' i}sica Interdisciplinar y Sistemas Complejos IFISC (CSIC-UIB), 07122 Palma de Mallorca, Spain}
\author{Thomas Jun Jewell}
\affiliation{Department of Physics and Astronomy, School of Natural Sciences,
	The University of Manchester, Manchester M13 9PL, United Kingdom}
\author{Christopher Ryder}
\affiliation{Department of Physics and Astronomy, School of Natural Sciences,
	The University of Manchester, Manchester M13 9PL, United Kingdom}
\author{Tobias Galla}
\email{tobias.galla@ifisc.uib-csic.es}
\affiliation{Instituto de F{\' i}sica Interdisciplinar y Sistemas Complejos IFISC (CSIC-UIB), 07122 Palma de Mallorca, Spain}
\affiliation{Department of Physics and Astronomy, School of Natural Sciences,
	The University of Manchester, Manchester M13 9PL, United Kingdom}

\begin{abstract}
The eigenvalue spectrum of a random matrix often only depends on the first and second moments of its elements, but not on the specific distribution from which they are drawn. The validity of this universality principle is often assumed without proof in applications. In this letter, we offer a pertinent counterexample in the context of the generalised Lotka--Volterra equations. Using dynamic mean-field theory, we derive the statistics of the interactions between species in an evolved ecological community. We then show that the full statistics of these interactions, beyond those of a Gaussian ensemble, are required to correctly predict the eigenvalue spectrum and therefore stability. Consequently, the universality principle fails in this system. 
We thus show that the eigenvalue spectra of random matrices can be used to deduce the stability of `feasible' ecological communities, but only if the emergent non-Gaussian statistics of the interactions between species are taken into account.
\end{abstract}

\maketitle

The theory of disordered systems enables one to deduce the behaviour of collections of many interacting constituents, whose interactions are assumed to be random, but fixed in time \cite{mezard1987}. A related discipline, random matrix theory (RMT), is concerned with the eigenvalue spectra of matrices with entries drawn from a joint probability distribution. Both fields have found numerous applications in physics \cite{wigner1958distribution, wigner1967random} (the study of spin glasses in particular \cite{mezard1987}),
and in other disciplines such as neural networks \cite{aljadeff2015transition, kuczala2016eigenvalue, coolen_kuehn_sollich, rajan2006eigenvalue, louart2018random}, economics \cite{laloux2000random, Bouchaud_chapter} and theoretical ecology \cite{may, may71, allesinatang2, opper1992phase, Galla_2018, birolibunin, altieri2021properties}.  

It is frequently assumed that the distribution of the randomness in RMT or disordered systems is Gaussian, possibly with correlations between different interaction coefficients or matrix entries. Reasons cited for this assumption include analytical convenience, maximum-entropy arguments and the observation that higher-order moments often do not contribute to the results of calculations \cite{Edwards_1975,mezard1987, Galla_Farmer_PNAS}.

In random matrix theory, this latter observation is referred to as the principle of {\em universality} \cite{taovu2010, taovukrishnapur2010, allesinatang1}. The principle states that results obtained for the spectra of Gaussian random matrices frequently also apply to matrix ensembles with non-Gaussian distributions. The conditions for universality to apply are usually mild (higher-order moments of the distribution must fall off sufficiently quickly with the matrix size \cite{taovu2010, taovukrishnapur2010}), and it is often tacitly assumed that these conditions will hold. 

In this letter, we offer a pertinent counterexample to the universality principle in RMT. We focus on the ecological community resulting from the dynamics of the generalised Lotka--Volterra equations with random interaction coefficients. The stability of this community is governed by the interactions between species that survive in the long run \cite{stone, barbier2021fingerprints}. This is a sub-matrix of the original interactions, which we will refer to as the `reduced interaction matrix'. 

Firstly, using dynamic mean-field theory \cite{dedominicis1978dynamics}, we obtain the statistics of the elements in the reduced interaction matrix. These turn out to be non-Gaussian (even when the original interaction matrix is Gaussian). Secondly, we analytically calculate the leading eigenvalue of this non-Gaussian ensemble of random matrices. We show that this eigenvalue is different from the one that we would obtain from a Gaussian ensemble with the same first and second moments as in the reduced interaction matrix. This demonstrates that the principle of universality fails, and it indicates that the Gaussian assumption should not be made lightly. 

Our findings have relevance to the random matrix approach to ecosystem stability, introduced by Robert May \cite{may,may71}. This approach assumes a random interaction structure between species in the community. One line of criticism of May's model is the observation that such interactions do not necessarily describe a feasible equilibrium (that is, an equilibrium for which all species abundances are positive) \cite{gilpin1975stability,namba,gibbs,stone,grilli2017}. The community of surviving species in the generalised Lotka--Volterra model on the other hand is feasible by construction, and we derive the statistics of the emergent random matrix ensemble that describes this community \cite{GOH197763, bunin2016interaction, servan2018coexistence, barbier2021fingerprints}. From this ensemble, we then recover the stability criteria that have previously been derived from the dynamic Lotka-Volterra model \cite{Galla_2018,bunin2017}. We thus show that one can construct a random matrix ensemble (in the sense of May) that correctly reflects the stability of a feasible community of coexistent species. This ensemble is non-Gaussian and quite intricate. In May's words, our work contributes to `elucidating the devious strategies of nature which make for stability in enduring natural systems' \cite{maybook}.

We start from the generalised Lotka-Volterra equations (GLVEs) \cite{Galla_2018, bunin2017} 
\begin{align}
\dot x_i = x_i \left( 1 - x_i + \sum_{ij}a_{ij} x_j \right), \label{gles}
\end{align}
where the $x_i\geq 0$ describe the abundances of species $i=1,\dots,N$. The interaction matrix elements in Eq.~(\ref{gles})  $a_{ij}$ are quenched random variables. We refer to these as the `original interaction matrix' elements. We assume that the mean of each matrix element is $\overline{a_{ij}} = \mu/N$ (we use an overbar to denote averages over the ensemble of interaction matrices), and that they have variance $\mathrm{Var}(a_{ij}) = \sigma^2/N$. We also allow for correlations between diagonally opposed matrix elements, $\mathrm{Corr}(a_{ij}, a_{ji}) = \Gamma$, ($-1\leq\Gamma\leq 1$) where $\mathrm{Corr}(a,b) = (\overline{ab}- \overline{a}\overline{b})/\sqrt{\mathrm{Var}(a)\mathrm{Var}(b)}$. 

The scaling with $N$ of the moments of $a_{ij}$ follows the standard conventions in disordered systems \cite{mezard1987} and guarantees a well-defined thermodynamic limit $N\to \infty$. All our results are independent of the higher moments of $a_{ij}$ as long as these moments decay sufficiently quickly with $N$. Further details can be found in Sec.~S1 of the Supplemental Material (SM).

\begin{figure}[h]
	\centering 
	\includegraphics[scale = 0.5]{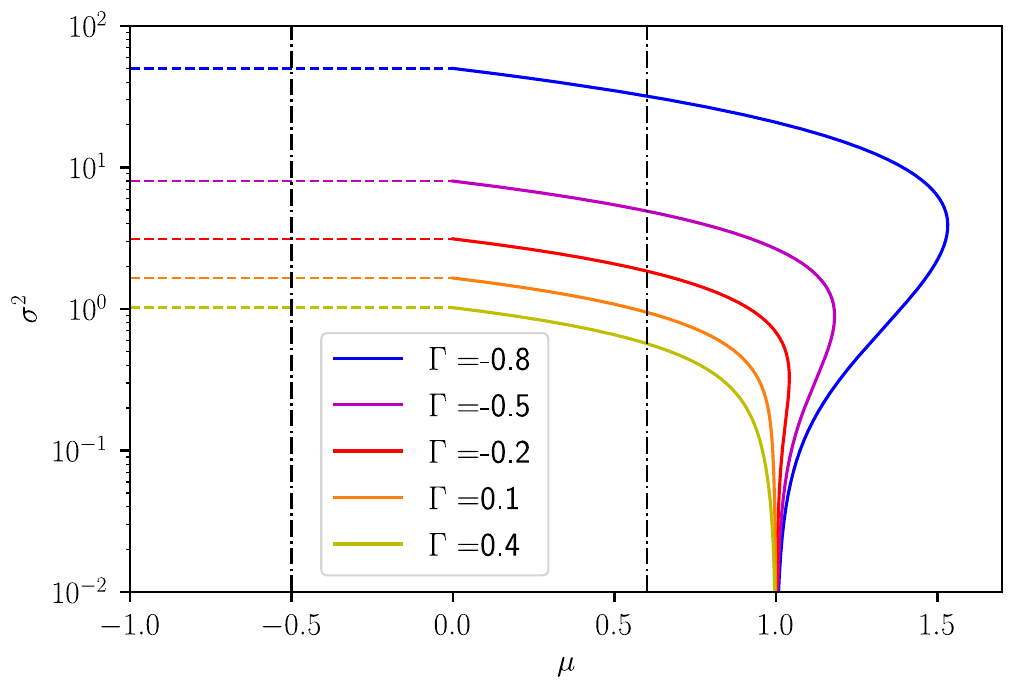}
	\caption{Stability diagram \cite{Galla_2018,bunin2017} of the GLVE system in the plane spanned by $\mu$ and $\sigma^2$ for fixed values of the correlation parameter $\Gamma$. Solid lines indicate the $M\to\infty$ transition, dashed horizontal lines the linear instability.  These lines were produced using Eqs.~(S22) and (S28) in the SM respectively. Vertical lines mark the values of $\mu$ used in the two panels of Fig. \ref{fig:lambdaversussigma}. The system has a unique stable fixed point below the dashed lines and to the left of the solid lines.}\label{fig:phasediagram}
\end{figure}

Previous analyses of this system  \cite{Galla_2018, bunin2017} in the thermodynamic limit have shown that there is a range of parameter combinations $\mu, \sigma^2$ and $\Gamma$ for which the dynamics reaches the a unique stable fixed point, independently of the starting conditions. This is the case in the region to the left and below the instability lines in the phase diagram in Fig.~\ref{fig:phasediagram}. 

When a fixed-point solution is reached, not all species survive, i.e. there are some species for which $x_i^\star>0$ and others with $x_i^\star=0$ (we use an asterisk to denote the fixed point). Using dynamic mean-field theory (DMFT), one can deduce these statistics of the species abundances at the fixed point. 

From the DMFT analysis, one can also find the combinations of system parameters at which the system is no longer able to support a unique stable fixed point. There are two types of transition: (1) the average species abundance can diverge [i.e., $M \to \infty$], or (2) the fixed-point solution can become linearly unstable to perturbations. Closed-form expressions for the critical sets of parameters ($\sigma$, $\Gamma$ and $\mu$) at which each of these transitions occur were derived in \cite{Galla_2018, bunin2017}. A selection of phase lines for different values of the correlation parameter $\Gamma$ are shown in Fig. \ref{fig:phasediagram}.

\begin{figure}[h]
	\centering 
	\includegraphics[scale = 0.5]{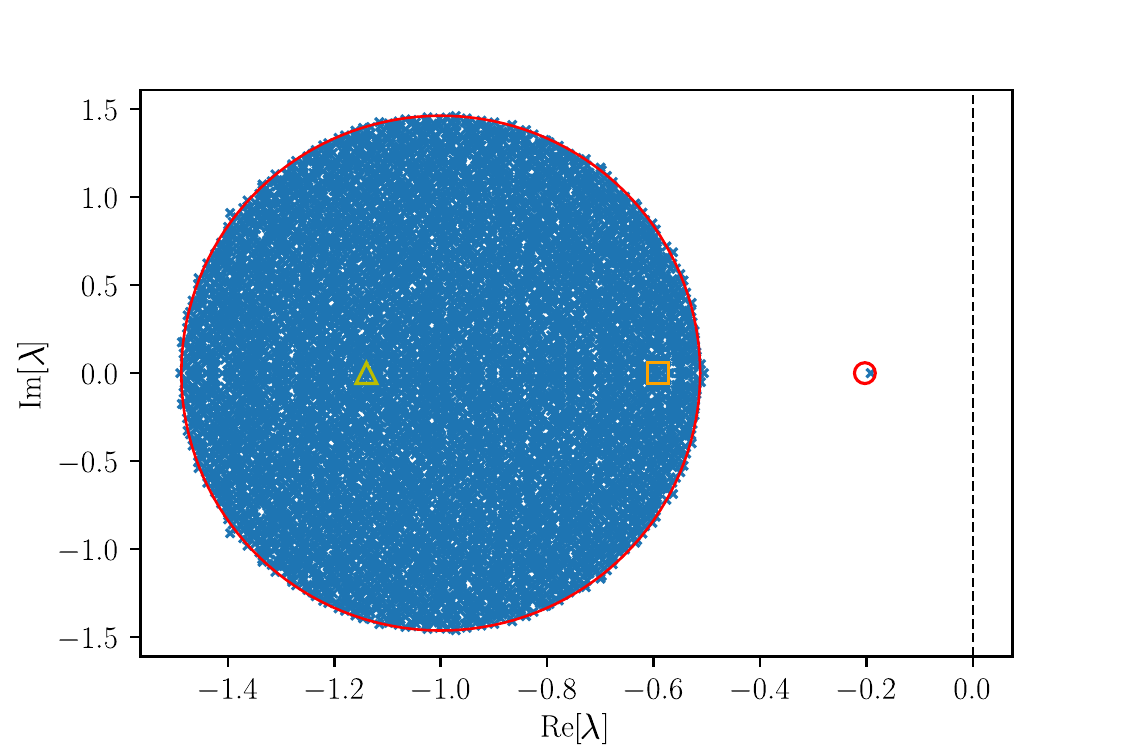}
	\caption{The eigenvalues of the reduced interaction matrix. Results from a computer simulation of the GLVE are shown as markers. The solid red curve and the hollow circle show the theoretical predictions for the bulk region and outlier eigenvalue in Eqs.~(\ref{bulkspectrum}) and Eqs.~(S71)--(S73) of the SM respectively. Two naive predictions for the outlier that do not take the full statistics of the reduced interaction matrix into account are shown as a yellow triangle ($\lambda_0$ in the text) and an orange square ($\lambda_1$ in the text). System parameters are $\sigma = 1.1$, $\mu = 0.9$, $\Gamma = -0.5$, simulation data is from a single realisation with $N = 10000$. }\label{fig:examplespectra}
\end{figure}

We now examine an alternative approach to analysing the stability of the GLVEs in Eq.~(\ref{gles}). Namely, we consider the reduced interaction matrix (the interaction matrix between the species in the surviving sub-community). More precisely, this is defined by
\begin{align}
a'_{ij} = a_{ij} - \delta_{ij} ,
\end{align}
where $i,j \in \mathcal{S}$ (with $\mathcal{S}$ the set of surviving species), and where the shift in the diagonal elements reflects the $-x_i$ term inside the brackets of Eq.~(\ref{gles}). It can be shown that a fixed point of the GLVEs is stable if and only if all of the eigenvalues of the reduced interaction matrix have negative real parts \cite{stone, birolibunin, barbier2021fingerprints} (see also Sec.~S2 in the SM).

We note that the statistics of the reduced interaction matrix elements are determined by the extinction dynamics in the GLVE system, and are consequently vastly different to those of the original interaction matrix \cite{bunin2016interaction, fraboul2021artificial}. For instance, they are non-Gaussian (even when the $a_{ij}$ are Gaussian), and there are correlations between elements sharing only one index (see SM Sec.~S6). This makes the calculation of the eigenvalue spectrum of the reduced interaction matrix a non-trivial task.

As is illustrated in Fig. \ref{fig:examplespectra}, the spectrum of the reduced interaction matrix consists of a bulk set of eigenvalues and a single outlier. Writing $z_{ij} = a_{ij} - \mu N^{-1}$ (where once again $i,j \in \mathcal{S}$), both the outlier eigenvalue $\lambda_{\mathrm{outlier}}$ and the bulk spectral density $\rho_{\mathrm{bulk}}(\lambda)$ can be obtained from the resolvent matrix $\underline{\underline{G}} =  \left[\omega\underline{\underline{\id}} - \underline{\underline{z}}\right]^{-1}$. The bulk density is calculated from the trace of $\underline{\underline{G}}$ via well-known relations \cite{sommers}. The outlier eigenvalue in turn fulfils \cite{orourke, BENAYCHGEORGES2011494, baron2021eigenvalues} 
\be
\mathcal{G}\left(1+\lambda_{\mathrm{outlier}}\right) = \frac{1}{\mu \phi}, \label{bulkoutfromres}
\ee
where $ \mathcal{G}\left(\omega\right) \equiv (N\phi)^{-1}\overline{\sum_{i,j \in \mathcal{S}}  G_{ij}(\omega)}$, and where $\phi$ is the fraction of surviving species at the fixed point. 

We first briefly discuss the bulk spectrum, for which the results do not run counter to the universality principle. We use a series expansion for a Hermitized version of the resolvent of the reduced interaction matrix. This standard approach accounts for the non-analytic nature of the resolvent in the bulk region \cite{JANIK1997603, feinberg1997non}.

\begin{figure*}[t]
	\centering 
	\includegraphics[scale = 0.5]{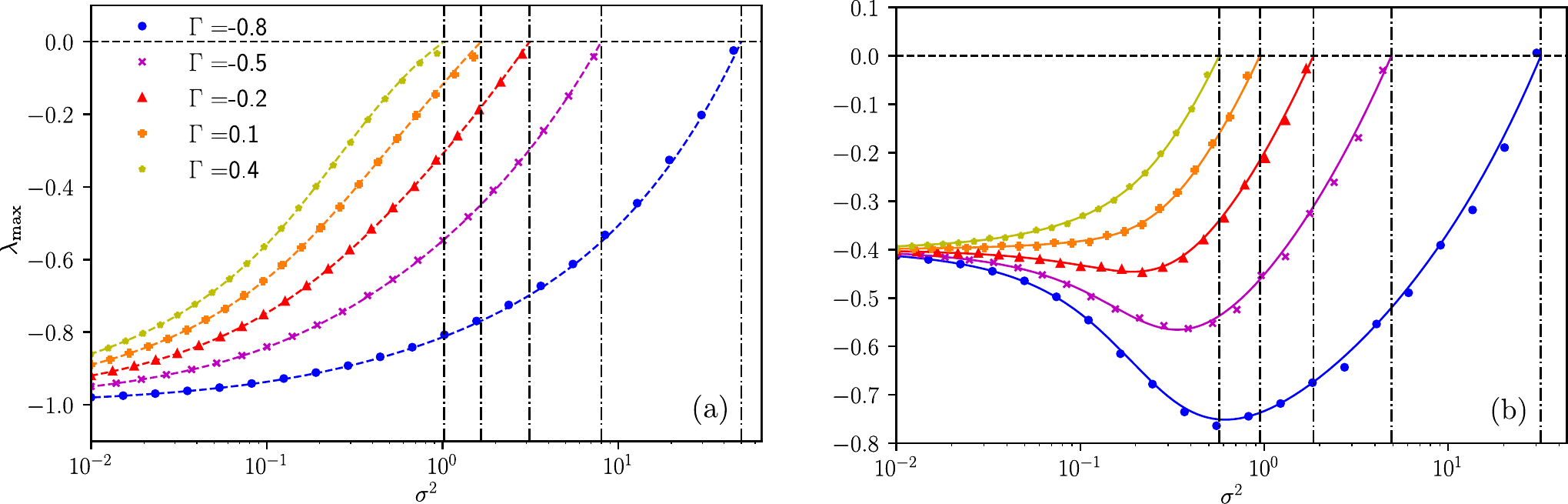}
	\caption{ Panel (a): Right edge of the bulk of the eigenvalue spectrum of the reduced interaction matrix versus $\sigma^2$ for different values of the system parameter $\Gamma$ and fixed $\mu = -0.5$. Markers are the result of averaging the results of $10$ simulations of the GLVE with 
	$N = 4000$. The dashed coloured lines are given by $\lambda_{\mathrm{edge}} = -1 + \sigma \sqrt{\phi} (1 + \Gamma)$, and the vertical dot-dashed lines are the  points where the linear instability occurs in the GLVE (see the dashed lines in Fig. \ref{fig:phasediagram}). Panel (b): Outlier eigenvalue of the reduced interaction matrix versus $\sigma^2$ at fixed $\mu = 0.6$ and for the same values of $\Gamma$ as in panel (a). Markers are the result of averaging the results of $10$ simulations with $N=4000$. The solid lines are the analytical result in Eqs.~(S71)--(S73) of the SM, and the vertical dot-dashed lines are the points where  $M\to\infty$ in the GLVE (see the solid lines in Fig. \ref{fig:phasediagram}).}\label{fig:lambdaversussigma}
\end{figure*}

We find that the resulting series for the trace of the resolvent matrix is identical to that of a Gaussian random matrix in the limit $N\to \infty$. That is, we show that the higher-order statistics of the reduced interaction matrix do not contribute to this series and, therefore, that the universality principle holds for the bulk region. The only statistics of the reduced interaction matrix that contribute are $(\sigma')^2 \equiv N_\ess \mathrm{Var}(a_{ij}') = \phi \sigma^2$ and $\Gamma' \equiv  \mathrm{Corr}(a_{ij}',a_{ji}')= \Gamma$ where $N_\ess$ is the number of surviving species (we calculate these statistics in Sec.~S6 of the SM). One obtains the familiar elliptic law 
\begin{align}
\rho_{\mathrm{bulk}}(\lambda) = 
\begin{cases}
\frac{1}{\pi (\sigma')^2 [1-(\Gamma')^2]} \, &\mathrm{if} \,\,\, \frac{(1+ x)^2}{(1+\Gamma')^2} + \frac{y^2}{(1-\Gamma')^2 }<  (\sigma')^2, \\
0 \, &\mathrm{otherwise},
\end{cases}\label{bulkspectrum}
\end{align}
where $\lambda = x+iy$. We can show (SM Sec.~S5C) that the bulk of the eigenvalue spectrum crossing the imaginary axis corresponds to the linear instability of the GLVEs, represented by the dashed horizontal lines in Fig.~\ref{fig:phasediagram} . This is verified in Fig. \ref{fig:lambdaversussigma}(a). 

We now move on to the outlier eigenvalue, which is a far less trivial matter. We first discuss two candidate expressions for the outlier eigenvalue based upon calculations for Gaussian random matrix ensembles. We show that neither of these expressions are accurate, and that the universality principle fails to predict the outlier eigenvalue. We subsequently derive an accurate expression for the outlier, which we show correctly predicts stability.

Noting previous work \cite{orourke, allesinatang1, allesinatang2, edwardsjones}, one might perhaps expect that $\mu'=N_\ess \overline{a_{ij}'}$ ($i \neq j$), together with $(\sigma')^2$ and $\Gamma'$ would be sufficient to predict the outlier eigenvalue of the reduced interaction matrix. Using an established formula for the outlier eigenvalues of Gaussian random matrices \cite{edwardsjones, orourke}, one then obtains $\lambda_{0} = -1 + \mu' + \Gamma' \sigma'^2/\mu'$. 

If we also include the effects of correlations between elements sharing only one index $\gamma' = N^2 \mathrm{Corr}(a'_{ij}, a'_{ki})$ (where $k \neq i$), we arrive at (using results from our previous work \cite{baron2021eigenvalues})
\be\label{eq:lambda1}
\lambda_{1} = -1 + \mu' + \frac{\mu'}{2} \left(1 + \frac{\Gamma'}{\gamma'}\right)\left[\sqrt{1 + \frac{4 \gamma' \sigma'^2}{(\mu')^2}} -1\right].
\ee
The approach leading to Eq.~(\ref{eq:lambda1}) takes into account all possible correlations for a Gaussian random matrix with statistical symmetry between different species. We note that correlations between elements in the same row or column also exist in the reduced interaction matrix (see SM Sec.~S6A), but these do not affect the location of the outlier \cite{baron2021eigenvalues}. 

\begin{figure}[b]
	\centering 
	\includegraphics[scale = 0.5]{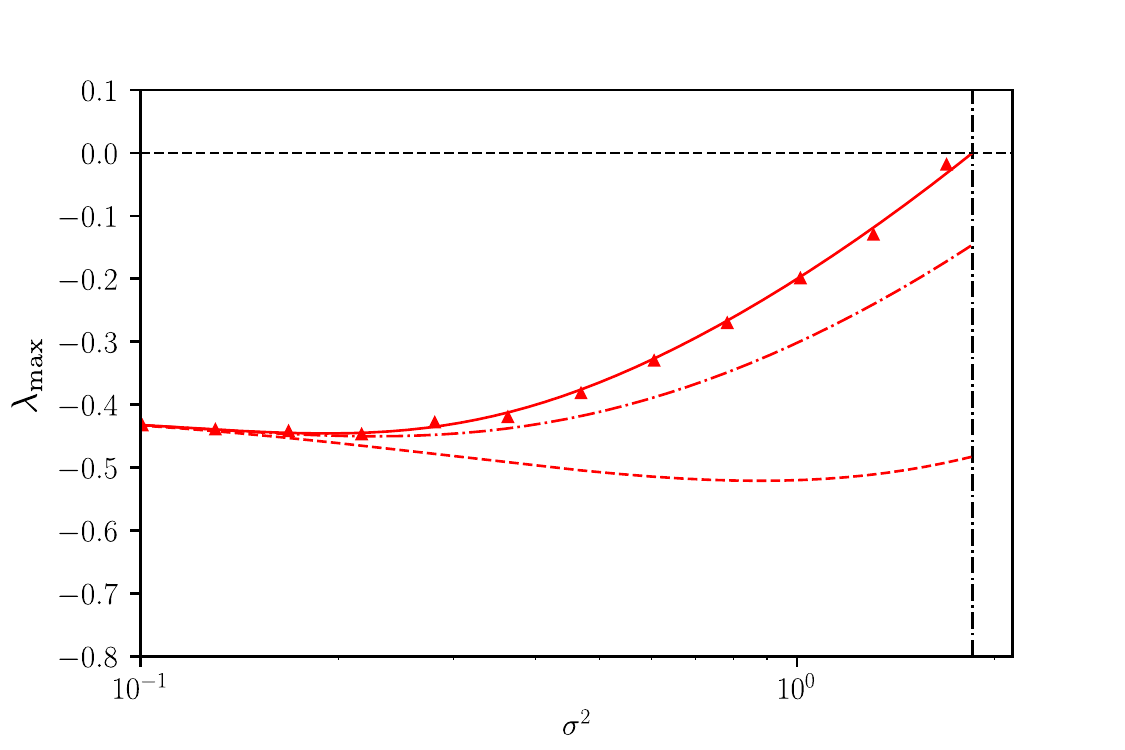}
	\caption{Outlier eigenvalue of the reduced interaction matrix as a function of $\sigma^2$, at fixed $\mu=0.6, \Gamma=-0.2$. Markers indicate the results of computer simulations ($N = 1000$, averaged over 10 trials). The solid line is from Eqs.~(S71)--(S73) of the SM, whereas the dashed line and dot-dashed lines are the two naive predictions $\lambda_0$ and $\lambda_1$ (respectively) given in the text. The vertical dot-dashed line marks the point at which $M\to \infty$ in the GVLE (see the solid lines in Fig. \ref{fig:phasediagram}).  }\label{fig:alttheories}
\end{figure}

If the universality principle were to apply to the reduced interaction matrix, then the Gaussian prediction $\lambda_1$ and the true outlier eigenvalue would coincide, whether or not the elements of the reduced interaction matrix were also Gaussian distributed. As can be seen in Fig.~\ref{fig:alttheories}, $\lambda_1$ is a better approximation than $\lambda_0$, but neither expression correctly predicts the outlier.

We now take into account the full statistics of the matrix elements $a'_{ij}$, as we did when calculating the bulk eigenvalue spectrum, and deduce the correct expression for the outlier eigenvalue. In the region of the complex plane outside the bulk (where the outlier resides), the resolvent can be expanded as a series in $1/\omega$ [Eq.~(S36) in the SM]. We evaluate each term in this series in terms of the statistics of species abundances, which are available to us via DMFT. This is accomplished via a generating-functional approach (SM Sec.~S4). 

Using diagrammatic techniques to recognise the self-similarity of the resulting series, we arrive at a compact formula for the resolvent [SM Eq.~(S69)]. Using Eq.~(\ref{bulkoutfromres}), we then obtain an implicit set of equations for the outlier eigenvalue in terms of the statistics of the surviving species abundances [see Eqs.~(S71)--(S73) in the SM]. We emphasise that in finding our final expression for the outlier, no approximations have been made other than assuming the thermodynamic limit. The simulation data in Figs. \ref{fig:lambdaversussigma} and \ref{fig:alttheories} verifies that the expression in Eqs.~(S71)--(S73) accurately predicts the outlier eigenvalue. 

We also demonstrate analytically (see SM Sec.~S4D) that this prediction for the outlier eigenvalue correctly predicts instability of the fixed point of the GLVE system. That is, $\lambda_{\mathrm{outlier}}$ crosses the imaginary axis precisely at locations in parameter space where the $M\to\infty$ transition occurs in the GLVEs. This is also verified in Figs. \ref{fig:lambdaversussigma} and \ref{fig:alttheories}. 

We thus conclude that stability cannot be predicted from the reduced interaction matrix using Gaussian random matrix results, even if all correlations are accounted for. This indicates that the extinction dynamics leads to some more intricate structure to the interactions in the surviving community. 

Advancing ideas in Refs. \cite{barbier2021fingerprints, bunin2016interaction}, we show in the SM (Sec.~S10) how one can generate the ensemble of reduced interaction matrices `from scratch' (i.e. without running the Lotka--Volterra dynamics and eliminating extinct species). This is achieved by first drawing a set of mock abundances from the known distribution of GLVE fixed-point abundances \cite{bunin2016interaction,Galla_2018}. Subsequently, one then draws interaction matrices from a carefully constructed distribution, which is dependent on the mock abundances. We verify in the SM that this bottom-up construction leads to non-Gaussian matrices with the same statistical properties and leading eigenvalue as the ensemble of true reduced interaction matrices. 

Having constructed the reduced interaction matrix ensemble in this way, we can thus see more clearly why universality fails to capture stability. The ensemble is manifestly non-Gaussian with complex interdependencies between matrix elements. By making a simple Gaussian assumption and ignoring the higher-order moments, one does not correctly take into account this intricate underlying structure.

Finally, we perform some additional tests of our results to demonstrate their robustness.  For example, realistic ecological communities might be composed of only a relatively small number of species. We have verified that our expression for the outlier in Eqs.~(S71)--(S73) of the SM is also a better predictor of stability than the more naive theories when $N=50$, leading to communities of about surviving $25$ species (Fig.~S4 in the SM). It has also been pointed out that heterogeneity of carrying capacities across species can significantly affect ecological equilibria \cite{bunin2016interaction, song2018will}. We show in Sec.~S9 of the SM that our conclusions continue to hold in such situations.

To conclude, we have deduced the stability of the generalised Lotka-Volterra system by calculating the eigenvalue spectrum of the interaction matrix of the surviving species. We have shown that results that are derived for Gaussian random matrices, which are often assumed also to apply to non-Gaussian ensembles, fail in this case. Instead, higher-order statistics of the reduced interaction matrix must be taken into account. We have therefore found a non-contrived class of random matrices for which the universality principle of RMT is not applicable. This demonstrates that there are limitations to results in RMT that are derived making an assumption of Gaussian interactions. Universality should therefore not be invoked without careful consideration.

Our results also have immediate relevance for the field of theoretical ecology. In the widely used approach pioneered by Robert May \cite{may, may71}, one supposes that the Jacobian governing small deviations of species abundances about a fixed point can be represented by a random matrix. May does not say what the dynamics are that lead to this Jacobian. One particular objection to this approach is hence that the statistics of May's random matrices do not necessarily correspond to `feasible' equilibria \cite{stone, gibbs, allesinatang1, gilpin1975stability}.

The fixed point of the GLVEs is feasible by construction. Therefore, our work shows that the stability of a feasible equilibrium in a complex ecosystem can be found by studying the eigenvalues of a random interaction matrix. Feasibility is reflected in the higher-order statistics of the interactions between species. Crucially, we find that these intricate statistics cannot be ignored if one is to correctly predict stability. 

\acknowledgements
JWB is grateful to M. A. Moore for insightful and helpful discussions. The authors also wish to thank to Guy Bunin and Lyle Poley for enlightening conversations. We acknowledge partial financial support from the Agencia Estatal de Investigaci\'on (AEI, MCI, Spain) and Fondo Europeo de Desarrollo Regional (FEDER, UE), under Project PACSS (RTI2018-093732-B-C21) and the Maria de Maeztu Program for units of Excellence in R\&D, Grant MDM-2017-0711 funded by MCIN/AEI/10.13039/501100011033.

\end{document}


\title{Breakdown of random-matrix universality in persistent Lotka--Volterra communities \\ \medskip \medskip{\Large--- Supplemental Material ---}}
	\author{Joseph W. Baron}
	\email{josephbaron@ifisc.uib-csic.es}
	\affiliation{Instituto de F{\' i}sica Interdisciplinar y Sistemas Complejos IFISC (CSIC-UIB), 07122 Palma de Mallorca, Spain}
	\author{Thomas Jun Jewell}
	\affiliation{Department of Physics and Astronomy, School of Natural Sciences,
		The University of Manchester, Manchester M13 9PL, United Kingdom}
	\author{Christopher Ryder}
	\affiliation{Department of Physics and Astronomy, School of Natural Sciences,
		The University of Manchester, Manchester M13 9PL, United Kingdom}
	\author{Tobias Galla}
	\email{tobias.galla@ifisc.uib-csic.es}
	\affiliation{Instituto de F{\' i}sica Interdisciplinar y Sistemas Complejos IFISC (CSIC-UIB), 07122 Palma de Mallorca, Spain}
	\affiliation{Department of Physics and Astronomy, School of Natural Sciences,
		The University of Manchester, Manchester M13 9PL, United Kingdom}

	\maketitle
	\tableofcontents
	\section{Dynamic mean-field theory and phase transitions}\label{section:dmft}
	For completeness, we show in this section how dynamic mean field theory can be used to deduce which sets of interaction statistics of the original Lotka--Volterra community can give rise to stability. This has previously been described in \cite{Galla_2018}, see in particular the Supplementary Material of this earlier work. In the course of this calculation, we introduce the generating functional formalism and some quantities of interest that will be necessary for quantifying the statistics of the reduced interaction matrix later.
	
	\subsection{Effective process}
	\noindent We begin with the generalised Lotka-Volterra equations \cite{malcai2002theoretical, BRENIG1988378}
	\begin{align}
		\dot x_i = x_i \left[ 1 - x_i + \sum_{j}a_{ij} x_j + h_i(t)\right],\label{gles}
	\end{align}
	where $h_i(t)$ is an external field, which is included for the purposes of the calculation, but which is later set to zero (the fields are therefore not a part of the model as such). The original interaction matrix elements have the following statistics
	\begin{align}
		\overline{ a_{ij} } &= \mu/N, \nonumber \\
		\overline{ (a_{ij} - \mu/N)^2 } &= \sigma^2/N, \nonumber \\
		\overline{ (a_{ij} - \mu/N)(a_{ji} - \mu/N) } &= \Gamma \sigma^2/N . \label{originalinteractionstats}
	\end{align}	
	The corresponding generating functional \cite{dedominicis1978dynamics}, from which the complete statistics of the process can be derived, is
	\begin{align}
		Z_0[\boldpsi] =& \int D[\mathbf{x}, \mathbf{\hat x}] \exp \left( i \sum_i \int dt \left[ \hat x_i(t) \left( \frac{\dot x_i(t)}{x_i(t)} - \left[ 1 - x_i(t) + \sum_j a_{ij} x_j(t) + h_i(t)\right]  \right) \right] \right) \nonumber \\
		&\times \exp\left(i \sum_i \int dt x_i(t) \psi_i(t) \right) . \label{genfunct0}
	\end{align}
	For later convenience, we define $\theta_i(t) = \Theta(x_i)$, where $\Theta(\cdot)$ is the Heaviside function. We also write $\theta_i=\lim_{t\to\infty} \theta_i(t)$ (in the phase where the system reaches a fixed point). Further, we introduce $\phi(t)$ as the fraction of species that are survive until time $t$ and $N_\ess$ the eventual number of surviving species respectively,
	\begin{align}
		\phi(t)&=\frac{1}{N}\sum_i \theta_i(t) , \nonumber \\
		N_{\ess} &= \lim_{t\to \infty} \sum_i \theta_i(t) . \label{phidef}
	\end{align}
	We write $\phi=N_{\cal S}/N$ for the asymptotic fraction of surviving species in the fixed point phase.

	Now, following for example Refs. \cite{sompolinsky1981dynamic, kirkpatrick1987p, opper1992phase} (especially Ref.~\cite{galla2018dynamically} in the context of the current problem), we perform a dynamic mean-field analysis. First one finds the disorder-averaged generating functional $\overline{Z_0[\boldpsi]}$, keeping only leading order terms in $N^{-1}$ in the exponent. We note that in taking the disorder average, we do not require $a_{ij}$ to be Gaussian random variables. Merely, we require that the higher moments of $a_{ij}$ decay sufficiently quickly with $N^{-1}$ so that we only need to include up to quartic order terms in $x_i$ and $\hat x_i$ \cite{baron2021eigenvalues, mezard1987}. 
	
	Then, by defining appropriate `order parameters' and performing a saddle-point approximation, which is valid in the thermodynamic limit $N\to\infty$, we find the following approximate expression for the generating functional
	\begin{align}
		\overline{Z_0[\boldpsi] }\approx& \prod_{i = 1}^N\Bigg[\int D[x_i,\widehat x_i] \exp\left(i\int dt~ \psi_i(t)x_i(t)\right)\nonumber \\
		&\times \exp\left(i\int dt ~\widehat x_i(t)  \left[\frac{\dot x_i(t)}{x_i(t)}-1+x_i(t) - \mu M(t) + \Gamma \sigma^2 \int dt' G(t,t') x(t') -h_i(t)\right]\right)\nonumber \\
		&\times \exp\left(-\frac{\sigma^2}{2}\int dt ~ dt'  C(t,t')\widehat x_i(t)\widehat x_i(t') \right)\Bigg], \label{nfoldgenfunct}
	\end{align}
	where we note that each species is now statistically equivalent. The quantities $M(t)$, $G(t,t')$ and $C(t,t')$ are defined self-consistently via
	\begin{align}
		M(t) &\equiv \langle x_i(t)\rangle, \nonumber \\
		R(t,t') &\equiv \left\langle \frac{\delta x_i(t)}{\delta h_i(t')}\right\rangle = -i\left\langle x_i(t)\hat x_i(t')\right\rangle , \nonumber \\
		C(t,t') &\equiv \langle \eta_i(t) \eta_i(t') \rangle = \langle x_i(t) x_i(t') \rangle,
	\end{align}
	where the angular brackets represent averages with respect to the disorder-averaged generating functional
	\begin{align}
		\langle \cdots \rangle =& \int D[\mathbf{x}, \mathbf{\hat x}] \left[ \cdots \right] \overline{\exp \left( i \sum_i \int dt \left[ \hat x_i(t) \left( \frac{\dot x_i(t)}{x_i(t)} - \left[ 1 - x_i(t) + \sum_j a_{ij} x_j(t) \right]  \right) \right] \right) } \nonumber \\
		\approx & \frac{1}{\Omega} \int D[\mathbf{x}, \mathbf{\hat x}]  \left[ \cdots \right]  \exp\left(-\frac{\sigma^2}{2}\int dt ~ dt' \sum_i C(t,t')\widehat x_i(t)\widehat x_i(t') \right) \nonumber \\
		&\times \exp\left(i\int dt ~ \sum_i \widehat x_i(t)  \left[\frac{\dot x_i(t)}{x_i(t)}-1+x_i(t) - \mu M(t) + \Gamma \sigma^2 \int dt' G(t,t') x(t') \right]\right) , \label{angularbracketdef}
	\end{align}
	where $\Omega$ is a normalisation constant. We also find that for large $N$
	\begin{align}
		\phi(t) &= \langle \theta_i(t) \rangle, \nonumber \\
		N_\ess &= N \lim_{t\to \infty} \phi(t).
	\end{align}
	It is in performing the saddle-point calculation that the necessity for the scaling with $N$ of the statistics in Eq.~(\ref{originalinteractionstats}) becomes apparent. If we had chosen instead, for example, $\overline{a_{ij}} = \mu$, then the term $\mu M$ in Eq.~(\ref{nfoldgenfunct}) would instead be $N \mu M$, which is of the order $N$. This would mean that this term would dominate the argument of the exponential in the limit $N\to \infty$.
	
However, once the saddle-point calculations have been performed, and using the observation that predictions formally derived in the limit $N\to\infty$ also hold to a good approximation for finite $N$ (see Sec.~\ref{sec:smaller_N}), one can map our choice of moments to that usually made in theoretical ecology as was done for example in Refs. \cite{baron2020dispersal, allesinatang2}. This is described briefly in Sec.~\ref{sec:comment} below.

	We thus see that in the thermodynamic limit, the disorder-averaged generating functional can be written as the product of $N$ identical generating functionals. From the form of these factors one can deduce that each species can be approximated as obeying a self-consistent stochastic process of the form
	\begin{align}
		\dot x_i &= x_i \left[ 1- x_i + \mu M(t) +\Gamma\sigma^2 \int dt' R(t,t') x_i(t') + \sigma \eta_i(t)  \right] , \quad \langle \eta_i(t) \eta_i(t') \rangle  = C(t,t'), \label{effectiveprocess}
	\end{align}
	where we use the fact that the angular brackets can also be thought of as averages over realisations of the coloured noise $\eta_i(t)$. Similar effective single-species dynamics have also been obtained using the cavity approach \cite{froy}. We note that the response function $R(t,t')$ and the average abundance $M(t)$ are also to be obtained self-consistently as averages over realisations of the process in Eq.~(\ref{effectiveprocess}). We note further that the site index $i$ serves no further purpose in Eq.~(\ref{effectiveprocess}), we will therefore drop this index from now on.
	
	For the sake of later analysis, we also define the following response functions
	\begin{align}
		T(t,t') &\equiv \left\langle \frac{\delta \theta(t) }{\delta h(t')} \right\rangle = -i\langle \theta_i(t) \hat x_i(t') \rangle , \nonumber \\
		T_2(t,t') &\equiv \left\langle \frac{\delta^2 \theta(t) }{\delta h(t') \delta h(t'')} \right\rangle = -\langle \theta(t) \hat x(t') \hat x(t'') \rangle . \label{additionalresponse}
	\end{align}
\subsection{Comment on scaling of moments of the interaction matrix elements with $N$}\label{sec:comment}

For the statistics of the original interaction matrix elements, we use a mean $\mu/N$ and a variance $\sigma^2/N$ [see Eq.~(\ref{originalinteractionstats})]. Many studies in theoretical ecology (see the discussion below for examples) would use $\mu_E$ and $\sigma^2_E$ without the scaling with $N$ (the subscript stands for `ecology'). There is therefore a direct mapping between the two parameterisations, $\mu =N\mu_{E}$ and $\sigma^2=N\sigma^2_{E}$. The parameter $\Gamma$ does not undergo any change, $\Gamma=\Gamma_{E}$ \cite{baron2020dispersal}.

In this way, any combination of $N, \mu_E$ and $\sigma_E^2$ can be mapped onto a set of parameters with the physics scaling ($\mu =N\mu_{E}$ and $\sigma^2=N\sigma^2_{E}$). Assuming that the theoretical predictions, formally derived in the limit $N\to\infty$, are valid as an approximation also for finite $N$ (we confirm this in Sec.~\ref{sec:smaller_N}), the phase diagram obtained in terms of the physics parameters (Fig.~1 in the main paper), can then be used to decide if the system is stable or not.

For example, for a given $\mu=N\mu_E$, the generating-functional calculation will generally indicate that the system becomes unstable at a value of $\sigma^2=\sigma_c^2$, with $\sigma_c^2$ some `critical' value, which will in general depend on $\Gamma$. Assuming that the results apply (as an approximation) to systems with finite $N$ one can then conclude that the system becomes unstable when $N\sigma_{E}^2=\sigma_c^2$.

Similar principles are indeed used by May \cite{may}, and Allesina and Tang \cite{allesinatang2}. Allesina and Tang for example make use of earlier results by Sommers et al in the physics literature \cite{sommers} on the spectra of random matrices. These results are also derived in the thermodynamic limit and using a scaling of the moments of the random matrix with $N$. Allesina and Tang then convert this into the parameter set that is commonly used in theoretical ecology, using a similar transformation as above, and assuming that the results in \cite{sommers} hold for finite $N$. This then generates the explicit factors of $N$ in the resulting stability criteria. 

\subsection{Fixed-point analysis}\label{sec:fp_analysis}
We now wish to construct the stability plot in Fig. 1 in the main text, following \cite{Galla_2018}. First, we note that the fixed point quantities defined in Eqs.~(2) and (3) of the main text are given by
\begin{align}
\phi &=  \lim_{N\to\infty}\frac{1}{N} \sum_i \overline{ \theta_i^\star }, \nonumber \\
M &=  \lim_{N\to\infty}\frac{1}{N} \sum_i  \overline{ x_i^\star }, \nonumber \\
q &=  \lim_{N\to\infty}\frac{1}{N} \sum_i  \overline{ x_i^\star x_i^\star }, \label{fixedpointquantities1}
\end{align}
and
\begin{align}
\chi &= \lim_{t\to\infty}  \lim_{N\to\infty}\frac{1}{N} \sum_i  \int_0^t dt'  \left.\frac{\delta \overline{x_i(t)}}{\delta h_i(t')}\right|_{h_i(t')=0} , \nonumber \\
\chi_{_T} &= \lim_{t\to\infty} \lim_{N\to\infty}\frac{1}{N} \sum_i  \int_0^t dt' \left.\frac{\delta \overline{\theta_i(t)}}{\delta h_i(t')}\right|_{h_i(t')=0}, \nonumber \\
\chi_2 &= \lim_{t\to\infty}  \lim_{N\to\infty}\frac{1}{N} \sum_i  \int_0^t \int_0^t dt' dt'' \left.\frac{\delta^2 \overline{\theta_i(t)}}{\delta h_i(t') \delta h_i(t'')} \right|_{h_i(t')=h_i(t'')=0}. \label{fixedpointquantities2}
\end{align}	
respectively. These quantities can also be written in terms of averages over realisations of the effective dynamics 
	\begin{align}
		\chi &= \lim_{t\to\infty} \int_0^t dt'\left\langle \frac{1}{\sigma}\frac{\delta x(t)}{\delta \eta(t')} \right\rangle, \nonumber \\
		\chi_{{}_T} &= \lim_{t\to\infty} \int_0^t dt'\left\langle \frac{1}{\sigma}\frac{\delta \theta(t)}{\delta \eta(t')} \right\rangle, \nonumber \\
		\chi_2 &= \lim_{t\to\infty}  \int_0^t \int_0^t dt' dt''\left\langle \frac{1}{\sigma^2}\frac{\delta^2 \theta(t)}{\delta \eta(t') \delta \eta(t'')} \right\rangle, \nonumber \\
		\phi &= \lim_{t\to\infty}\langle \theta(t) \rangle, \nonumber \\
		q &= \lim_{t\to \infty}\langle x(t) x(t') \rangle, \nonumber \\
		M &= \lim_{t\to \infty}\langle x(t)  \rangle . \label{fixedpointquantities}
	\end{align}
	Setting $\dot x = 0$ in Eq.~(\ref{effectiveprocess}) after dropping the index $i$, we thus obtain the following expression for the fixed points of the surviving species
	\begin{align}
		x^\star = \frac{1 + \mu M + \sigma \sqrt{q} z  }{1 - \Gamma \sigma^2 \chi} \Theta\left( \frac{1 + \mu M + \sigma \sqrt{q} z  }{1 - \Gamma \sigma^2 \chi}   \right), \label{fp}
	\end{align}
	where $z$ is a Gaussian random variable with zero mean and unit variance. Following \cite{Galla_2018}, this then leads to the self-consistency relations (with $Dz=\frac{dz}{\sqrt{2\pi}}e^{-z^2/2}$)
	\BE\label{eq:sc}
	\chi&=&\frac{1}{1-\Gamma\sigma^2\chi}\int_{-\infty}^\Delta Dz, \nonumber \\
	M&=& \frac{\sqrt{q}\sigma}{1-\Gamma\sigma^2\chi}\int_{-\infty}^\Delta Dz ~(\Delta-z),\nonumber\\
	1&=&\frac{\sigma^2}{(1-\Gamma\sigma^2\chi)^2}\int_{-\infty}^\Delta Dz~ (\Delta-z)^2, \label{fpquants}
	\EE
	where $\Delta=\frac{1+\mu M}{\sqrt{q}\sigma}$. We also have 
	\begin{align}
		\phi&=\int_{-\infty}^\Delta Dz, \nonumber \\
		\chi_{{}_T} &= \frac{d \phi}{dh} = \frac{1}{\sigma \sqrt{2 \pi q}} e^{-\Delta^2/2}, \nonumber \\
		\chi_2 &= \frac{d^2 \phi}{dh^2} = -\frac{\Delta}{\sigma^2 q \sqrt{2 \pi }} e^{-\Delta^2/2} . \label{addintrep}
	\end{align}
	For positive integers $\ell$ we now define the following truncated Gaussian integrals
	\be\label{eq:w_general}
	w_\ell(\Delta)\equiv \int_{-\infty}^\Delta Dz~ (\Delta-z)^\ell.
	\ee
	Explicitly, we have
	\BE\label{eq:w}
	w_0(\Delta) & = & \frac{1}{2}\left[1+\erf\left(\frac{\Delta}{\sqrt{2}}\right)\right], \nonumber \\
	w_1(\Delta) & = & \frac{1}{2}\left[e^{-\Delta^2/2}\sqrt{\frac{2}{\pi}}+\Delta\left(1+\erf\left(\frac{\Delta}{\sqrt{2}}\right)\right)\right], \nonumber \\
	w_2(\Delta) & = & \frac{1}{2} \left(1+\Delta^2\right) \left[1+\erf\left(\frac{\Delta}{\sqrt{2}}\right)\right] +\frac{1}{\sqrt{2\pi}}e^{-\Delta^2/2}\Delta.
	\EE
	One also has the relation
	\be\label{eq:w_relation}
	w_2(\Delta)  =  w_0(\Delta)+\Delta w_1(\Delta).
	\ee
	After some algebra, we derive from Eqs.~(\ref{fpquants}) a single equation that we can solve to find $\Delta$ for a given $(\mu, \sigma, \Gamma)$ [the interaction statistics of the original community -- see Eq.~(\ref{originalinteractionstats})]. That is, we solve the following numerically for $\Delta$
	\begin{align}
		\sigma^2&=\frac{w_2(\Delta)}{[w_2(\Delta)+\Gamma w_0(\Delta)]^2}. \label{sigmasol}
	\end{align}
	We see therefore that $\Delta$ is independent of $\mu$. We can then obtain the remaining fixed-point order parameters by substituting this value of $\Delta$ into
	\BE
	\chi&=&w_0+\Gamma\frac{w_0^2}{w_2}, \nonumber \\
	\frac{1}{M}&=&\frac{\Delta}{w_1}\frac{w_2}{w_2+\Gamma w_0}-\mu, \nonumber \\
	q&=&\left(\frac{M}{\sigma w_1} \frac{w_2}{w_2 + \Gamma w_0} \right)^2,\nonumber\\ 
	\phi&=&w_0,\nonumber \\
	\chi_{{}_T} &=& \frac{1}{\sigma \sqrt{q}} (w_1 - \Delta w_0),\nonumber \\
	\chi_2 &=& -\frac{\Delta}{\sigma^2 q} (w_1 - \Delta w_0). \label{summaryeqs}
	\EE
	\subsection{Transitions}
	The validity of the fixed point solution can break down in two different ways, indicating the onset of instability.
	\subsubsection{Diverging abundances}\label{section:minftrans}
	One transition occurs when the average fixed-point abundance diverges, i.e. $M \to \infty$. Consulting Eqs.~(\ref{sigmasol}) and (\ref{summaryeqs}), the sets of points at which this transition occurs (for a fixed $\Gamma$) obey
	\begin{align}
		\mu &= \frac{\Delta}{w_1}\frac{w_2}{w_2 + \Gamma w_0}, \nonumber \\
		\sigma^2 &= \frac{w_2}{(w_2 + \Gamma w_0)^2} . \label{ferromagnetictransition}
	\end{align}
	These can be viewed as a parametric set of equations in $\Delta$ for the phase transition line in the $\mu$--$\sigma$ plane (with $\Gamma$ held fixed). From these equations, the solid lines in Fig. 1 in the main text can be produced.
	
	\subsubsection{Linear instability}\label{section:oppertrans}
	The other transition occurs when the fixed point becomes linearly unstable to perturbations. Linearising the effective process in Eq.~(\ref{effectiveprocess}) about its fixed point, we obtain for small perturbations $\epsilon(t) = x(t) - x^\star$ and $\delta\eta(t) = \eta(t) - \sigma\sqrt{q}z$ that arise from an external white noise $\xi(t)$
	\begin{align}
		\dot \epsilon &= x^\star \left[ - \epsilon +\Gamma\sigma^2 \int dt' G(t,t') \epsilon(t') + \sigma \delta\eta(t) + \xi  \right] ,
	\end{align}
	where $x^\star$ satisfies Eq.~(\ref{fp}) and $\langle \xi(t)\xi(t')\rangle = \delta(t-t')$ and $\langle \xi \rangle = 0$. Taking the Fourier transform (indicated by a tilde in the following), rearranging and taking the limit $\omega\to 0$ \cite{opper1992phase}, we find
	\begin{align}
		\lim_{\omega \to 0}\langle \vert \tilde\epsilon(\omega) \vert^2 \rangle &= \frac{1 }{(1 - \Gamma \sigma^2 \chi)^2/\phi- \sigma^2}.
	\end{align}
	The object on the right diverges when
	\begin{align}
		(1-\Gamma \sigma^2 \chi )^2 =& \phi \sigma^2,
	\end{align}
	indicating that our solution no longer holds and that the system becomes unstable to perturbations. Using Eqs.~(\ref{fpquants}) we therefore deduce
	\begin{align}
		\phi &= \frac{1}{\sigma^2(1+\Gamma)^2}, \nonumber \\
		\chi &= \frac{1}{\sigma^2(1+\Gamma)}. \label{chaoticcond}
	\end{align}
	Finally, using Eqs.~(\ref{summaryeqs}), we see that
	\begin{align}
		\chi \sigma^2 &= \frac{\phi}{w_2 + \Gamma \phi} = \frac{1}{1+\Gamma}, \nonumber \\
		\Rightarrow w_2 &= \phi, \nonumber \\
		\Rightarrow \phi &= 1/2 \nonumber \\
		\Rightarrow \Delta &= 0. \label{delta0}
	\end{align}
	So finally, substituting $\phi = 1/2$ into the first of Eqs.~(\ref{chaoticcond}), we see that an instability occurs when 
	\begin{align}
		\sigma^2 = \frac{2}{(1+\Gamma)^2}, \label{chaotictransition}
	\end{align}
	as previously derived in \cite{Galla_2018}. Using Eq.~(\ref{chaotictransition}), one obtains the dashed horizontal lines in Fig. 1 in the main text. 
	
	\section{Reduced interaction  matrix and Jacobian matrix}
	\subsection{Definitions of the matrices}
	We now introduce and discuss several different matrices. One is the full (or original) $N\times N$ interaction matrix $\underline{\underline{a}}$, with elements $a_{ij}$. A second matrix is what we call the `reduced interaction matrix', $\underline{\underline{a}}'$. This is obtained from the original interaction matrix by removing all rows and columns corresponding to extinct species, and by carrying out a shift of the diagonal elements by $-1$ to capture the $-x_i$ term inside the square bracket of the generalised Lotka-Volterra Eqs. (1) in the main paper. This reduced matrix is of size $N_S\times N_S$, where $N_S$ is the number of surviving species in the long run [see Eq.~(\ref{phidef})]. We denote the reduced interaction matrix elements by $a'_{ij} = (a_{ij} - \delta_{ij})$ for $i, j \in \mathcal{S}$, where $\mathcal{S}$ is the set of surviving species as $t\to \infty$. 
	
	Similarly, we also define the full and reduced Jacobian matrices of the GLVEs,  $\underline{\underline{J}}$ and $\underline{\underline{J}}'$ respectively. The (full) Jacobian of the system (about the fixed point $\underline{x}^\star$) takes the form
	\begin{align}\label{eq:jac}
		J_{ij} = \delta_{ij}\left[1 - x^\star_i + \sum_j a_{ij} x^\star_j\right] + x_i^\star \left(a_{ij} - \delta_{ij}\right),
	\end{align}
	where $i,j=1,\dots,N$. 
	
	We now imagine that (in a particular realisation) we re-arrange the species indices such that $i=1,\dots,N_S$ are the surviving species, and $i=N_S+1,\dots, N$ the extinct species. This can always be done retrospectively without loss of generality. The Jacobian can then be written in block form
	\be
	\underline{\underline{J}}=\left(\begin{array}{cc} \underline{\underline{J}}' & \underline{\underline{B}} \\ \underline{\underline{0}} & \underline{\underline{D}} \end{array}\right),
	\ee
	
	The reduced Jacobian $\underline{\underline{J}}'$ makes up the upper left $N_S\times N_S$ block. We label the lower right-hand $(N-N_\ess) \times (N - N_\ess)$ block $\underline{\underline{D}}$. The upper-right block is labelled $\underline{\underline{B}}$. For an extinct species $i$ we have $J_{ij}=0$ for all $j\neq i$ [Eq.~(\ref{eq:jac})]. Hence the block on the lower left is zero, and the matrix $\underline{\underline{D}}$ is diagonal.
	
	We hence have
	\begin{align}
		\mathrm{det}\left(\underline{\underline{J}}-\lambda\id_{N}\right) = \mathrm{det}(\underline{\underline{J}}'-\lambda\id_{N_S})\mathrm{det}(\underline{\underline{D}}-\lambda\id_{N-N_S}),
	\end{align}
	(where $\id_N$ is the identity matrix of size $N\times N$), and the eigenvalues of $\underline{\underline{J}}$ are given by the combined eigenvalues of $\underline{\underline{J}}'$ and $\underline{\underline{D}}$.

	We focus first on a species $i$ that goes extinct ($x_i^*=0$). For such a species, one finds $J_{ii}=1 -x_i^* + \sum_j a_{ij} x^\star_j<0$. Hence $\underline{\underline{D}}$ is a diagonal matrix with only negative diagonal entries, and so we need only consider the eigenvalues of $\underline{\underline{J'}}$ to determine stability.
	
\begin{figure}[H]
		\centering 
		\includegraphics[scale = 0.42]{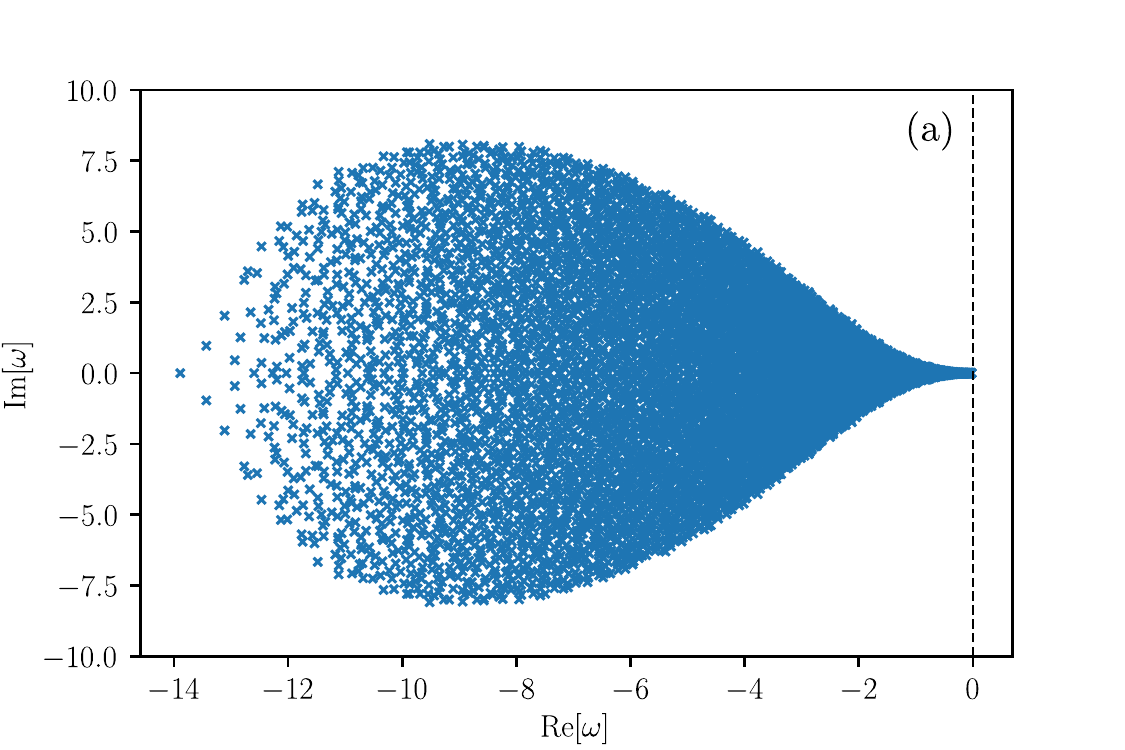}
		\includegraphics[scale = 0.42]{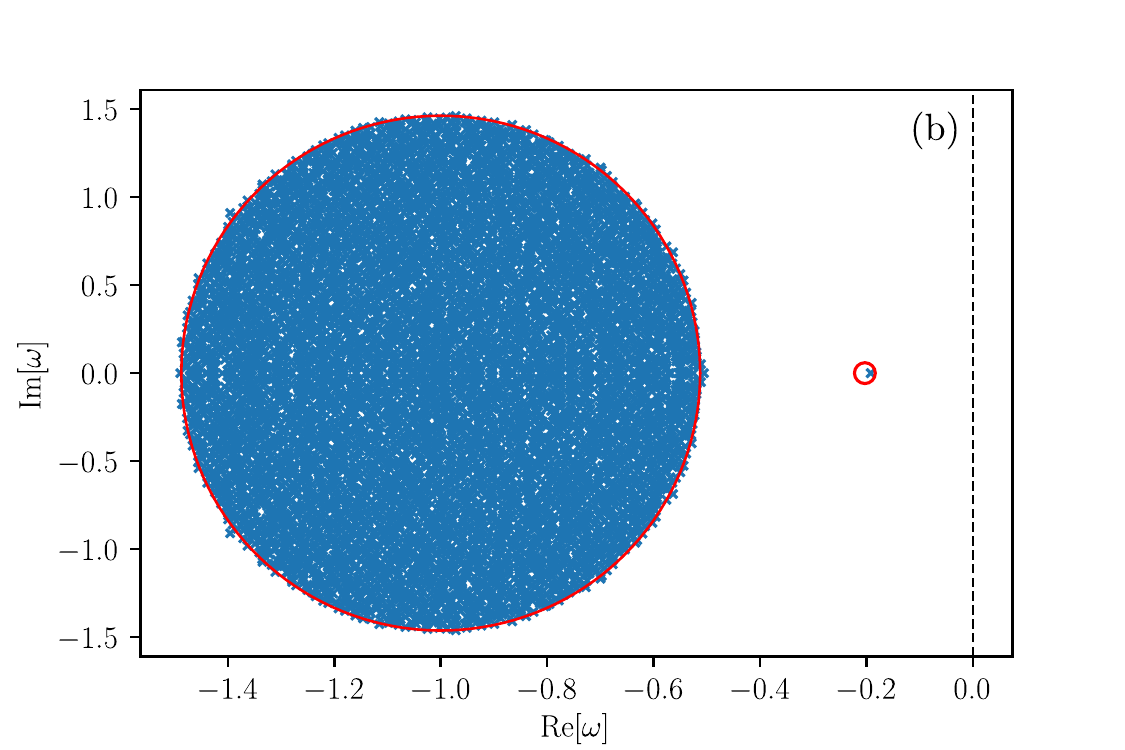}
		\caption{Panel (a): Example eigenvalue spectrum of the reduced Jacobian. Panel (b): Example spectrum of the reduced interaction matrix. The red line and circle in panel (b) show the analytical predictions for the bulk region and outlier eigenvalue in Eqs.~(\ref{bulkspectrum}) and (\ref{gsol}) respectively. Parameters are $\sigma = 1.1$, $\mu = 0.9$, $\Gamma = -0.5$, $N = 4000$. }\label{fig:examplespectra}
	\end{figure}

	Now we consider the reduced Jacobian $\underline{\underline{J}}'$. Given that $1 - x^\star_i + \sum_j a_{ij} x^\star_j = 0$ for values of $i$ corresponding to surviving species, we see from Eq.~(\ref{eq:jac}) that the reduced Jacobian matrix of the Lotka-Volterra system takes the simple form
	\begin{align}\label{eq:j_prod}
		J_{ij}' = x_i^\star a_{ij}' , 
	\end{align}
	where $i,j\in{\cal S}$. Examples of eigenvalue spectra of the reduced Jacobian and the reduced interaction matrix  are given in Fig. \ref{fig:examplespectra}.  
	
	\subsection{Reduced Jacobian is not practical for determining stability}
	One notes that the eigenvalue spectrum of the reduced Jacobian comes arbitrarily close to the imaginary axis. This is observed for all values of the model parameters in the phase with a unique fixed point. This is due to the fact that the distribution of fixed-point abundances $x_i^\star$ of the surviving species comes arbitrarily close to zero. (The distribution of abundances of surviving species is a Gaussian clipped at zero, see e.g. \cite{Galla_2018}.) For this reason, it is not helpful to study the spectrum of the full or reduced Jacobian when determining stability -- one cannot identify points in parameter space at which one eigenvalue first touches the imaginary axis, or crosses into the right half of the complex plane. 
	
	\subsection{Spectrum of reduced interaction matrix determines stability}
	We now argue as to why we need only consider the eigenvalue spectrum of the reduced interaction matrix $\underline{\underline{a}}'$ when determining stability instead of the reduced Jacobian. We note for the following discussion that the leading eigenvalues of both the reduced Jacobian and the reduced interaction matrix are real.
	
	Eq.~(\ref{eq:j_prod}) indicates that the reduced Jacobian matrix can be written as the product of the reduced interaction matrix and a diagonal matrix of species abundances. The determinant of $\underline{\underline{J}}'$ is therefore the product of the determinants of these two matrices. The abundances in the diagonal matrix are strictly positive, and therefore $\mbox{sgn}( \det\,\underline{\underline{D}})=1$. Hence, $\mbox{sgn}( \det\,\underline{\underline{J}}')=\mbox{sgn}(\det\,\underline{\underline{a}}')$. If the determinant of the reduced Jacobian changes sign as parameters are varied (indicating loss of stability), so must therefore that of the reduced interaction matrix and vice versa. \smallskip
	
	Imagine now we start in a region of parameter space for which the fixed point is stable and that we then vary the model parameters. The fixed point becomes unstable at the point where the leading eigenvalue of the reduced interaction matrix becomes positive. Therefore, we can deduce the stability of the system by examining the eigenvalue spectrum of the reduced interaction matrix only. A similar argument to this was given in Ref. \cite{stone}. Crucially (as we will see), the leading eigenvalue of the reduced interaction matrix is genuinely negative in the stable regime (i.e., not infinitessimally close to the imaginary axis like that of the reduced Jacobian), and only reaches the axis at the point of instability. 
	
	\subsection{Components of the spectrum of the reduced interaction matrix}
	As illustrated in Fig. \ref{fig:examplespectra} (b), there is an elliptic `bulk' region of the complex plane, to which the majority of the eigenvalues of the reduced interaction matrix are confined, and a single outlier. We therefore write the eigenvalue density of the reduced interaction matrix in the form
	\begin{align}
		\rho(\omega) = \rho_{\mathrm{bulk}}(\omega) + \frac{1}{N_\ess}\delta(\omega - \lambda_{\mathrm{outlier}}).
	\end{align}
	In the following sections, we deduce both the bulk eigenvalue density and the location of the outlier eigenvalue. We show that the point in parameter space at which the outlier crosses the imaginary axis is given by Eq.~(\ref{ferromagnetictransition}). We also demonstrate that the  bulk spectrum crosses into the right half of the complex plane at the point described by Eq.~(\ref{chaotictransition}).
	
	\section{Finding the outlier eigenvalue -- general approach}
 
	The outlier eigenvalue, $\lambda_{\mathrm{outlier}}$, of the reduced interaction matrix by definition must obey
	\begin{align}
		\det\left( \lambda_{\mathrm{outlier}} \id_{N_\ess} - \underline{\underline{a}}'  \right) =0,
	\end{align}
	where $ \id_{N_\ess}$ is the identity matrix of size $N_\ess\times N_\ess$.
	
	Suppose we introduce a uniform matrix $\underline{\underline{u}}$ with all entries equal to $\nu/N_\ess$. Following Refs. \cite{orourke, BENAYCHGEORGES2011494} and using Sylvester's determinant identity, one finds
	\begin{align}
		\det\left( \id_{N_\ess} - \frac{\nu}{N_\ess}  \underline{\underline{G}}\right) =  1 - \frac{\nu}{N_\ess} \sum_{ij} G_{ij}(1+\lambda_{\mathrm{outlier}}) = 0, \label{outliercalcgeneral}
	\end{align}
	where we have introduced the resolvent matrix $\underline{\underline{G}} (1 + \lambda_{\mathrm{outlier}}) = [ \lambda_{\mathrm{outlier}} \id_{N_\ess} - (\underline{\underline{a}}'  - \underline{\underline{u}})]^{-1}$. Thus, to find the outlier eigenvalue, one has to find the resolvent matrix and solve Eq.~(\ref{outliercalcgeneral})  for $\lambda_{\mathrm{outlier}}$. We stress that all elements of the resolvent are required in Eq.~(\ref{outliercalcgeneral}), not only the diagonal entries.
	
	We note that we have the freedom to choose the value of $\nu$, as long as it is non-zero and $ \lambda_{\mathrm{outlier}} \id_{N_\ess} - (\underline{\underline{a}}'  - \underline{\underline{u}})$ remains invertible. We exploit this freedom to simplify the calculation of the resolvent in the next section.
	
	\section{Using the generating functional to find the resolvent of the reduced interaction matrix} \label{section:outlierresolvent}
	\subsection{Series expansion for the resolvent of the reduced interaction matrix}
	To simplify our calculation of the resolvent matrix, we choose $\nu = \phi \mu$ in Eq.~(\ref{outliercalcgeneral}). Letting $z_{ij} = \lim_{t\to\infty}[a_{ij} - \mu N^{-1}]\theta_i \theta_j$ [see the discussion preceding Eq.~(\ref{phidef}) for a definition of $\theta_i$], we see that the disorder-averaged resolvent matrix that we must evaluate to find the outlier can be expressed as the following series
	\begin{align}
		\mathcal{G}[\omega] &\equiv \overline{N_\ess^{-1}\sum_{i,j \in \mathcal{S}}  G_{ij}(\omega)}  = \overline{N_\ess^{-1}  \sum_{i,j \in \mathcal{S}} \left[ \delta_{ij}\omega - z_{ij}\right]^{-1} } \nonumber \\
		&= \overline{N_\ess^{-1}\sum_{ij} \left[ \omega^{-1} \delta_{ij} \theta_i \theta_j + \omega^{-2} z_{ij} + \omega^{-3} \sum_k z_{ik} z_{kj} + \cdots \right]}, \label{resolventexpansion0}
	\end{align}
	where sums over $i,j \in \mathcal{S}$ denote a sum over the reduced interaction matrix elements, whereas sums over $ij$ indicate a sum over all elements of the original interaction matrix. 
	
	To find the terms of this series, we now construct the following generating functional
	\begin{align}
		Z[\boldpsi, \blambda] &= \int D[\mathbf{x}, \mathbf{\hat x}] \exp \left( i \sum_i \int dt \left[ \hat x_i(t) \left( \frac{\dot x_i(t)}{x_i(t)} - \left[ 1 - x_i(t) + \sum_j a_{ij} x_j(t) + h_i(t)\right]  \right) \right] \right) \nonumber \\
		&\times \exp\left( -i \int dt \, \sum_{ij} \lambda_{ij}(t) [a_{ij}- \mu N^{-1} ] \theta_i(t) \theta_j(t)\right)\exp\left(i \sum_i \int dt x_i(t) \psi_i(t) \right) . \label{genfunct}
	\end{align}
	This generating functional has the same form as in Eq.~(\ref{genfunct0}), with the addition of another source term containing the auxiliary variables $\lambda_{ij}(t)$, which we introduce in this step. The dynamics of $x_i(t)$ are still constrained to follow the Lotka-Volterra equations in Eq.~(\ref{gles}), but by functionally differentiating with respect to $\lambda_{ij}(t)$, we can obtain the terms in the series in Eq.~(\ref{resolventexpansion0}). For example,
	\begin{align}
		\overline{\frac{\delta Z}{\delta \lambda_{ij}(t)} }\bigg\rvert_{\boldpsi = 0, \lambda = 0} = -i \overline{   [a_{ij}- \mu N^{-1} ] \theta_i(t) \theta_j(t) }=-i\overline{z_{ij}}.
	\end{align}
	
	We now find for the disorder-averaged resolvent [from which we can find the outlier eigenvalue via Eq.~(\ref{outliercalcgeneral})]
	\begin{align}
		\mathcal{G}[\omega] =  \overline{N_\ess^{-1}\sum_{ij} \lim_{t \to \infty}\left[\omega^{-1} \delta_{ij} \theta_i \theta_j + i\omega^{-2} \frac{\delta Z}{\delta \lambda_{ij}(t)} - \omega^{-3} \sum_k \frac{\delta ^2 Z}{\delta \lambda_{ik}(t) \delta \lambda_{kj}(t)} + \cdots \right]} \bigg\rvert_{\boldpsi = 0, \lambda = 0} .\label{resolventexpansion}
	\end{align}

	\subsection{Evaluating the series for the resolvent}\label{section:resolventserieseval}
	\subsubsection*{Setup and strategy for evaluation of the series}
	To find the terms of the series in Eq.~(\ref{resolventexpansion}), we begin by calculating the following average that appears in the expression for $\overline{Z[\boldpsi, \blambda]}$
	\begin{align}
		A &= \overline{\exp \left( i \sum_i \int dt \left[ - \sum_j a_{ij} \hat x_i(t) x_j(t) \right] \right)\exp\left( -i \int dt \, \sum_{ij} \lambda_{ij}(t) [a_{ij} -\mu/N]\theta_i(t) \theta_j(t)\right)} \nonumber \\
		&= \exp \left( -i \sum_i \int dt \left[  \sum_j \frac{\mu}{N} \hat x_i(t) x_j(t) \right] \right)\nonumber \\
		&\times \exp\left(- \frac{\sigma^2}{2 N} \sum_{ij} \left[\int dt \, \hat x_i(t) x_j(t) + \lambda_{ij}(t) \theta_i(t) \theta_j(t) \right]^2 \right)\nonumber \\
		&\times\exp\left(- \frac{\sigma^2}{2 N} \sum_{ij} \Gamma \left[ \int dt \, \hat x_i(t) x_j(t) + \lambda_{ij}(t) \theta_i(t) \theta_j(t)\right] \left[ \int dt \, \hat x_j(t) x_i(t) + \lambda_{ji}(t) \theta_i(t) \theta_j(t) \right]\right).
	\end{align}
	One thus finds that the derivatives in Eq.~(\ref{resolventexpansion}) can be written as, for example,
	\begin{align}
		\overline{\frac{\delta ^2 Z}{\delta \lambda_{ik} \delta \lambda_{kj}}} \bigg\vert_{\psi = 0, \lambda = 0} =  \left\langle \frac{1}{A} \frac{\delta ^2 A}{\delta \lambda_{ik} \delta \lambda_{kj}} \right\rangle \bigg\vert_{\lambda = 0} ,
	\end{align}
	where we note that there are two kinds of averages here: an average over realisations of the interaction coefficients represented by $\overline{\cdots}$ and an average over the dynamics enforced by the disorder averaged generating functional denoted by angular brackets $\langle \cdots \rangle$ [see Eq.~(\ref{angularbracketdef})].
	
	The series in Eq.~(\ref{resolventexpansion}) can therefore be rewritten
	\begin{align}
		\mathcal{G}[\omega] &=  \frac{1}{\phi N}\sum_{ij} \lim_{t \to \infty}\left[\frac{ \delta_{ij} \theta_i \theta_j}{\omega} + \frac{i}{\omega^2} \left\langle\frac{1}{A}\frac{\delta A}{\delta \lambda_{ij}(t)} \right\rangle - \frac{1}{\omega^3} \sum_k \left\langle \frac{1}{A}\frac{\delta ^2 A}{\delta \lambda_{ik}(t) \delta \lambda_{kj}(t)} \right\rangle + \cdots \right]  \bigg\vert_{\lambda = 0}. \label{resolventexpansion2}
	\end{align}
	Let us now begin to construct the series for the resolvent in Eq.~(\ref{resolventexpansion2}). Consider the derivatives of $A$:
	\begin{align}
		B_{ij}(t) &\equiv \frac{1}{A}\frac{\delta A}{\delta \lambda_{ij}(t)} \nonumber \\
		&= - \Bigg[ \theta_i(t) \theta_j(t) \frac{\sigma^2}{N} \int dt' \hat x_i(t') x_j(t') +  \theta_i(t) \theta_j(t) \frac{\sigma^2}{N} \int dt' \theta_i(t') \theta_j(t') \lambda_{ij}(t') \nonumber \\
		&+  \theta_i(t) \theta_j(t) \frac{\Gamma\sigma^2}{N} \int dt' \hat x_j(t') x_i(t')+ \theta_i(t) \theta_j(t) \frac{\Gamma\sigma^2}{N} \int dt' \theta_i(t') \theta_j(t') \lambda_{ji}(t') \Bigg] , \nonumber \\
		\frac{\delta^2 A}{\delta \lambda_{ik}(t)\delta \lambda_{kj}(t)} &= \left[\frac{\delta B_{ik}(t)}{\delta \lambda_{kj}(t)} + B_{ik} B_{kj} \right] A, \nonumber \\
		\frac{\delta^3 A}{\delta \lambda_{ik}(t)\delta \lambda_{kl}(t)\lambda_{lj}(t)} &= \left[ B_{ik}\frac{\delta B_{kl}(t)}{\delta \lambda_{lj}(t)}  + \frac{\delta B_{ik}(t)}{\delta \lambda_{kl}(t)} B_{lj}+ B_{ik} B_{kl}B_{lj} \right] A, \nonumber \\
		\frac{\delta^4 A}{\delta \lambda_{ik}(t)\delta \lambda_{kl}(t)\lambda_{lm}(t)\lambda_{mj}(t)} &= \Bigg[ \frac{\delta B_{ik}(t)}{\delta \lambda_{kl}(t)} \frac{\delta B_{lm}(t)}{\delta \lambda_{mj}(t)} + \frac{\delta B_{ik}(t)}{\delta \lambda_{mj}(t)} \frac{\delta B_{lm}(t)}{\delta \lambda_{kl}(t)}+ B_{ik} B_{kl}\frac{\delta B_{lm}(t)}{\delta \lambda_{mj}(t)}  \nonumber \\
		& + B_{ik} \frac{\delta B_{kl}(t)}{\delta \lambda_{lm}(t)}B_{mj} + \frac{\delta B_{ik}(t)}{\delta \lambda_{kl}(t)}  B_{lm}B_{mj}  + B_{ik} B_{kl}B_{lm}B_{mj} \Bigg] A, \label{derivatives}
	\end{align}
	where terms of higher order in $N^{-1}$ have been omitted. 
	
	The series for the resolvent in Eq.~(\ref{resolventexpansion2}) is a complicated mixture of terms with $B_{ij}$ and its derivatives appearing in various combinations. Manifestly, all second order or higher order derivatives of $B_{ij}$ evaluate to zero since $B_{ij}$ is linear in $\lambda_{ij}$ and $\lambda_{ji}$, but some terms with first derivatives are non-vanishing in the thermodynamic limit. 
	
	Our strategy for evaluating the series for the resolvent is as follows. We first consider the terms involving derivatives of $B_{ij}$ with respect to $\lambda_{kl}$ and use diagrammatic methods to understand the structure of the surviving terms in the thermodynamic limit. We use this to show that the series in Eq.~(\ref{resolventexpansion2}) can be rewritten partly in terms of the resolvent of an ensemble of random matrices with an elliptic spectrum of the type described in Ref. \cite{sommers}. The complexity of the series can therefore be greatly simplified, see Eq.~(\ref{simplifiedseries}) below. In particular, the resulting expression for the series contains averaged products of the objects $B_{ij}$ only.
	
	In a second step, we show that these surviving terms can be written in terms of the fixed-point quantities in Eqs.~(\ref{fixedpointquantities}). We then construct an auxiliary diagrammatic formalism to aid us in spotting the self-similarity of the series. This ultimately enables us to perform the summation [see Eq.~(\ref{dequations})] and find a compact expression for the outlier eigenvalue in terms of the fixed point quantities [see Eq.~(\ref{gsol})].
	
	\subsubsection*{Terms with derivatives of $B_{ij}$ with respect to $\lambda_{kl}$}\label{subsection:withderivatives}
	Now, we are tasked with evaluating the derivatives of $A$ with respect to $\lambda_{ij}(t)$ in Eq.~(\ref{derivatives}). First consider the following expression that arises from $A^{-1}\frac{\delta^2 A}{\delta \lambda_{ik} \delta \lambda_{kj}}$
	\begin{align}
		\frac{1}{N \phi} \sum_{i,k, j} \left\langle  \frac{\delta B_{ik}}{\delta \lambda_{kj}}\right\rangle \bigg\vert_{\lambda = 0} &= -\frac{\Gamma \sigma^2}{\phi N^2} \sum_{i,k, j} \left\langle \delta_{ij} \theta_i \theta_k  \right\rangle + O(N^{-1}) = -\phi \Gamma \sigma^2 .
	\end{align}
	One notes that this is the same as $-\phi \Gamma \sigma^2 \times\frac{1}{N\phi} \sum_{ij} \theta_i \delta_{ij}$. Consider also the term that arises from $A^{-1}\frac{\delta^3 A}{\delta \lambda_{ik} \delta \lambda_{kl} \delta \lambda_{lj}}$
	\begin{align}
		\frac{1}{N \phi} \sum_{i,k,l, j} \left\langle  B_{ik} \frac{\delta B_{kl}}{\delta \lambda_{lj}}\right\rangle \bigg\vert_{\lambda = 0}\nonumber \\
		=& \frac{\Gamma \sigma^4}{\phi N^3}\sum_{i,k,l, j} \left\langle  \left[  \delta_{kj} \theta_i(t)\theta_l(t)  \theta_k(t) \left(\int dt' \hat x_i(t') x_k(t') + \Gamma \int dt' \hat x_k(t') x_i(t') \right)\right]\right\rangle \nonumber \\
		&+ O(N^{-1})\nonumber \\
		=& \frac{1}{\phi N}\sum_{i, j, k} (-\phi \Gamma \sigma^2) \delta_{kj}\left\langle B_{ik} \right\rangle \vert_{\lambda = 0} + O(N^{-1})\nonumber \\
		=& -\Gamma \sigma^2 \phi \times \frac{1}{N \phi} \sum_{i, j} \left\langle  B_{ij}\right\rangle \vert_{\lambda = 0} + O(N^{-1}).
	\end{align}

	We begin to see a pattern emerging: if $\frac{\delta B_{kl}}{\delta \lambda_{lj}}$ appears inside the angular brackets, it gives rise to a Kronecker delta function and a multiplicative factor. We note that terms like $\frac{1}{N \phi} \sum_{i,k,l, j} \left\langle  B_{kl} \frac{\delta B_{ik}}{\delta \lambda_{lj}}\right\rangle $ do not survive in the thermodynamic limit, since they give rise to too many Kronecker delta functions, meaning that the factors of $1/N$ are not cancelled when we perform the sums.
	
	Now let us examine examples of terms with more than one factor of $\frac{\delta B_{kl}}{\delta \lambda_{lj}}$. Consider for example the following terms that appear in $A^{-1}\frac{\delta^4 A}{\delta \lambda_{ik} \delta \lambda_{kl} \delta \lambda_{lm} \delta \lambda_{mj}}$
	\begin{align}
		\frac{1}{N\phi} \sum_{ijklm}\left\langle \frac{\delta B_{ik}(t)}{\delta \lambda_{kl}(t)} \frac{\delta B_{lm}(t)}{\delta \lambda_{mj}(t)} \right\rangle\bigg\vert_{\lambda = 0} &= \frac{1}{N\phi} \sum_{ijklm}\delta_{il} \frac{\Gamma \sigma^2}{N} \theta_i \theta_k\delta_{lj} \frac{\Gamma \sigma^2}{N} \theta_l\theta_m = \phi^2 \Gamma^2 \sigma^4 ,\nonumber \\
		\frac{1}{N\phi} \sum_{ijklm} \left\langle\frac{\delta B_{ik}(t)}{\delta \lambda_{mj}(t)} \frac{\delta B_{lm}(t)}{\delta \lambda_{kl}(t)} \right\rangle \bigg\vert_{\lambda = 0} &= \frac{1}{N\phi} \sum_{ijklm}\delta_{ij} \frac{\Gamma \sigma^2}{N} \theta_i \theta_k\delta_{ll} \frac{\Gamma \sigma^2}{N} \theta_l\theta_m = \phi^2 \Gamma^2 \sigma^4 . \label{4thorderexample}
	\end{align}
	Both of these contributions give rise to terms that survive in the thermodynamic limit. 
	
	We can understand which terms survive more easily with the aid of so-called rainbow diagrams \cite{brezin1994correlation, kuczala2016eigenvalue, JANIK1997603}. Representing each pair of indices that appear in the same object [e.g. $(i,k)$ in $B_{ik}$] with a pair of dots and joining indices that are constrained to be the same with lines, the above two terms in Eq.~(\ref{4thorderexample}) can be represented diagrammatically (respectively)
	\begin{figure}[H]
		\centering 
		\includegraphics[scale = 0.45]{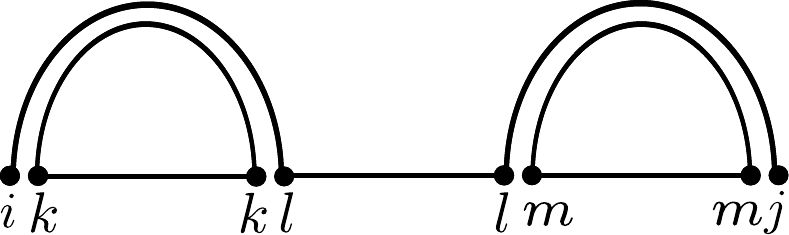}\\
		\vspace{4mm}
		\includegraphics[scale = 0.45]{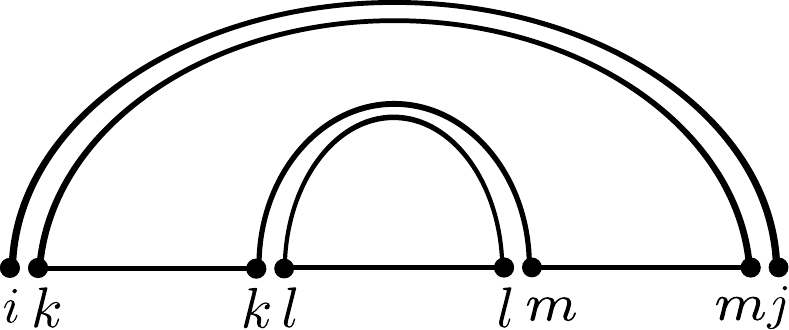}
		\captionsetup{labelformat=empty}\label{fig:feynman1and2}
	\end{figure}
	Horizontal lines join indices that are the same by construction. Arcs connect indices that are constrained to be the same by Kronecker deltas that arise from the derivatives $\frac{\delta B_{ik}(t)}{\delta \lambda_{mj}(t)} \propto \delta_{ij} \delta_{km}$. Only diagrams that are of a planar structure (i.e. those without intersecting arcs) survive in the thermodynamic limit. This is known as t'Hooft's theorem \cite{brezin1994correlation, HOOFT1974461, JANIK1997603, kuczala2019dynamics}.
	
	\subsubsection*{Summation of terms with derivatives of $B_{ij}$ with respect to $\lambda_{ij}$}\label{subsection:derivativessummation}
	Let us consider the sum of all the surviving terms in the series for $G(\omega) = \frac{1}{N\phi}\sum_{ij}\overline{G_{ij}(\omega)}$ that contain only derivatives of $B_{ij}$
	\begin{align}
		g(\omega) &\equiv \omega^{-1} -  \omega^{-3}\frac{1}{N \phi}\sum_{ijk} \left\langle \frac{\delta B_{ik}}{\delta \lambda_{kj}} \right\rangle\bigg\vert_{\lambda = 0} +  \omega^{-5}\frac{1}{N \phi}\sum_{ijklm} \left[ \left\langle \frac{\delta B_{ik}}{\delta \lambda_{kl}} \frac{\delta B_{lm}}{\delta \lambda_{mj}} \right\rangle+ \left\langle \frac{\delta B_{ik}}{\delta \lambda_{mj}} \frac{\delta B_{lm}}{\delta \lambda_{kl}}  \right\rangle  \right] \bigg\vert_{\lambda = 0} + \cdots \nonumber \\
		&= \frac{1}{\omega} -  \frac{ \phi\Gamma \sigma^2}{\omega^3}\frac{1}{N }\sum_{ijk} \delta_{ij}  +  \frac{\phi^2 \Gamma^2 \sigma^4}{\omega^5}\frac{1}{N }\sum_{ijklm} \left[\delta_{il} \delta_{lj} + \delta_{ij} \delta_{km} \delta_{mk}  \right] + \cdots. \label{derivativeseries}
	\end{align}
	This sum can be represented diagrammatically as the sum of all so-called planar diagrams
	\begin{figure}[H]
		\centering 
		\includegraphics[scale = 0.3]{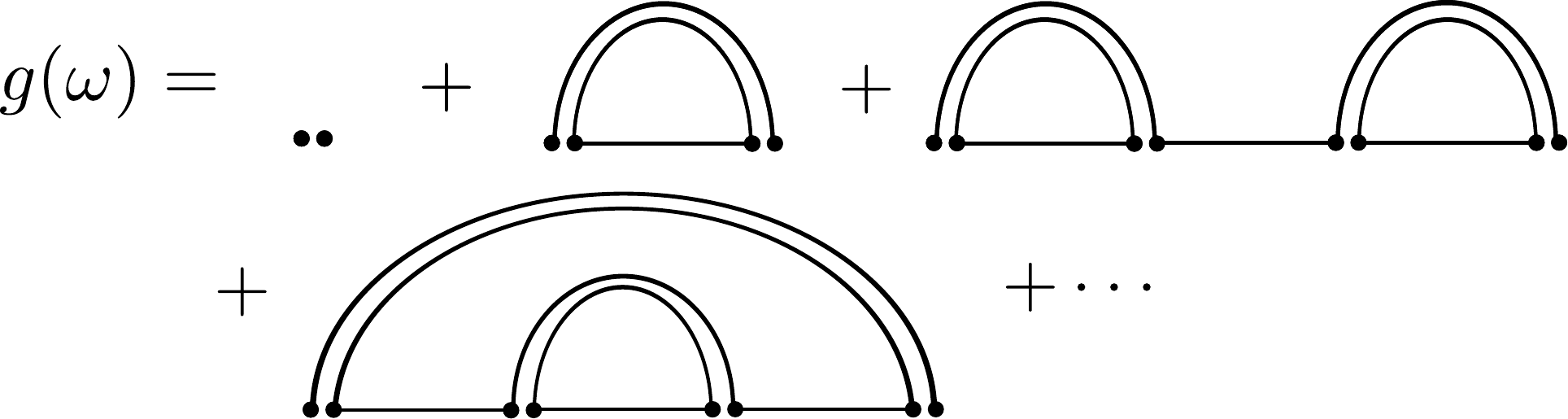}
		\captionsetup{labelformat=empty}\label{fig:series}
	\end{figure}
	This is exactly the same series of diagrams \cite{JANIK1997603, kuczala2019dynamics} as for the resolvent of the random matrices investigated by Ginibre with elliptic eigenvalue spectra \cite{ginibre1965} (once terms that vanish in the thermodynamic limit have been removed). That is, 
	\begin{align}
		g(\omega) = \frac{1}{\omega} - \frac{1}{\omega^3} \sum_{ik} \overline{y_{ik} y_{ki}}  + \frac{1}{\omega^5} \sum_{iklm}\overline{y_{ik} y_{kl} y_{lm} y_{mi}}  + \cdots
	\end{align}
	where $\overline{y_{ik}} = 0$ and $\overline{y_{ik}y_{lm}} =\frac{\phi\sigma^2}{N}\left[\Gamma \delta_{im}\delta_{kl} + \delta_{il}\delta_{km}\right]$ and $y_{ik}$ are Gaussian random variables. This resolvent can be shown to obey \cite{JANIK1997603}
	\begin{align}
		g(\omega) &= \frac{1}{\omega - \phi \sigma^2 \Gamma g(\omega)}, \nonumber \\
		\Rightarrow g(\omega) &= \frac{1}{2\phi\Gamma\sigma^2}\left[\omega - \sqrt{\omega^2 - 4 \phi \Gamma\sigma^2} \right] = \frac{2}{\omega + \sqrt{\omega^2 - 4 \phi\Gamma \sigma^2}} . \label{gofomega}
	\end{align}
	This same series appears repeatedly in the expression for the full resolvent $\mathcal{G}(\omega)$, which means we can gather terms with the same power $B_{ij}$. For example, we can gather terms that are linear in $B_{ij}$ [see Eq.~(\ref{resolventexpansion2}) and (\ref{derivatives})] and do not vanish in the thermodynamic limit in the following way
	\begin{align}
		&\frac{i}{\phi N \omega }\sum_{ij}\bigg \langle\frac{1}{\omega} B_{ij}  - \frac{1}{\omega^3} \sum_{kl} B_{ik} \frac{\delta B_{kl}(t)}{\delta \lambda_{lj}(t)} + \frac{1}{\omega^5}\sum_{klmn} B_{ik} \left( \frac{\delta B_{kl}(t)}{\delta \lambda_{lm}(t)} \frac{\delta B_{mn}(t)}{\delta \lambda_{nj}(t)} + \frac{\delta B_{kl}(t)}{\delta \lambda_{nj}(t)} \frac{\delta B_{mn}(t)}{\delta \lambda_{lm}(t) } \right)  + \cdots \bigg\rangle \nonumber \\
		&+\frac{i}{\phi N }\sum_{ijkl}\bigg \langle\frac{-1}{\omega^3}\frac{\delta B_{ik}}{\delta \lambda_{kl}}\bigg[\frac{1}{\omega} B_{lj}  - \frac{1}{\omega^3} \sum_{mn} B_{lm} \frac{\delta B_{mn}(t)}{\delta \lambda_{nj}(t)} \nonumber \\
		&\quad\quad+ \frac{1}{\omega^5}\sum_{mnqr} B_{lm} \left( \frac{\delta B_{mn}(t)}{\delta \lambda_{nq}(t)} \frac{\delta B_{qr}(t)}{\delta \lambda_{rj}(t)} + \frac{\delta B_{mn}(t)}{\delta \lambda_{rj}(t)} \frac{\delta B_{qr}(t)}{\delta \lambda_{nq}(t) } \right)  + \cdots \bigg]\bigg\rangle + \cdots \nonumber \\
		&= \frac{i}{\phi N }\sum_{ijk} \bigg[ \bigg \langle\frac{1}{\omega} B_{ik} \delta_{kj} g(\omega) \bigg\rangle - \sum_{kl}\bigg \langle\frac{1}{\omega^3}\frac{\delta B_{ik}}{\delta \lambda_{kl}} B_{lj}  g(\omega) \bigg\rangle  \nonumber \\
		&\quad \quad+ \sum_{klmn}\bigg \langle\frac{1}{\omega^5} \left( \frac{\delta B_{ik}(t)}{\delta \lambda_{kl}(t)} \frac{\delta B_{lm}(t)}{\delta \lambda_{mn}(t)}+ \frac{\delta B_{ik}(t)}{\delta \lambda_{mn}(t)} \frac{\delta B_{lm}(t)}{\delta \lambda_{kl}(t) } \right) B_{nj} g(\omega) \bigg\rangle + \cdots \bigg]\nonumber \\
		&= \frac{i g(\omega)^2}{\phi N  }\sum_{ij} \left\langle B_{ij}  \right\rangle .
	\end{align}
	Taking into account similar considerations for the higher-order terms in $B_{ij}$, we obtain following series for the full resolvent
	\begin{align}
		\mathcal{G}(\omega) = g(\omega) + \frac{i g(\omega)^2}{N\phi } \sum_{ij} \langle B_{ij} \rangle\vert_{\lambda = 0} -  \frac{ g(\omega)^3}{N\phi } \sum_{ijk } \langle B_{ik} B_{kj}\rangle\vert_{\lambda = 0} - \frac{i g(\omega)^4}{N\phi} \sum_{ijkl} \langle B_{ik} B_{kl} B_{lj}\rangle\vert_{\lambda = 0} + \cdots . \label{simplifiedseries}
	\end{align}
	
	\subsubsection*{An auxiliary diagrammatic convention}\label{subsection:diagramrules}
	Now that we have simplified the problem by collecting the terms in the series with the same multiples of the matrix $B$, we can proceed to evaluate the series as a whole. To aid us in spotting the self-similarity of the series, we introduce a second set of diagrammatic conventions.
	
	Each factor of $B_{ij}$ (when the limit $\lambda_{ij} \to 0$ is taken) has two terms [see the first of Eqs.~(\ref{derivatives})]. When we expand a product of $m$ matrices $B$ and take the ensemble average, we generate $2^m$ terms, each one containing a product of $m$ of the summands of $B$.
	
	Consider for example the second-order term $-  \frac{ g(\omega)^3}{N\phi } \sum_{ijk } \langle B_{ik} B_{kj}\rangle$. Referencing the definition of $B_{ik}$ in Eq.~(\ref{derivatives}), we obtain the following terms upon evaluating the ensemble average in the limit $\lambda_{ij} \to 0$
	\begin{align}
		-  \frac{ g(\omega)^3}{N\phi } \sum_{ijk } \langle B_{ik} B_{kj}\rangle \vert_{\lambda = 0} &= -  \frac{ g(\omega)^3 \sigma^4}{N^3\phi } \sum_{ijk } \bigg[ \left\langle \theta_i(t) \theta_k(t) \theta_j(t) \int dt' dt'' \hat x_i(t') x_k(t') \hat x_k(t'') x_j(t'') \right\rangle \nonumber \\
		&+ \Gamma \left\langle \theta_i(t) \theta_k(t) \theta_j(t) \int dt' dt'' x_i(t') \hat x_k(t')  \hat x_k(t'')  x_j(t'')  \right\rangle \nonumber \\
		&+ \Gamma \left\langle \theta_i(t) \theta_k(t) \theta_j(t) \int dt' dt'' \hat x_i(t') x_k(t') x_k(t'') \hat x_j(t'')  \right\rangle \nonumber \\
		&+ \Gamma^2 \left\langle \theta_i(t) \theta_k(t) \theta_j(t) \int dt' dt'' x_i(t') \hat x_k(t') x_k(t'') \hat x_j(t'')  \right\rangle \bigg]. \label{exampleterm}
	\end{align}
	Let us take the specific example of the first bracket. First, we observe from Eq.~(\ref{angularbracketdef}) that since the different species decouple in the thermodynamic limit, the sums factorise
	\begin{align}
		&\lim_{t\to\infty, N\to \infty} \frac{(i)^2 \sigma^4 g(\omega)^3}{\phi N^3}\sum_{ijk }\left\langle \theta_i(t) \theta_k(t) \theta_j(t) \int dt' dt'' \hat x_i(t') x_k(t') \hat x_k(t'') x_j(t'') \right\rangle \nonumber \\
		&= \lim_{t\to\infty, N\to \infty} \frac{(i)^2 \sigma^4 g(\omega)^3}{\phi N^3}    \int dt' dt'' \sum_i\left\langle \theta_i(t)\hat x_i(t') \right\rangle \sum_k\left\langle \theta_k(t) x_k(t') \hat x_k(t'') \right\rangle \sum_j \left\langle  \theta_j(t) x_j(t'')\right\rangle .
	\end{align}
	Then taking the limit $t \to \infty$ and assuming that time-translational invariance applies, we find 
	\begin{align}
		&\lim_{t\to\infty, N\to \infty} \frac{(i)^2 \sigma^4 g(\omega)^3}{\phi N^3}\sum_{ijk }\left\langle \theta_i(t) \theta_k(t) \theta_j(t) \int dt' dt'' \hat x_i(t') x_k(t') \hat x_k(t'') x_j(t'') \right\rangle \nonumber \\
		&= \lim_{t\to\infty} \frac{ \sigma^4 g(\omega)^3}{\phi} \int dt' dt'' T(t, t') G(t', t'') M(t'') = \frac{ \sigma^4 g(\omega)^3}{\phi} \chi_{{}_T} \chi M,
	\end{align}
	where we have used the final value theorem for Laplace transforms $\lim_{t\to \infty} f(t) = \lim_{u \to 0} u \mathcal{L}_t[f(t)](u)$, and Eqs.~(\ref{effectiveprocess} -- \ref{fixedpointquantities}) and (\ref{addintrep}) to deduce the last equality. The other terms can be evaluated in a similar manner. We thus obtain
	\begin{align}
		-  \frac{ g(\omega)^3}{N\phi } \sum_{ijk } \langle B_{ik} B_{kj}\rangle \vert_{\lambda = 0} =  \frac{ \sigma^4 g(\omega)^3}{\phi} \left[ \chi_{{}_T} \chi M   + \Gamma \chi_2 M^2+ \Gamma \chi_{{}_T}^2 q + \Gamma^2 \chi_{{}_T} \chi M \right] .
	\end{align}
	It is possible to generalise this approach and to represent each term with a diagram. We assign a node to each summation index. The direction of the arrows between nodes indicates which of the two summands is chosen from each factor of $B_{ij}$. Nodes connecting two edges involve two variables ($x_k$ or $\hat x_k$) whereas nodes connected only to one edge are associated with a sum over one variable. 
	
	With the above example in mind, we construct diagrams with the following rules: 
	\begin{align}
		\begin{tikzpicture}
			\node (A) {};
			\node[right=1.5cm of A] (B) {};
			\Edge[Direct](B)(A) 
		\end{tikzpicture} &= \sigma^2, \nonumber \\
		\begin{tikzpicture}
			\node (A) {};
			\node[right=1.5cm of A] (B) {};
			\Edge[Direct](A)(B)
		\end{tikzpicture} &= \Gamma \sigma^2, \nonumber \\
		\begin{tikzpicture}[baseline=-1mm]
			\Vertex{A} at (-1,-15) 
			\node[right=1.5cm of A] (B) {};
			\Edge[Direct](B)(A) 
		\end{tikzpicture}  &= g(\omega)\chi_{{}_T}, \nonumber \\
		\begin{tikzpicture}[baseline=-1mm]
			\Vertex{A} \node[right=1.5cm of A] (B) {};
			\Edge[Direct](A)(B)
		\end{tikzpicture} &= g(\omega) M, \nonumber \\
		\begin{tikzpicture}[baseline=-1mm]
			\Vertex{A} 
			\node[right=1.5cm of A] (B) {};
			\node[left=1.5cm of A] (C) {};
			\Edge[Direct](A)(B)
			\Edge[Direct](A)(C)
		\end{tikzpicture} &= g(\omega) q, \nonumber \\
		\begin{tikzpicture}[baseline=-1mm]
			\Vertex{A} 
			\node[right=1.5cm of A] (B) {};
			\node[left=1.5cm of A] (C) {};
			\Edge[Direct](B)(A)
			\Edge[Direct](C)(A)
		\end{tikzpicture} &= g(\omega) \chi_2, \nonumber \\
		\begin{tikzpicture}[baseline=-1mm]
			\Vertex{A} 
			\node[right=1.5cm of A] (B) {};
			\node[left=1.5cm of A] (C) {};
			\Edge[Direct](A)(B)
			\Edge[Direct](C)(A)
		\end{tikzpicture} &= 
		\begin{tikzpicture}[baseline=-1mm]
			\Vertex{A} 
			\node[right=1.5cm of A] (B) {};
			\node[left=1.5cm of A] (C) {};
			\Edge[Direct](B)(A)
			\Edge[Direct](A)(C)
		\end{tikzpicture} = g(\omega) \chi. \label{Feynman}
	\end{align}
	We can therefore write for the terms in Eq.~(\ref{exampleterm})
	\begin{align}
		-  \frac{ g(\omega)^3}{N} \sum_{ijk } \langle B_{ik} B_{kj}\rangle\vert_{\lambda = 0}  &= 
		\begin{tikzpicture}[baseline=-1mm]
			\Vertex[y= 0]{A} 
			\Vertex[x = 1.5]{E}
			\Vertex[x=-1.5]{D} 
			\node[right=1.cm of A] (B) {};
			\node[left=1.cm of A] (C) {};
			\Edge[Direct](B)(A)
			\Edge[Direct](A)(C)
		\end{tikzpicture} +
		\begin{tikzpicture}[baseline=-1mm]
			\Vertex{A} 
			\Vertex[x = 1.5]{E}
			\Vertex[x=-1.5]{D} 
			\node[right=1.cm of A] (B) {};
			\node[left=1.cm of A] (C) {};
			\Edge[Direct](B)(A)
			\Edge[Direct](C)(A)
		\end{tikzpicture} \nonumber  \\
		& +
		\begin{tikzpicture}[baseline=-1mm]
			\Vertex{A} 
			\Vertex[x = 1.5]{E}
			\Vertex[x=-1.5]{D} 
			\node[right=1.cm of A] (B) {};
			\node[left=1.cm of A] (C) {};
			\Edge[Direct](A)(B)
			\Edge[Direct](A)(C)
		\end{tikzpicture} +
		\begin{tikzpicture}[baseline=-1mm]
			\Vertex{A} 
			\Vertex[x = 1.5]{E}
			\Vertex[x=-1.5]{D} 
			\node[right=1.cm of A] (B) {};
			\node[left=1.cm of A] (C) {};
			\Edge[Direct](A)(B)
			\Edge[Direct](C)(A)
		\end{tikzpicture} . 
	\end{align}
	The series in Eq.~(\ref{resolventexpansion2}) can thus be written as
	\begin{align}
		\frac{ 1}{N \phi } \sum_{i,j } G_{ij} = g(\omega)  + \frac{1}{\phi}\mathcal{T}, \label{diagramsum}
	\end{align}
	where $\mathcal{T}$ is the sum over all possible such diagrams. 
	\subsubsection*{Sum over all possible diagrams}\label{subsection:sum}
	The challenge now is to perform the sum over all possible diagrams. By `all possible diagrams', we mean diagrams with any number of nodes and any configuration of edge directions. We do this by categorising each diagram by the directions of its outermost two edges. In this way, the sum over all diagrams can then be decomposed in a self-similar fashion. 
	
	Forgetting for now about the contributions from the outermost nodes, consider the sum over all possible diagrams with two outer edges of the type  \begin{tikzpicture}
		\node (A) {};
		\node[right=0.5cm of A] (B) {};
		\Edge[Direct](B)(A)
	\end{tikzpicture}. We call this sum $D_1$ and denote it diagrammatically as
	\begin{align}
		D_1 &= \begin{tikzpicture}[baseline=-1mm]
			\Vertex[shape=rectangle]{A} 
			\node[right=1.5cm of A] (B) {};
			\node[left=1.5cm of A] (C) {};
			\Edge[Direct](B)(A)
			\Edge[Direct](A)(C)
		\end{tikzpicture} \nonumber \\
		&\equiv \begin{tikzpicture}
			\node (A) {};
			\node[right=1.5cm of A] (B) {};
			\Edge[Direct](B)(A)
		\end{tikzpicture}+
		\begin{tikzpicture}[baseline=-1mm]
			\Vertex{A} 
			\node[right=1.5cm of A] (B) {};
			\node[left=1.5cm of A] (C) {};
			\Edge[Direct](B)(A)
			\Edge[Direct](A)(C)
		\end{tikzpicture} \nonumber \\
		&+\begin{tikzpicture}[baseline=-1mm]
			\Vertex{A} 
			\Vertex[x = 1.5]{E}
			\Vertex[x=-1.5]{D} 
			\node[right=1.cm of A] (B) {};
			\node[left=1.cm of A] (C) {};
			\Edge[Direct](B)(A)
			\Edge[Direct](A)(C)
			\node[right=1.0cm of E] (F) {};
			\node[left=1.0cm of D] (G) {};
			\Edge[Direct](D)(G)
			\Edge[Direct](F)(E)
		\end{tikzpicture} +
		\begin{tikzpicture}[baseline=-1mm]
			\Vertex{A} 
			\Vertex[x = 1.5]{E}
			\Vertex[x=-1.5]{D} 
			\node[right=1.cm of A] (B) {};
			\node[left=1.cm of A] (C) {};
			\Edge[Direct](B)(A)
			\Edge[Direct](C)(A)
			\node[right=1.0cm of E] (F) {};
			\node[left=1.0cm of D] (G) {};
			\Edge[Direct](D)(G)
			\Edge[Direct](F)(E)
		\end{tikzpicture} \nonumber  \\
		& +
		\begin{tikzpicture}[baseline=-1mm]
			\Vertex{A} 
			\Vertex[x = 1.5]{E}
			\Vertex[x=-1.5]{D} 
			\node[right=1.cm of A] (B) {};
			\node[left=1.cm of A] (C) {};
			\Edge[Direct](A)(B)
			\Edge[Direct](A)(C)
			\node[right=1.0cm of E] (F) {};
			\node[left=1.0cm of D] (G) {};
			\Edge[Direct](D)(G)
			\Edge[Direct](F)(E)
		\end{tikzpicture} +
		\begin{tikzpicture}[baseline=-1mm]
			\Vertex{A} 
			\Vertex[x = 1.5]{E}
			\Vertex[x=-1.5]{D} 
			\node[right=1.cm of A] (B) {};
			\node[left=1.cm of A] (C) {};
			\Edge[Direct](A)(B)
			\Edge[Direct](C)(A)
			\node[right=1.0cm of E] (F) {};
			\node[left=1.0cm of D] (G) {};
			\Edge[Direct](D)(G)
			\Edge[Direct](F)(E)
		\end{tikzpicture} \nonumber \\
		&+\cdots. \label{d1def}
	\end{align}
	Equally, we define the sums over all possible diagrams with other combinations of outer edge pairs
	\begin{align}
		\begin{tikzpicture}[baseline=-1mm]
			\Vertex[shape=diamond]{A} 
			\node[right=1.5cm of A] (B) {};
			\node[left=1.5cm of A] (C) {};
			\Edge[Direct](B)(A)
			\Edge[Direct](C)(A)
		\end{tikzpicture} = D_2 ,\label{d2def}
	\end{align}
	\begin{align}
		\begin{tikzpicture}[baseline=-1mm]
			\Vertex[shape=diamond]{A} 
			\node[right=1.5cm of A] (B) {};
			\node[left=1.5cm of A] (C) {};
			\Edge[Direct](A)(B)
			\Edge[Direct](A)(C)
		\end{tikzpicture} = D_3 ,\label{d3def}
	\end{align}
	\begin{align}
		\begin{tikzpicture}[baseline=-1mm]
			\Vertex[shape=rectangle]{A} 
			\node[right=1.5cm of A] (B) {};
			\node[left=1.5cm of A] (C) {};
			\Edge[Direct](A)(B)
			\Edge[Direct](C)(A)
		\end{tikzpicture} = D_4 . \label{d4def}
	\end{align}
	
	With this in mind, the sum over all diagrams in Eq.~(\ref{diagramsum}) can be rewritten
	\begin{align}
		\mathcal{T} &=
		\begin{tikzpicture}[baseline=-1mm]
			\Vertex[shape=rectangle]{A} 
			\Vertex[x = 1.5]{E}
			\Vertex[x=-1.5]{D} 
			\node[right=1.cm of A] (B) {};
			\node[left=1.cm of A] (C) {};
			\Edge[Direct](B)(A)
			\Edge[Direct](A)(C)
		\end{tikzpicture} +
		\begin{tikzpicture}[baseline=-1mm]
			\Vertex[shape=diamond]{A} 
			\Vertex[x = 1.5]{E}
			\Vertex[x=-1.5]{D} 
			\node[right=1.cm of A] (B) {};
			\node[left=1.cm of A] (C) {};
			\Edge[Direct](B)(A)
			\Edge[Direct](C)(A)
		\end{tikzpicture} \nonumber  \\
		& +
		\begin{tikzpicture}[baseline=-1mm]
			\Vertex[shape=diamond]{A} 
			\Vertex[x = 1.5]{E}
			\Vertex[x=-1.5]{D} 
			\node[right=1.cm of A] (B) {};
			\node[left=1.cm of A] (C) {};
			\Edge[Direct](A)(B)
			\Edge[Direct](A)(C)
		\end{tikzpicture} +
		\begin{tikzpicture}[baseline=-1mm]
			\Vertex[shape=rectangle]{A} 
			\Vertex[x = 1.5]{E}
			\Vertex[x=-1.5]{D} 
			\node[right=1.cm of A] (B) {};
			\node[left=1.cm of A] (C) {};
			\Edge[Direct](A)(B)
			\Edge[Direct](C)(A)
		\end{tikzpicture}  , \label{alldiagrams}
	\end{align}
	so Eq.~(\ref{diagramsum}) becomes
	\begin{align}
		\frac{1}{N \phi } \sum_{i,j } G_{ij}(\omega) = g(\omega)  + \frac{[\sigma^2 g(\omega)]^2}{\phi}\left[ \chi_{{}_T} D_1 M + \Gamma M D_2 M + \Gamma \chi_{{}_T} D_3 \chi_{{}_T} +\Gamma^2 M D_4 \chi_{{}_T}  \right] . \label{diagramsum2}
	\end{align}
	We now make the crucial observation that the infinite sums $D_1$, $D_2$, $D_3$ and $D_4$ can be expressed in terms of one another due to the self-similarity of the series. Diagrammatically, we have
	\begin{align}
		\begin{tikzpicture}[baseline=-1mm]
			\Vertex[shape=rectangle]{A} 
		\end{tikzpicture}=& \,
		\begin{tikzpicture}[baseline=-1mm]
			\Vertex[shape=rectangle]{A} 
			\Vertex[x = 1.5]{E}
			\Vertex[x=-1.5]{D} 
			\node[right=1.cm of A] (B) {};
			\node[left=1.cm of A] (C) {};
			\Edge[Direct](B)(A)
			\Edge[Direct](A)(C)
		\end{tikzpicture} +
		\begin{tikzpicture}[baseline=-1mm]
			\Vertex[shape=diamond]{A} 
			\Vertex[x = 1.5]{E}
			\Vertex[x=-1.5]{D} 
			\node[right=1.cm of A] (B) {};
			\node[left=1.cm of A] (C) {};
			\Edge[Direct](B)(A)
			\Edge[Direct](C)(A)
		\end{tikzpicture} \nonumber  \\
		& +
		\begin{tikzpicture}[baseline=-1mm]
			\Vertex[shape=diamond]{A} 
			\Vertex[x = 1.5]{E}
			\Vertex[x=-1.5]{D} 
			\node[right=1.cm of A] (B) {};
			\node[left=1.cm of A] (C) {};
			\Edge[Direct](A)(B)
			\Edge[Direct](A)(C)
		\end{tikzpicture} +
		\begin{tikzpicture}[baseline=-1mm]
			\Vertex[shape=rectangle]{A} 
			\Vertex[x = 1.5]{E}
			\Vertex[x=-1.5]{D} 
			\node[right=1.cm of A] (B) {};
			\node[left=1.cm of A] (C) {};
			\Edge[Direct](A)(B)
			\Edge[Direct](C)(A)
		\end{tikzpicture}  +  \begin{tikzpicture}[baseline=-1mm]
			\Vertex{A} 
		\end{tikzpicture} + \emptyset, \label{squarediagram}
	\end{align}
	where the last two terms account for diagrams with one inner node and no inner nodes respectively. Similarly we have
	\begin{align}
		\begin{tikzpicture}[baseline=-1mm]
			\Vertex[shape=diamond]{A} 
		\end{tikzpicture}=& \,
		\begin{tikzpicture}[baseline=-1mm]
			\Vertex[shape=rectangle]{A} 
			\Vertex[x = 1.5]{E}
			\Vertex[x=-1.5]{D} 
			\node[right=1.cm of A] (B) {};
			\node[left=1.cm of A] (C) {};
			\Edge[Direct](B)(A)
			\Edge[Direct](A)(C)
		\end{tikzpicture} +
		\begin{tikzpicture}[baseline=-1mm]
			\Vertex[shape=diamond]{A} 
			\Vertex[x = 1.5]{E}
			\Vertex[x=-1.5]{D} 
			\node[right=1.cm of A] (B) {};
			\node[left=1.cm of A] (C) {};
			\Edge[Direct](B)(A)
			\Edge[Direct](C)(A)
		\end{tikzpicture} \nonumber  \\
		& +
		\begin{tikzpicture}[baseline=-1mm]
			\Vertex[shape=diamond]{A} 
			\Vertex[x = 1.5]{E}
			\Vertex[x=-1.5]{D} 
			\node[right=1.cm of A] (B) {};
			\node[left=1.cm of A] (C) {};
			\Edge[Direct](A)(B)
			\Edge[Direct](A)(C)
		\end{tikzpicture} +
		\begin{tikzpicture}[baseline=-1mm]
			\Vertex[shape=rectangle]{A} 
			\Vertex[x = 1.5]{E}
			\Vertex[x=-1.5]{D} 
			\node[right=1.cm of A] (B) {};
			\node[left=1.cm of A] (C) {};
			\Edge[Direct](A)(B)
			\Edge[Direct](C)(A)
		\end{tikzpicture}  +  \begin{tikzpicture}[baseline=-1mm]
			\Vertex{A} 
		\end{tikzpicture}, \label{diamonddiagram}
	\end{align}
	Substituting iteratively the expression in Eqs.~(\ref{squarediagram}) and (\ref{diamonddiagram}) into Eq.~(\ref{alldiagrams}) produces the summation over all diagrams that we desire. Using Eqs.~(\ref{squarediagram}) and (\ref{diamonddiagram}) and the definitions in Eqs.~(\ref{d1def})-(\ref{d4def}), we find the following set of simultaneous equations for the quantities $D_1$, $D_2$, $D_3$ and $D_4$,
	\begin{align}
		D_1 &= [\sigma^2 g(\omega)]^2 \left( \chi D_1 \chi + \Gamma q D_2 \chi + \Gamma \chi D_3 \chi_2 + \Gamma^2 q D_4 \chi_2\right) + \chi g(\omega) + \frac{1 }{\sigma^2} , \nonumber \\
		D_2 &= [\sigma^2 g(\omega)]^2 \left( \chi_2 D_1 \chi + \Gamma \chi D_2 \chi + \Gamma \chi_2 D_3 \chi_2 + \Gamma^2 \chi D_4 \chi_2\right) +  \chi_2 g(\omega), \nonumber \\
		D_3 &= [\sigma^2 g(\omega)]^2 \left( \chi D_1 q + \Gamma q D_2 q + \Gamma \chi D_3 \chi + \Gamma^2 q D_4 \chi \right) +  q g(\omega), \nonumber \\
		D_4 &= [\sigma^2 g(\omega)]^2 \left( \chi_2 D_1 q + \Gamma \chi D_2 q + \Gamma \chi_2 D_3 \chi + \Gamma^2 \chi D_4 \chi\right) + \chi g(\omega)  + \frac{1 }{\Gamma\sigma^2}. \label{dequations}
	\end{align}
	
	\subsection{Final expression for the outlier}\label{subsection:outlierexpression}
	We now are left with the relatively simple task of solving the linear Eqs.~(\ref{dequations}) for $D_1(\omega)$, $D_2(\omega)$, $D_3(\omega)$ and $D_4(\omega)$ to obtain the disorder-averaged resolvent $\overline{\frac{1}{N_\ess} \sum_{i,j } G_{ij}(\omega)}$.  We find that the functions $D_1(\omega)$, $D_2(\omega)$, $D_3(\omega)$ and $D_4(\omega)$ are given by
	\begin{align}
		D_1(\omega) &= \frac{1}{\sigma^2 D[g(\omega)]} \left[1-  g(\omega)\chi \Gamma \sigma^2  \right],\nonumber \\
		D_2(\omega) &= \frac{\chi_2 g(\omega)}{  D[g(\omega)]},\nonumber \\
		D_3(\omega) &= \frac{q g(\omega)}{  D[g(\omega)]},\nonumber \\
		D_4(\omega) &= \frac{1}{\Gamma\sigma^2 D[g(\omega)]} \left[1-  g(\omega) \chi  \sigma^2  \right], \nonumber \\
		D[g(\omega)] &=  1 - (1 + \Gamma) \sigma^2 \chi g(\omega)  + \left[ (\chi)^2 - \chi_2 q  \right] [g(\omega)]^2 \Gamma \sigma^4 . \label{dequations2}
	\end{align}
	Substituting these expressions into Eq.~(\ref{diagramsum2}), one obtains 
	\begin{align}
		\mathcal{G}[\omega] &= g(\omega) + \frac{ \sigma^2 [g(\omega)]^2}{\phi D[g(\omega)]} \left[ \chi_{{}_T} M (1 + \Gamma) + \Gamma \sigma^2\left(M^2 \chi_2 + \chi_{{}_T}^2 q - 2 \chi \chi_{{}_T} M \right) g(\omega) \right].
	\end{align}
	Finally, now that we have the function $\mathcal{G}[\omega]$, the outlier eigenvalue we seek is then given by the solution $\lambda_{\mathrm{outlier}}$ to [c.f. Eq.~(\ref{outliercalcgeneral})] 
	\begin{align}
		\mathcal{G}[1 + \lambda_{\mathrm{outlier}}] &= \frac{1}{\mu \phi}. \label{gsol}
	\end{align}
	
	\subsubsection*{Solution strategy}
	The solution $\lambda_{\mathrm{outlier}}$ for a given set $(\mu, \sigma^2, \Gamma)$ can be obtained efficiently from Eq.~(\ref{gsol}) by adopting the following parametric solution strategy. First, one obtains the fixed-point quantities $\chi$, $\chi_{{}_T}$, $\chi_2$, $q$, $M$ and $\phi$ from Eqs.~(\ref{sigmasol}) and (\ref{summaryeqs}). Then, one solves the following for $g$ 
	\be
	\mathcal{F}[g] \equiv g + \frac{ \sigma^2 g^2}{\phi D[g]} \left[ \chi_{{}_T} M (1 + \Gamma) + \Gamma \sigma^2\left(M^2 \chi_2 + \chi_{{}_T}^2 q - 2 \chi \chi_{{}_T} M \right) g \right] = \frac{1}{\mu \phi}, \label{gparametric} \ee
	where
	\be
	D[g] =  1 - (1 + \Gamma) \sigma^2 \chi g  + \left[ (\chi)^2 - \chi_2 q  \right] g^2 \Gamma \sigma^4. 
	\ee
	Eq.~(\ref{gparametric}) is a cubic equation and can be solved readily.
	
	Then one plugs the resulting value of $g$ into the following to obtain the outlier
	\begin{align}
		\lambda_{\mathrm{outlier}}(g)&= -1+ \frac{1}{g} + \phi \sigma^2 \Gamma g . \label{lambdafromg}
	\end{align}
	This last relation results from the expression for $g(\omega)$ in the first line of Eq.~(\ref{gofomega}).
	
	\subsubsection*{Validity of the solutions}
	When solving the cubic Eq.~(\ref{gparametric}) for $g$, we obtain a maximum of three possible solutions for the outlier eigenvalue. We thus seek a criterion by which to rule out the two unphysical solutions. This is accomplished by realising that $g$ is actually the trace of the resolvent matrix in the thermodynamic limit. 
	
	Let us examine again the series in Eq.~(\ref{resolventexpansion2}), but now with the sum over all elements $\sum_{ij}$ replaced by a trace (i.e. setting $i = j$ and summing over the single index $i$). We then see that most terms in this modified series no longer survive in the thermodynamic limit. The only ones that do survive are those proportional to $\delta_{ij}$. Therefore, the trace of the resolvent is simply given by those terms consisting only of products of derivatives like $\frac{\delta B_{ik}}{\delta \lambda_{kj}}$, which means that $\frac{1}{N}\sum_{i}G_{ii}(\omega)=g(\omega)$.  
	
	One can show as in Ref. \cite{baron2021eigenvalues} (the calculation follows along very similar lines and we do not reproduce it here), that the trace of the resolvent matrix can be related to the response function of a carefully constructed linear process. By requiring that the power spectrum of fluctuations of this linear process be positive, we can deduce that the modulus squared of this response function (which is equivalent to $g$) must be greater than the reciprocal of the variance of the random matrix elements. We hence obtain the following constraint on $g$ [analogous to Eq.~(S46) of Ref. \cite{baron2021eigenvalues}]
	\begin{align}
		\vert g \vert^2 < \frac{1}{\phi \sigma^2}. \label{validitycriterion}
	\end{align}
	We note that when $g = 1/\sqrt{\phi \sigma^2}$, $\lambda_{\mathrm{outlier}} = -1 +(1+\Gamma)\sqrt{\phi \sigma^2}$ and the one valid solution for the outlier is absorbed into the bulk of the eigenvalue spectrum [which is given in Eq.~(\ref{bulkspectrum})].
	
	\subsubsection*{Special case: $\Gamma = 0$}
	In this special case, Eq.~(\ref{gsol}) becomes quadratic, allowing us to obtain a more compact expression for the outlier. Writing  $\lambda$ instead of $\lambda_{\mathrm{outlier}}$ (to keep the resulting relation compact) we have
	\begin{align}
		\frac{1}{1 + \lambda} + \frac{\sigma^2}{\phi(1+\lambda)} \frac{\chi_{{}_T} M }{1 + \lambda - \sigma^2 \chi} = \frac{1}{\phi \mu} ,\label{outliergamma0}
	\end{align}
	from which one finds the pleasingly succinct expression
	\begin{align}
		\lambda = -1 + \frac{\phi}{2}\left[ \mu + \sigma^2 + \sqrt{(\mu-\sigma^2)^2 + 4 \chi_{{}_T} M \mu \sigma^2/\phi^2 } \right] ,
	\end{align}
	where we have used $\chi = \phi$ at the fixed point for $\Gamma = 0$.  
	
	\subsection{The diverging abundance transition ($M\to \infty$) corresponds to the outlier crossing the imaginary axis}
	We now proceed to show that when the $M \to \infty$ transition occurs, the outlier eigenvalue given in Eq.~(\ref{gsol}) hits the imaginary axis.
	
	Multiplying both sides of the first of Eqs.~(\ref{gparametric}) by $\mu \phi D[g]$, we obtain the following cubic 
	\begin{align}
		D[g] \{\mu \phi \mathcal{F}[g] - 1\}=&\left[ 1 - (1 + \Gamma) \sigma^2 \chi g + \Gamma \sigma^4 \left( \chi^2 - \chi_2 q\right) g^2\right] [1-\mu\phi g] \nonumber \\
		&- \mu\sigma^2 g^2 \left[ \chi_{{}_T} M (1 + \Gamma) + \Gamma \sigma^2\left(M^2 \chi_2 + \chi_{{}_T}^2 q - 2 \chi \chi_{{}_T} M \right) g \right] = 0. \label{outlierminf}
	\end{align}
	We now note that if $\lambda_{\mathrm{outlier}} = 0$ is indeed a solution to Eqs.~(\ref{gsol}), then $\Gamma \sigma^2 \phi g^2 - g + 1$ must be a factor of the left-hand side of Eq.~(\ref{outlierminf})  [one can see this by setting $\lambda_{\mathrm{outlier}} = 0$ in Eq.~(\ref{lambdafromg})].  If this is the case, then we must be able to factorise the cubic in Eq.~(\ref{outlierminf}) to give an expression of the form
	\begin{align}
		(\Gamma \sigma^2 \phi g^2 - g + 1)(1 + b g) = 0, \label{factorisationansatz}
	\end{align}
	where $b$ is a coefficient to be found.  Equating coefficients in the two cubic expressions in Eqs.~(\ref{outlierminf}) and (\ref{factorisationansatz}), one obtains three expressions for $b$ which must all be equal if Eq.~(\ref{factorisationansatz}) is a valid factorisation of Eq.~(\ref{outlierminf}). These expressions are
	\begin{align}
		b_1 &= 1 - (1 + \Gamma) \sigma^2 \chi - \mu \phi,\nonumber \\
		b_2&= \Gamma \sigma^2 \phi - (1+\Gamma)\sigma^2\chi \mu \phi - \Gamma \sigma^4 \chi^2 + \mu \sigma^2 \chi_{{}_T} M,\nonumber \\
		b_3&= -\frac{\mu \sigma^2}{\phi} \left[ \chi_2 M^2 + \chi_{{}_T}^2 q- 2 \chi \chi_{{}_T} M  + \phi (\chi^2 - \chi_2 q)\right].
	\end{align}
	If we can show that $b_1 = b_2 = b_3$ when $M\to \infty$, then we will have proved that $\lambda_{\mathrm{outlier}} = 0$ is a possible solution when $M\to \infty$. We can see that this is indeed the case by writing each of the above expressions $b_1$, $b_2$ and $b_3$ in terms of only functions of $\Delta$ and $\Gamma$. We first note from the relations in Eq.~(\ref{summaryeqs}) that when $M\to \infty$ we have
	\begin{align}
		\sigma^2 \mu \chi_{{}_T} M &= \frac{w_2\Delta (w_1 - \Delta w_0)}{(w_2 + \Gamma w_0)^2}, \nonumber \\
		\chi \sigma^2 &= \frac{w_0 }{w_2 + \Gamma w_0}, \nonumber \\
		\Gamma \sigma^2 \phi &= \frac{\Gamma w_0 w_2 }{ (w_2 + \Gamma w_0)^2}, \nonumber \\
		w_2 &= w_0 + \Delta w_1.
	\end{align}
	We therefore find that the first two expression for $b$ are equal [recalling Eq.~(\ref{eq:w_relation})]
	\begin{align}
		b_1 &= \frac{w_1[w_2 + \Gamma w_0 - (1+\Gamma) w_0 ] - \Delta w_0  w_2}{w_1 (w_2 + \Gamma w_0)} = \Delta\frac{w_1^2 - w_0 w_2}{w_1 (w_2 + \Gamma w_0)}, \nonumber \\
		b_2 &= \frac{ \Gamma w_0 w_2 w_1 - (1+\Gamma) w_0^2 \Delta w_2 - \Gamma w_1 w_0^2 + \Delta w_2 w_1(w_1 - \Delta w_0) }{w_1 (w_2 + \Gamma w_0)^2} \nonumber \\
		&= \Delta\frac{ \Gamma w_0 w_1^2 -\Gamma w_0^2 w_2 + w_2 w_1^2 - w_2^2 w_0 }{w_1 (w_2 + \Gamma w_0)^2} = b_1 .
	\end{align}
	
	\noindent Noting further the following equalities in the limit $M\to \infty$
	\begin{align}
		- \mu \sigma^2 \chi_2 M^2 &= \Delta^2\frac{(w_1 - \Delta w_0)w_1}{w_2 + \Gamma w_0}, \nonumber \\
		- \chi_{{}_T}^2 q \mu \sigma^2 &= - \Delta\frac{w_2 (w_1 - \Delta w_0)^2 }{w_1 (w_2 + \Gamma w_0)} , \nonumber \\
		2 \chi \chi_{{}_T} M \mu \sigma^2 &= \frac{2 w_0 \Delta (w_1 - \Delta w_0)}{(w_2 + \Gamma w_0)}, \nonumber \\
		- \mu \sigma^2 \phi \chi^2 &= -\frac{\Delta w_0^3 }{w_1 (w_2 + \Gamma w_0)}, \nonumber \\
		\mu \sigma^2 \chi_2 q \phi &= -\frac{\Delta^2w_0 w_2 (w_1 - \Delta w_0)}{w_1 (w_2 + \Gamma w_0)} ,
	\end{align}
	we obtain for the final expression
	\begin{align}
		b_3 =& \frac{\Delta}{w_0 w_1 (w_2 + \Gamma w_0)} \big[\Delta w_1^2 (w_1 -\Delta w_0) - w_2 (w_1 - \Delta w_0)^2 \nonumber \\
		&+ 2 w_0 w_1 (w_1 - \Delta w_0) - w_0^3 - \Delta w_0 w_2 (w_1 - \Delta w_0) \big]\nonumber \\
		=& \frac{\Delta}{w_0 w_1 (w_2 + \Gamma w_0)} \left[w_0 w_1^2 - w_0^2 w_2 \right] = b_1 .
	\end{align}
	Hence we have shown that $b_1 = b_2 = b_3$. This means that when $M\to \infty$, we can write 
	\begin{align}
		D[g] \{\mu \phi \mathcal{F}[g] - 1\} = (\Gamma \sigma^2 \phi g^2 - g + 1)\{1 + [1 - (1+\Gamma) \sigma^2 \chi -\mu \phi] g\} = 0. \label{minfg}
	\end{align}
	Hence, $\lambda = 0$ is a solution to this equation when $M \to \infty$. 
	
	Let us examine the alternative solution to Eq.~(\ref{minfg}) $g = -[1 - (1+\Gamma) \sigma^2 \chi -\mu \phi]^{-1} = -b_1^{-1}$ to see if it satisfies the criterion in Eq.~(\ref{validitycriterion}). One can examine the function $r(\Delta) = \phi\sigma^2/\vert b_1 \vert^2 -1$ (which turns out to be independent of $\Gamma$). In order for the transition $M\to \infty$ to occur, we must have that $\mu >0$ (see Fig.~1 in the main text, and also \cite{galla2018dynamically}), and hence that $\Delta>0$ [as can be seen from the second of Eqs.~(\ref{summaryeqs}) when $M\to \infty$. For $\Delta >0$, it can be verified that $r(\Delta)>0$]. Thus, one finds that the only valid solution to Eqs.~(\ref{minfg}) is the one that corresponds to $\lambda = 0$.
	
	\section{Bulk spectrum: derivation using the Hermitized resolvent}
	\subsection{Hermitized resolvent}
	In Section \ref{section:outlierresolvent}, we evaluated a series expansion of the resolvent matrix so that we could find the outlier eigenvalue. In the region of the complex plane in which the outlier resides, the resolvent is analytic, which is why we could use the expansion in Eq.~(\ref{resolventexpansion}). In order to find the bulk eigenvalue density, we also need to evaluate the resolvent matrix (in this case its trace, rather than the sum of all its elements). However, in the region of the complex plane occupied by the bulk of the eigenvalue spectrum, the resolvent is no longer analytic. So that we can proceed, we must construct an alternative series expansion for the resolvent that takes this non-analytic nature into account. We follow the method of Ref. \cite{feinberg1997non}, which involves constructing a `hermitized' resolvent.
	
	We have the following identity
	\be
	\rho(x, y) = \frac{1}{\pi} \bar \partial G(\omega, \omega^\star) \vert_{\omega = 1 + \lambda}, \label{densityfromhresolv}
	\ee
	relating the disorder-averaged resolvent 
	\be
	G(\omega, \omega^\star) \equiv \overline{\left\langle \frac{1}{N \phi} \mathrm{Tr}\left[ \frac{1}{\omega \id_{N_\ess} - \underline{\underline{z}}}\right]\right\rangle } \ee
	to the eigenvalue density $\rho(x,y)$. We have again written $z_{ij} = (a_{ij}- \mu N^{-1})\theta_i \theta_j$, as well as $\omega = x + i y$ and $\bar\partial = (\partial_x + i \partial_y)/2$. From this, by the Cauchy-Riemann equations of complex analysis, we see immediately that the eigenvalue density is non-zero if and only if the resolvent is non-analytic. 
	
	We now define the $2N_\ess \times 2N_\ess$ Hermitian matrix 
	\begin{align}
		H = \begin{bmatrix}
			0 & \underline{\underline{z}} - \omega \id_{N_\ess} \\
			(\underline{\underline{z}} - \omega \id_{N_\ess})^\dagger & 0
		\end{bmatrix} ,
	\end{align}
	and the Hermitized Green's function 
	\begin{align}
		\mathcal{H}(\eta, \omega, \omega^\star) = \overline{\left\langle \frac{1}{\eta - H} \right\rangle } .
	\end{align}
	From these definitions we see that we can recover the resolvent we seek via
	\begin{align}
		G(\omega, \omega^\star) = \frac{1}{N\phi} \mathrm{Tr} \left[  \mathcal{H}^{21} (0, z, z^\star)\right],
	\end{align}
	where the indices of $\mathcal{H}$ refer to its blocks. Hence, if we define
	\begin{align}
		\mathcal{H}_0^{-1} &\equiv \begin{bmatrix}
			\eta \id_{N_\ess} & \omega \id_{N_\ess} \\
			\omega^\star \id_{N_\ess} & \eta \id_{N_\ess}
		\end{bmatrix} , \nonumber \\
		\mathcal{J} &\equiv \begin{bmatrix}
			0& \underline{\underline{z}} \\
			\underline{\underline{z}}^\dagger & 0
		\end{bmatrix},
	\end{align}
	then we obtain the following Dyson series for $\mathcal{H}$, 
	\begin{align}
		\mathcal{H} = \mathcal{H}_0 + \overline{\langle \mathcal{H}_0 \mathcal{J} \mathcal{H}_0 \rangle} +  \overline{ \langle \mathcal{H}_0 \mathcal{J} \mathcal{H}_0 \mathcal{J} \mathcal{H}_0 \rangle} + \cdots, \label{hermitizedseries}
	\end{align}
	which then yields the resolvent we desire.
	\subsection{The series for the bulk spectrum is that of a Gaussian random matrix}
	Let us consider for example the first two non-trivial terms in Eq.~(\ref{hermitizedseries}). We have
	\begin{align}
		\left[ \mathcal{H}_0 \mathcal{J} \mathcal{H}_0\right]^{21} &= \frac{1}{(\eta^2 - \vert \omega\vert^2)^2} \left[(\omega^\star)^2 \underline{\underline{z}}  + \eta^2 \underline{\underline{z}}^\dagger \right], \nonumber \\
		\left[ \mathcal{H}_0 \mathcal{J} \mathcal{H}_0 \mathcal{J} \mathcal{H}_0 \right]^{21} &= \frac{1}{(\eta^2 - \vert \omega\vert^2)^3}\left\{- \eta \underline{\underline{z}}^\dagger (\omega \eta \underline{\underline{z}}^\dagger + \eta \omega^\star \underline{\underline{z}}) - \omega^\star \underline{\underline{z}} [(\omega^\star)^2 \underline{\underline{z}} + \eta^2 \underline{\underline{z}}^\dagger]\right\} .
	\end{align}
	In order to find the eigenvalue density of the bulk region, we take the trace of these terms.  That is, we must find quantities such as $\overline{\frac{1}{N_\ess}\sum_{i}z_{ii}}$ and  $\overline{\frac{1}{N_\ess}\sum_{ik}z_{ik} z^\dagger_{ki}}$.  This is notably different to the calculation of the outlier eigenvalue. In that case, we instead had to sum all elements of the resolvent and we therefore needed to calculate objects like $\overline{\frac{1}{N_\ess}\sum_{ij}z_{ij}}$ and $\overline{\frac{1}{N_\ess}\sum_{ijk}z_{ik} z_{kj}}$. We will now show that the resulting series for the bulk spectrum is far simpler by virtue of this difference. Many terms that were important for the calculation of the outlier eigenvalue vanish in thermodynamic limit in the calculation of the bulk spectrum. 
	
	Let us examine the quantity $\overline{\frac{1}{N_\ess}\sum_{ik}z_{ik} z^\dagger_{ki}}$. This can once again be derived from the generating functional as 
	\begin{align}
		\frac{1}{N \phi}\sum_{ik}\overline{z_{ik} z^\dagger_{ki}} = -\frac{1}{N \phi}\sum_{ik} \left\langle \frac{1}{A}\frac{\delta ^2 A}{\delta \lambda_{ik}^2} \right\rangle \bigg\vert_{\lambda = 0} = -\frac{1}{N \phi}\sum_{ik} \left[\left\langle\frac{\delta B_{ik}(t)}{ \delta \lambda_{ik}(t)} \right\rangle + \left\langle B_{ik}^2 \right\rangle  \right]\bigg\vert_{\lambda = 0}  .\label{squaredexample}
	\end{align}
	Examining the latter quantity we find
	\begin{align}
		-\frac{1}{N \phi}\sum_{ik} \left\langle B_{ik}^2 \right\rangle &= -\frac{1}{N^3 \phi}\sum_{ik}\bigg[ \left\langle \theta_i(t) \theta_k(t) \theta_i(t) \int dt' dt'' \hat x_i(t') x_k(t') \hat x_i(t'') x_k(t'') \right\rangle \nonumber \\
		&+ \Gamma \left\langle \theta_i(t) \theta_k(t) \theta_i(t) \int dt' dt'' x_i(t') \hat x_k(t')  \hat x_i(t'')  x_k(t'')  \right\rangle \nonumber \\
		&+ \Gamma \left\langle \theta_i(t) \theta_k(t) \theta_i(t) \int dt' dt'' \hat x_i(t') x_k(t') x_i(t'') \hat x_k(t'')  \right\rangle \nonumber \\
		&+ \Gamma^2 \left\langle \theta_i(t) \theta_k(t) \theta_i(t) \int dt' dt'' x_i(t') \hat x_k(t') x_i(t'') \hat x_k(t'')  \right\rangle \bigg]  = O(N^{-1}) .
	\end{align}
	Immediately, we see that the factor of $N^3$ in the denominator is not cancelled by the factor of $N^2$ that arises from carrying out the sums over $i$ and $k$. Therefore, this term vanishes in the thermodynamic limit. However, considering the other term in Eq.~(\ref{squaredexample}) we see that
	\begin{align}
		-\frac{1}{N \phi}\sum_{ik} \left\langle\frac{\delta B_{ik}(t)}{ \delta \lambda_{ik}(t)} \right\rangle = \frac{1}{N \phi}\sum_{ik} \left\langle \theta_i \theta_k \frac{\sigma^2}{N} \right\rangle = \sigma^2 \phi.
	\end{align}
	In general, only the terms containing solely derivatives of $B_{ij}$ with respect to $\lambda_{kl}$ survive in the thermodynamic limit. So, in a similar way to Section \ref{section:resolventserieseval} [see the discussion around Eq.~(\ref{derivativeseries}) in particular], we find that the series for the trace of the resolvent can be represented by the same series of diagrams \cite{JANIK1997603} as for the resolvent of the kinds of random matrices investigated by Ginibre, which had elliptic eigenvalue spectra \cite{ginibre1965} (once terms that vanish in the thermodynamic limit have been removed).

	That is, if we were to represent the series in Eq.~(\ref{hermitizedseries}) with diagrams, it would take the same form as that depicted after Eq.~(\ref{gofomega}), except now the edges would carry two indices: a block index (from the hermitization) and the usual species index \cite{feinberg1997non, kuczala2019dynamics, JANIK1997603}. We hence arrive at the result \cite{sommers, JANIK1997603}
	\begin{align}
		G(\omega_x, \omega_y) = \begin{cases}
			\frac{\omega}{2 \phi \Gamma \sigma^2} \left[1 - \sqrt{1 - 4 \phi \Gamma \sigma^2/\omega^2 } \right] \,\,\,\,\, &\mathrm{for} \,\,\,\,\, \left(\frac{\omega_x}{1 + \Gamma}\right)^2 + \left(\frac{\omega_y}{1-\Gamma}\right)^2 > \phi \sigma^2,  \\ 
			\frac{\omega_x}{\phi\sigma^2(1 + \Gamma)} - \frac{i\omega_y}{\phi \sigma^2 (1 - \Gamma)}  \,\,\,\,\, &\mathrm{for} \,\,\,\,\, \left(\frac{\omega_x}{1 + \Gamma}\right)^2 + \left(\frac{\omega_y}{1-\Gamma}\right)^2 < \phi \sigma^2 ,
		\end{cases}
	\end{align}
	where $\omega = \omega_x + i \omega_y$.  Consulting Eq.~(\ref{densityfromhresolv}), the resulting eigenvalue density of the bulk region is 
	\begin{align}
		\rho_{\mathrm{bulk}}(x,y)  = \begin{cases}
			\frac{1}{ \pi \phi \sigma^2 (1- \Gamma^2)}\,\,\,\,\, &\mathrm{for} \,\,\,\,\, \left(\frac{1+x}{1 + \Gamma}\right)^2 + \left(\frac{y}{1-\Gamma}\right)^2 > \phi \sigma^2,  \\ 
			0 \,\,\,\,\, &\mathrm{for} \,\,\,\,\, \left(\frac{1+x}{1 + \Gamma}\right)^2 + \left(\frac{y}{1-\Gamma}\right)^2 < \phi \sigma^2 ,
		\end{cases} \label{bulkspectrum}
	\end{align}
	where $\lambda = x + i y$.
	\subsection{Linear instability occurs when the bulk region crosses the imaginary axis}
	\noindent The rightmost point on the edge of the bulk spectrum is given by 
	\begin{align}
		\lambda_{\mathrm{bulk}} = -1 + (1 + \Gamma) \sqrt{\phi} \sigma . \label{bulkedge}
	\end{align}
	When the bulk of the eigenvalue spectrum first crosses the imaginary axis, we thus have
	\begin{align}
		\sigma^2 = \frac{1}{\phi (1 + \Gamma)^2}.
	\end{align}
	Comparing with Eq.~(\ref{chaoticcond}), we see readily that this corresponds to the point at which the linear instability of the generalised Lotka-Volterra dynamics occurs.
	
	\section{Modified interaction statistics}\label{section:modifiedstatistics}
	As a result of removing the rows and columns associated with extinct species from the interaction matrix, the statistics of the reduced interaction matrix elements $a'_{ij}$ differ from those of the original interaction matrix. We can deduce the modified interaction statistics by evaluating the ensemble averaged derivatives of the generating functional in Eq.~(\ref{genfunct}) with respect to $\lambda_{ij}$.
	\subsection{Modified mean, variance and second-order correlations}
	The statistics of the reduced interaction matrix can be obtained from derivatives like those in Eq.~(\ref{derivatives}). We denote the modified statistics with a dash. For the modified (scaled) mean, we have 
	\begin{align}
		\mu' \equiv \frac{1}{\phi N}\sum_{ij} \overline{ a_{ij}\theta_i \theta_j } &= \phi \mu + \frac{i}{\phi N}\sum_{ij}\left\langle \frac{1}{A}\frac{\delta A}{\delta \lambda_{ij}}\right\rangle \bigg\vert_{\lambda = 0} , \nonumber \\
		&= \phi \mu + \frac{\sigma^2}{\phi} (1 + \Gamma) \chi_{{}_T} M . \label{scaledmean}
	\end{align}
	Similarly, for the variance and the second-order correlations between transpose pairs, we obtain respectively
	\begin{align}
		\sigma'^2 &\equiv \frac{1}{\phi N}\sum_{ij} \overline{\left( a_{ij} - \frac{\mu'}{N_\ess} \right)^2 \theta_i \theta_j} = \frac{1}{\phi N}\sum_{ij} \overline{ \left( a_{ij}\theta_i \theta_j  \right)^2 } + O(N^{-1}) \nonumber \\
		&\approx -\frac{1}{\phi N}\sum_{ij} \left\langle \frac{1}{A}\frac{\delta^2 A}{\delta \lambda_{ij}^2} \right \rangle\bigg\vert_{\lambda = 0} =  \phi \sigma^2 , \nonumber \\
		\Gamma' &\equiv \frac{1}{\phi N (\sigma')^2}\sum_{ij} \overline{ \left( a_{ij}- \frac{\mu'}{N_\ess} \right)\left( a_{ji} - \frac{\mu'}{N_\ess} \right)\theta_i \theta_j} \nonumber \\
		&\approx -\frac{1}{\phi N (\sigma')^2}\sum_{ij} \left\langle\frac{1}{A}\frac{\delta^2 A}{\delta \lambda_{ij}\delta\lambda_{ji}} \right\rangle\bigg\vert_{\lambda=0} = \Gamma,
	\end{align}
	where the approximation is valid for large $N$. The removal of extinct species gives rise to additional correlations between elements that only share one index (as was also pointed out by Bunin  \cite{bunin2016interaction}), despite no such correlations being present in the original ensemble for the full $N\times N$ interaction matrix $a_{ij}$. These correlations can be shown to greatly affect the location of outlier eigenvalue \cite{baron2021eigenvalues}. We find the following correlations between elements that share only a single index
	\begin{align}
		r' \equiv& \frac{1}{\phi N} \sum_{(ijk)} \overline{ \left( a_{ki}- \frac{\mu'}{N_\ess} \right)\left( a_{kj} - \frac{\mu'}{N_\ess} \right)\theta_i \theta_j \theta_k}  \nonumber \\
		&=  \frac{\sigma^4}{ \phi} \left[ \chi_2 M^2 + 2 \Gamma \chi \chi_{{}_T} M + \Gamma^2 q \chi_{{}_T}^2 \right] - (\phi \mu - \mu')^2, \nonumber \\
		c' \equiv& \frac{1}{\phi N} \sum_{(ijk)} \overline{ \left( a_{ik}- \frac{\mu'}{N_\ess} \right)\left( a_{jk} - \frac{\mu'}{N_\ess} \right)\theta_i \theta_j \theta_k } \nonumber \\
		&=  \frac{\sigma^4}{ \phi} \left[ q \chi_{{}_T}^2 + 2 \Gamma \chi \chi_{{}_T} M + \Gamma^2 \chi_2 M^2 \right] - (\phi \mu - \mu')^2,  \nonumber \\
		\gamma' \equiv& \frac{1}{\phi N} \sum_{(ijk)} \overline{ \left( a_{ik}- \frac{\mu'}{N_\ess} \right)\left( a_{kj} - \frac{\mu'}{N_\ess} \right)\theta_i \theta_j \theta_k} \nonumber \\
		&= -\frac{1}{\phi N} \sum_{(ijk)} \left\langle \frac{1}{A} \frac{\delta^2 A}{\delta \lambda_{ik} \delta\lambda_{kj}}\right\rangle \bigg\vert_{\lambda= 0} - (\phi \mu - \mu')^2 \nonumber \\
		&=  \frac{\sigma^4}{ \phi} \left[  \chi_{{}_T} \chi  M  + \Gamma \chi_{{}_T}^2 q + \Gamma \chi_2 M^2 + \Gamma^2 \chi_{{}_T} \chi  M  \right] - (\phi \mu - \mu')^2, \nonumber \\
		\label{correlations}
	\end{align}
	where the notation $(ijk)$ indicates that none of the set $i$, $j$ and $k$ can take the same value.  We note that the first coefficient ($r'$) in Eq.~(\ref{correlations}) captures correlations between elements in the same row of the reduced interaction matrix. The second coefficient ($c'$) describes in-column correlations. The coefficient $\gamma'$ describes correlations between one elements whose first index equals that of the second index of another element. 
	
	Correlations between elements of the reduced interaction matrix that have no indices in common vanish in the thermodynamic limit, that is
	\begin{align}
		\frac{1}{(\phi N)^2} \sum_{(ijkl)} \overline{ \left( a_{ij}- \frac{\mu'}{N_\ess} \right)\left( a_{kl} - \frac{\mu'}{N_\ess} \right)\theta_i \theta_j \theta_k } &= -\frac{1}{(\phi N)^2} \sum_{(ijkl)} \left\langle \frac{1}{A}\frac{\delta^2 A}{\delta \lambda_{ij} \delta\lambda_{kl}}\right\rangle\bigg\vert_{\lambda = 0} - (\phi \mu - \mu')^2 \nonumber \\
		&=  \frac{\sigma^4}{ \phi^2}(1 + \Gamma)^2 M^2 \chi_{{}_T}^2  - (\phi \mu - \mu')^2 \nonumber \\
		&= 0 .
	\end{align}
	\begin{figure}[H]
		\centering 
		\includegraphics[scale = 0.47]{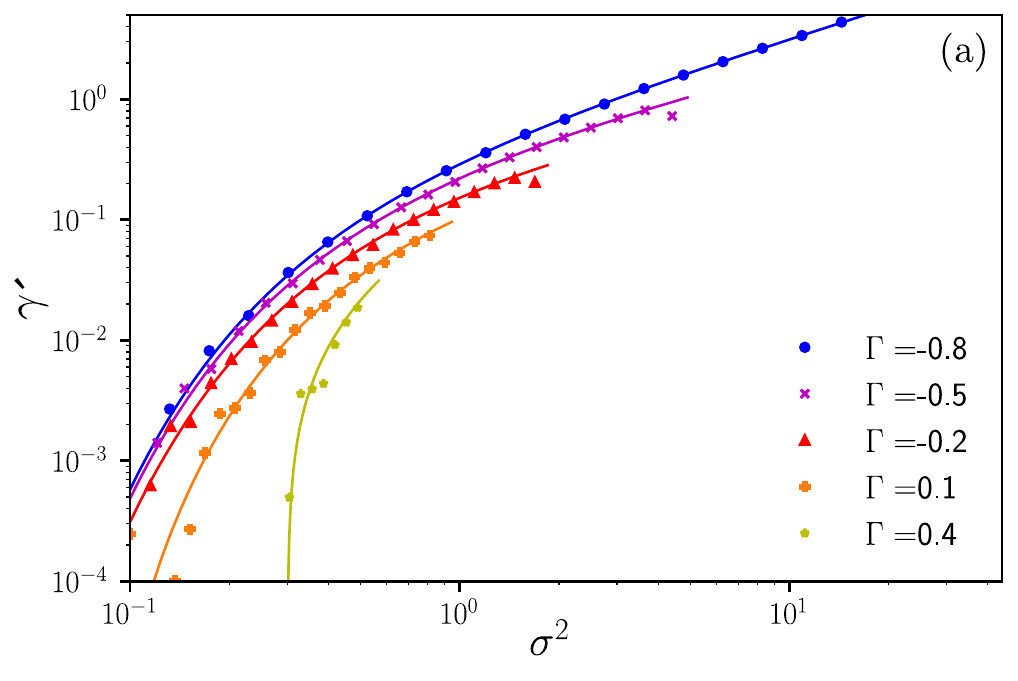}
		\includegraphics[scale = 0.47]{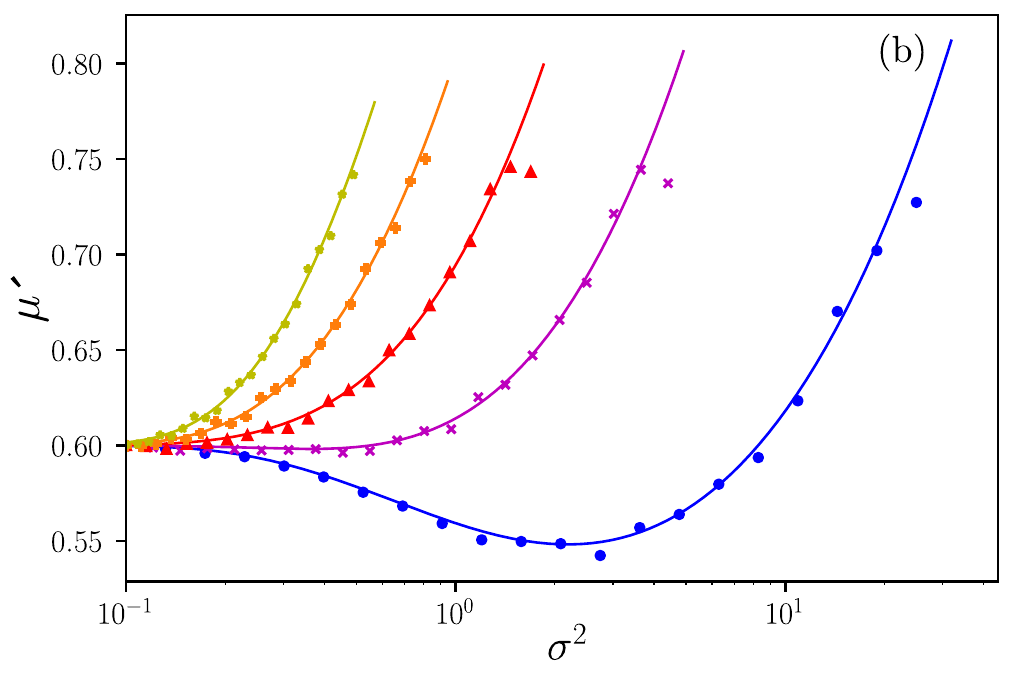}
		\caption{(a) The correlations between elements $a_{ij}$ and $a_{ki}$ [defined in Eq.~(\ref{correlations})] and (b) the scaled mean of the reduced interaction matrix elements [see Eq.~(\ref{scaledmean})]. The remaining system parameters are $\mu= 0.6$, $N = 4000$ and the results represented by points were averaged over 10 trials. }\label{fig:statsversussigma}
	\end{figure}
	\subsection{Non-Gaussian statistics}\label{sec:non_gauss_statistics}
	Let us now consider some of the higher-order statistics of the reduced interaction matrix that are relevant for the calculation of the eigenvalue spectrum. For example, consider the quantity
	\begin{align}
		S_3 = \frac{1}{(\phi N)^2}\sum_{ijkl} \overline{ (a_{ik} - \mu'/N_\ess) (a_{kl} - \mu'/N_\ess) (a_{lj} - \mu'/N_\ess)  \theta_i \theta_j \theta_k \theta_l }, \label{s3}
	\end{align}
	This can be related to the quantities that appear in the series for the resolvent in Eq.~(\ref{resolventexpansion})
	\begin{align}
		S_3 &= -i\frac{1}{(\phi N)^2} \sum_{(ijkl)} \left\langle \frac{1}{A}\frac{\delta^3 A}{\delta \lambda_{ik} \delta\lambda_{kl} \delta \lambda_{lj}}\right\rangle\bigg\vert_{\lambda = 0} \nonumber \\
		&+ \phi^3 \mu^3 - (\mu')^3 - 3 \phi^2 \mu^2 \mu' + 3 (\mu')^3 + (\phi \mu - \mu') \left[ (\mu')^2 + \gamma' + 2 \phi \sigma^2 \Gamma +  2 (\mu')^2\right] , \label{s31}
	\end{align}
	where we have
	\begin{align}
		-i\frac{1}{(\phi N)^2}& \sum_{(ijkl)} \left\langle \frac{1}{A}\frac{\delta^3 A}{\delta \lambda_{ik} \delta\lambda_{kl} \delta \lambda_{lj}}\right\rangle\bigg\vert_{\lambda=0} = \frac{\sigma^4}{\phi} \bigg[ (1 + \Gamma) \sigma^2 \Gamma \chi \chi_2 M^2 + (1+\Gamma) \Gamma q \chi_{{}_T}^2 \sigma^2 \chi \nonumber \\
		&+ \chi_{{}_T} M \Gamma^2 (2 \phi + \chi^2 \Gamma \sigma^2 + \chi_2 q \sigma^2) + \chi_{{}_T} M (2 \Gamma \phi + \chi^2 \sigma^2 + \chi_2 \Gamma q \sigma^2) \bigg] . \label{s32}
	\end{align}
	If the matrix elements $z_{ij}$ (and hence $a'_{ij}$) were Gaussian random variables, then the quantity $S_3$ would vanish. We see that $S_3$ does not vanish, even when the elements of the original interaction matrix are Gaussian random variables (see Fig. \ref{fig:s3versussigma} below). 
	\begin{figure}[H]
		\centering 
		\includegraphics[scale = 0.55]{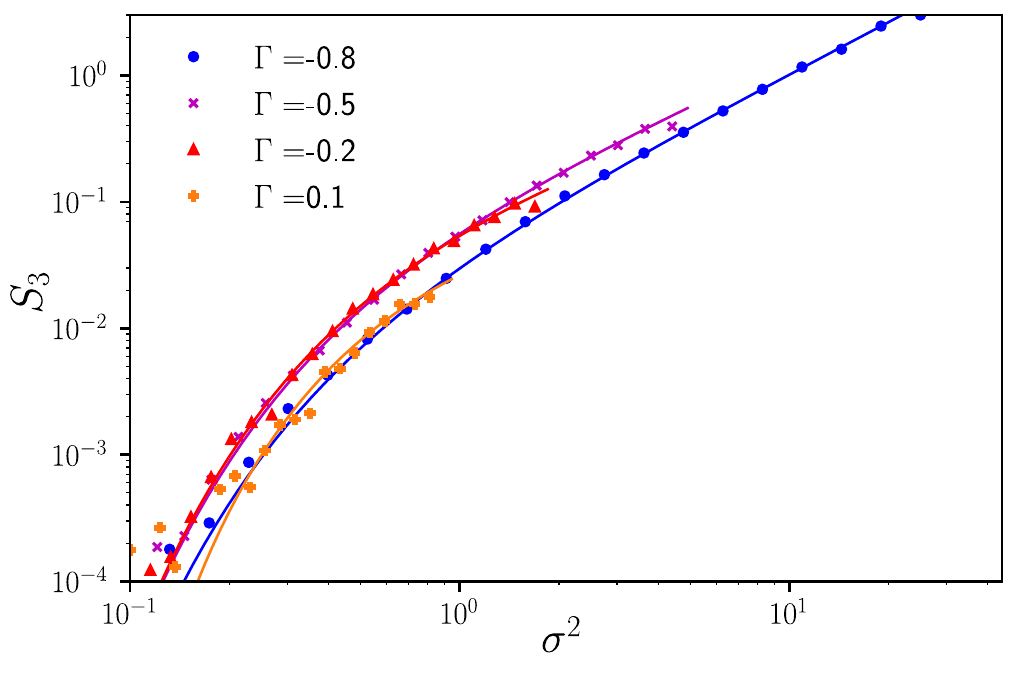}
		\caption{Demonstrating that the statistics of the reduced interaction matrix elements are non-Gaussian. The quantity $S_3$ would be zero if $z_{ij}$ were Gaussian random numbers. The remaining system parameters are $\mu= 0.6$, $N = 4000$ and the results represented by points were averaged over 10 trials. The results for $\Gamma = 0.4$ are too small to be visible. }\label{fig:s3versussigma}
	\end{figure}

\section{Smaller numbers of species}\label{sec:smaller_N}
The theory that we developed is formally derived in the thermodynamic limit ($N\to\infty$). In this section we verify that the predictions from the theory are also a good approximation for realistic ecological community sizes. To this end, we have conducted simulations for systems with an initial pool of $N=50$ species, noting that this results on surviving communities of approximately $25$ species, depending on parameters.
	
\begin{figure}[H]
	\centering 
	\includegraphics[scale = 0.5]{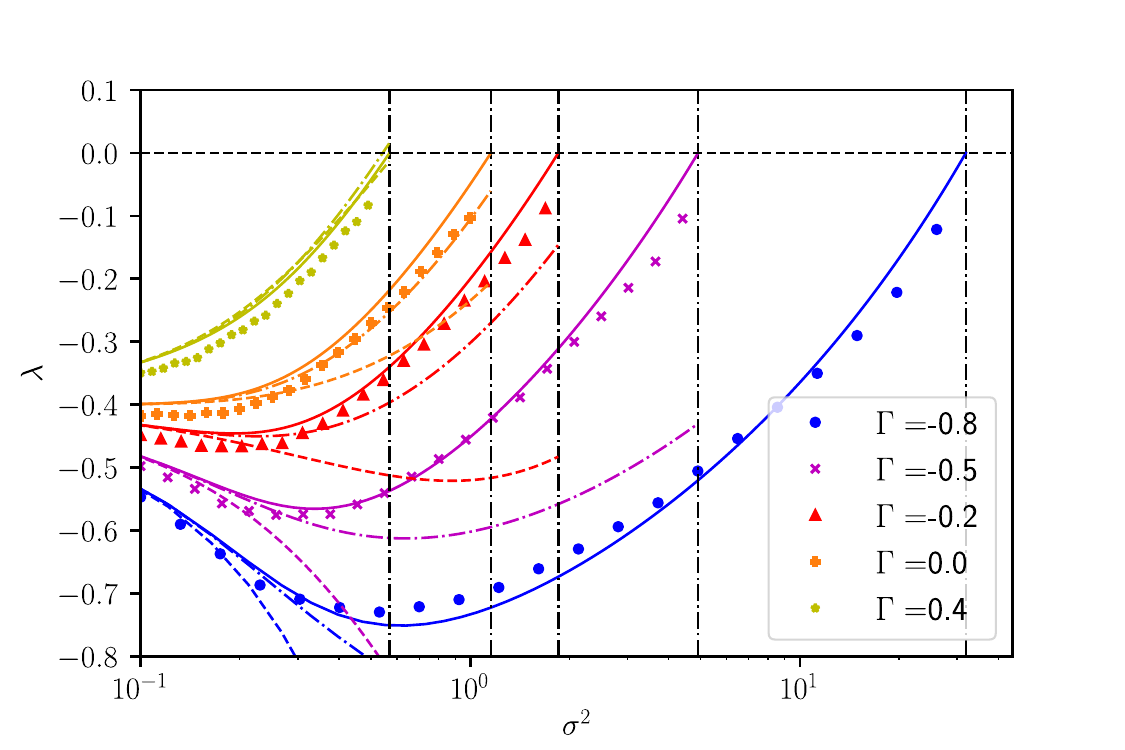}
	\caption{Analog of Fig.~3 of the main text, but markers are now from simulations for $N = 50$ (leading to $\approx 25$ surviving species). Solid lines are the predictions of our non-Gaussian theory [Eq.~(8) in the main paper]. We also show the predictions $\lambda_0$ and $\lambda_1$ from the Gaussian theory (see main paper) for comparison (dashed lines are $\lambda_0$, dot dashed lines are $\lambda_1$). We include $\Gamma = 0$, to demonstrate that there is no anomaly for this choice.  The figure demonstrates that our calculation, which takes into account the full non-Gaussian statistics of the surviving community, remains a better predictor of the outlier eigenvalue and therefore stability.}\label{fig:n50_2}
	\end{figure}

We generally find that, at such values of $N$, there are small quantitative deviations between the theory and the simulations, as would be expected. Nevertheless, as seen in Fig.~\ref{fig:n50_2}, when the full non-Gaussian theory makes predictions that are substantially different to the Gaussian approach, then the former remains a far better predictor of the leading eigenvalue. Hence, the conclusion that it is necessary to take into account non-Gaussian interaction statistics to correctly predict the stability of a complex ecosystem is also true for communities with a smaller numbers of species than in the figures in the main text.
	\section{Variation of the leading eigenvalue with $\Gamma$ }
\begin{figure}[H]
		\centering 
		\includegraphics[scale = 0.55]{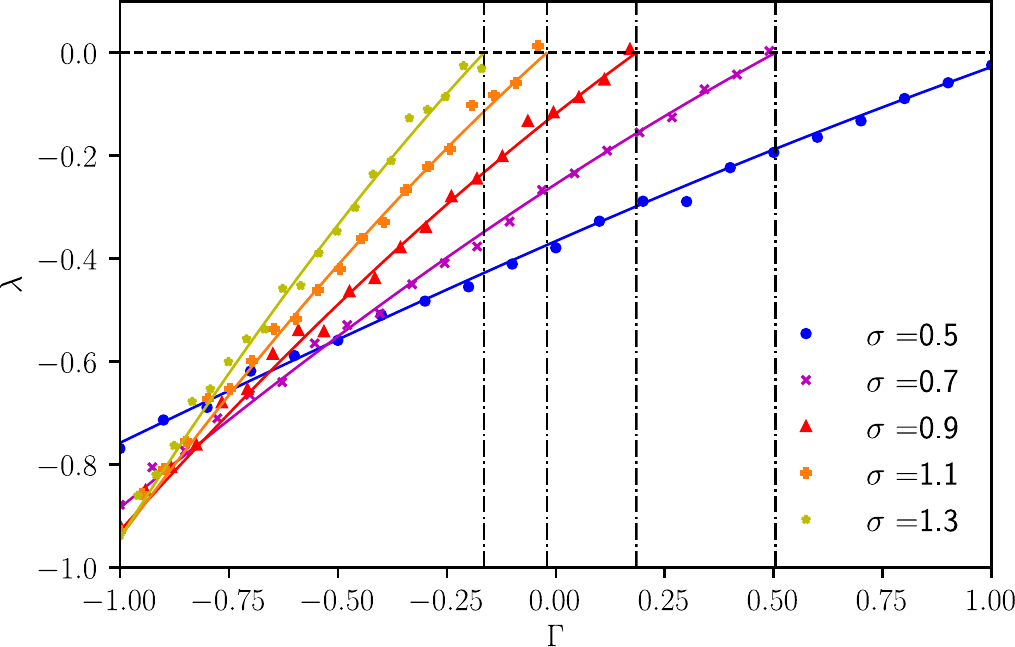}
		\caption{Leading eigenvalue of the reduced interaction matrix as a function of the correlation parameter $\Gamma$ for various values of $\sigma$, at fixed $\mu = 0.6$. Solid lines are from Eq.~(8) in the main text, markers are from computer simulations ($N=400$, averaged over 10 trials.) }\label{fig:lambdaversusgamma}
	\end{figure}

	To further verify the formula for the outlier eigenvalue given in Eq.~(8) of the main text, we plot the outlier as a function of $\Gamma$ in Fig. \ref{fig:lambdaversusgamma}. The figure demonstrates good agreement between the results of computer simulation and our theory prediction.
	\section{The effect of varying the intrinsic growth rate/carrying capacity}
		\subsection{Generalisation of relations for the order parameters at a stable fixed point}
		\subsubsection{Fixed point relations for order paramaters}
	We now allow for the possibility of different carrying capacities $k_i$ for each species. That is, Eqs.~(\ref{gles}) become
	\begin{align}
		\dot x_i = x_i \left[ k_i - x_i + \sum_{j}a_{ij} x_j + h_i(t)\right] ,
	\end{align}

	where the coefficients $k_i$ are drawn independently for each species from a distribution $\gamma(k)$, with support contained in the interval $[k_{\rm min},k_{\rm max}]$. The analysis in Section \ref{section:dmft} remains largely unchanged and we arrive at an alternative version of Eq.~(\ref{fp})
	\begin{align}
	x^\star = \frac{k + \mu M + \sigma \sqrt{q} z  }{1 - \Gamma \sigma^2 \chi} \Theta\left( \frac{k + \mu M + \sigma \sqrt{q} z  }{1 - \Gamma \sigma^2 \chi}   \right), 
	\end{align}
	where now we have
	\begin{align}
	\chi=&\frac{1}{1-\Gamma\sigma^2\chi}\int dk \, \gamma(k) \int_{-\infty}^{\Delta_k} Dz, \nonumber \\
	M=& \frac{\sqrt{q}\sigma}{1-\Gamma\sigma^2\chi}\int dk \, \gamma(k)\int_{-\infty}^{\Delta_k} Dz ~(\Delta_k-z),\nonumber\\
	1=&\frac{\sigma^2}{(1-\Gamma\sigma^2\chi)^2}\int dk \, \gamma(k)\int_{-\infty}^{\Delta_k} Dz~ (\Delta_k-z)^2,  \label{eq:aux_k_1}
	\end{align}
	with $\Delta_k = \frac{k + \mu M}{\sigma \sqrt{q}}$. 
	
\subsubsection{Solution procedure}
We introduce the following shorthand
	\begin{align}
		\langle w_r \rangle_k = \int dk \, \gamma(k)\int_{-\infty}^{\Delta_k} Dz~ (\Delta_k-z)^r,
	\end{align}
and use the substitution $k = \sigma \sqrt{q} \Delta_k - \mu M$ such that $dk = \sigma \sqrt{q} d\Delta_k$. We assume that the support of $\gamma(k)$ ranges from $k_{\rm min}$ to $k_{\rm max}$, and define $\Delta_{k_{\rm min}}$ and $\Delta_{k_{\rm max}}$ via the relations
\begin{align}
k_{\rm min} = \sigma \sqrt{q} \Delta_{k_{\rm min}} - \mu M, \nonumber \\
k_{\rm max} = \sigma \sqrt{q} \Delta_{k_{\rm max}} - \mu M. \label{eq:aux_k_2}
\end{align}
The objects $\langle w_r \rangle_k$ can be written in terms of $\Delta_{k_\mathrm{max}}$ and $\Delta_{k_\mathrm{min}}$ (for given $\sigma^2, \mu$ and $\Gamma$, as well as $k_{\rm min}$ and $k_{\rm max}$). This can be seen from
\begin{align}
\langle w_r \rangle_k  &=\sigma\sqrt{q} \int^{\Delta_{k_\mathrm{max}}}_{\Delta_{k_\mathrm{min}}}d\Delta~ \gamma\left(\sigma\sqrt{q}\Delta-\mu M\right) w_r(\Delta),
\end{align}
and the fact that the relations in Eq.~(\ref{eq:aux_k_2}) can be used to express $q$ and $M$ in terms of $\Delta_{k_\mathrm{max}}$ and $\Delta_{k_\mathrm{min}}$
\begin{align}
	\sigma \sqrt{q} &= \frac{k_{\mathrm{max}} - k_{\mathrm{min}}}{\Delta_{k_{\mathrm{max}}} - \Delta_{k_{\mathrm{min}}}}, \nonumber \\
	\mu M &= \frac{k_{\mathrm{min}} \Delta_{k_\mathrm{max}}- k_{\mathrm{max}} \Delta_{k_\mathrm{min}}}{\Delta_{k_\mathrm{min}}-\Delta_{k_\mathrm{max}}}.
\end{align}

We thus deduce that $\Delta_{k_\mathrm{max}}$ and $\Delta_{k_\mathrm{min}}$ are determined (for given $\mu$, $\sigma^2, \Gamma$, $k_{\rm min}$ and $k_{\rm max}$) by the following two equations
	\begin{align}
		\sigma^2 &= \frac{\langle w_2 \rangle_k}{(\langle w_2 \rangle_k + \Gamma \langle w_0 \rangle_k)^2}, \nonumber \\
		\mu \left(\frac{1}{k_\mathrm{max}} -\frac{1}{k_\mathrm{min}}  \right) &=\left(\frac{\Delta_{k_\mathrm{min}}}{k_\mathrm{max}} -\frac{\Delta_{k_\mathrm{max}}}{k_\mathrm{min}}  \right) \frac{1}{\langle w_1 \rangle_k} \frac{\langle w_2 \rangle_k}{\langle w_2 \rangle_k + \Gamma \langle w_0 \rangle_k} . \label{solvefordeltas}
	\end{align}
The first of these is analogous to Eq.~(\ref{sigmasol}), and can be obtained directly from Eqs.~(\ref{eq:aux_k_1}). The second relation is derived from subtracting the two relations in Eq.~(\ref{eq:aux_k_2}) from one another and 
\begin{align}
\frac{\sigma\sqrt{q}}{M}=\frac{1}{\langle w_1 \rangle_k} \frac{\langle w_2 \rangle_k}{\langle w_2 \rangle_k + \Gamma \langle w_0 \rangle_k},
\end{align}
which in turn is obtained from Eqs.~(\ref{eq:aux_k_1}).

The values of $\Delta_{k_\mathrm{max}}$ and $\Delta_{k_\mathrm{min}}$ that satisfy Eqs.~(\ref{solvefordeltas}) can then be substituted into the following equations to yield the order parameters of interest
	\begin{align}
		\chi=&\langle w_0 \rangle_k+\Gamma\frac{\langle w_0 \rangle_k^2}{ \langle w_2 \rangle_k}, \nonumber \\
		\frac{k_{\mathrm{min}}}{M}=&\frac{\Delta_{k_{\mathrm{min}}}}{ \langle w_1 \rangle_k}\frac{ \langle w_2 \rangle_k}{ \langle w_2 \rangle_k+\Gamma \langle w_0 \rangle_k}-\mu, \nonumber \\
				\frac{k_{\mathrm{max}}}{M}=&\frac{\Delta_{k_{\mathrm{max}}}}{ \langle w_1 \rangle_k}\frac{ \langle w_2 \rangle_k}{ \langle w_2 \rangle_k+\Gamma \langle w_0 \rangle_k}-\mu, \nonumber \\
		q=&\left(\frac{M}{\sigma \langle w_1 \rangle_k} \frac{ \langle w_2 \rangle_k}{ \langle w_2 \rangle_k + \Gamma \langle w_0 \rangle_k} \right)^2,\nonumber\\ 
		\phi=& \langle w_0 \rangle_k,\nonumber \\
		\chi_{{}_T} =& \frac{1}{\sigma \sqrt{q}} \langle w_1 - \Delta_k w_0 \rangle_k,\nonumber \\
		\chi_2 =& -\frac{1}{\sigma^2 q} \langle \Delta_k (w_1 - \Delta_k w_0) \rangle_k . \label{orderparameterscc}
	\end{align}
Only one of the second and third relations is required to find $M$.
	\subsection{Onset of instability}
	Following similar reasoning to Section \ref{section:minftrans}, we see that the mean abundance diverges ($M\to\infty$) when
	\begin{align}
	\mu = \frac{\Delta_{k_{\mathrm{min}}}}{ \langle w_1 \rangle_k}\frac{ \langle w_2 \rangle_k}{ \langle w_2 \rangle_k+\Gamma \langle w_0 \rangle_k}=\frac{\Delta_{k_{\mathrm{max}}}}{ \langle w_1 \rangle_k}\frac{ \langle w_2 \rangle_k}{ \langle w_2 \rangle_k+\Gamma \langle w_0 \rangle_k}.
	\end{align}
	From this, we deduce that at the transition point, $\Delta_{k_{\mathrm{min}}} = \Delta_{k_{\mathrm{max}}}$. This means that the transition point in the parameter space $(\mu, \Gamma, \sigma)$ is exactly that given by Eq.~(\ref{ferromagnetictransition}). The prediction that the point in parameter space where the mean abundance diverges remains unaffected by the introduction of heterogeneous carrying capacities is verified in Fig. \ref{fig:carryingcapacities}.
			
	However, in the case of the linear instability described in Section \ref{section:oppertrans}, things are not as simple. Following the reasoning in Section \ref{section:oppertrans}, one arrives at the following simultaneous expressions, which can be solved to yield $\Delta_{k_\mathrm{max}}$  and $\Delta_{k_\mathrm{min}}$ (and hence all other order parameters) at the point of instability 
	\begin{align}
		\langle w_2 \rangle_k &= \langle w_0 \rangle_k, \nonumber \\
		\mu \left(\frac{1}{k_\mathrm{max}} -\frac{1}{k_\mathrm{min}}  \right) &=\left(\frac{\Delta_{k_\mathrm{min}}}{k_\mathrm{max}} -\frac{\Delta_{k_\mathrm{max}}}{k_\mathrm{min}}  \right) \frac{1}{\langle w_1 \rangle_k} \frac{\langle w_2 \rangle_k}{\langle w_2 \rangle_k + \Gamma \langle w_0 \rangle_k} .
	\end{align}	
	We note here that in general $\phi = \langle w_0 \rangle_k \neq 1/2$. This means that the linear instability does not necessarily occur at $\sigma = \sqrt{2}/(1 + \Gamma)$, as was the case when the carrying capacities were all the same, i.e. $k_i = 1$. 
	\subsection{Leading eigenvalues}
	The expressions for the boundary of the bulk of the eigenvalue spectrum and the outlier eigenvalue in Eqs.~(\ref{bulkedge}) and (\ref{gsol}) respectively are given entirely in terms of the order parameters listed in Eqs.~(2) and (3) of the main text. Eqs.~(\ref{bulkedge}) and (\ref{gsol}) do not change when heterogeneous carrying capacities are introduced. That is, what one has to do in the case where the carrying capacities are heterogeneous is to calculate the quantities $\chi$, $\chi_T$, $\chi_2$, $M$, $q$ and $\phi$ using Eqs.~(\ref{orderparameterscc}) and substitute these values into Eq.~(\ref{bulkedge}) for the edge of the bulk spectrum, and into Eq.~(\ref{gsol}) for the outlier [equivalently into Eq.~(8) in the main text].
	
	We demonstrate the efficacy of our theory for reproducing the correct leading eigenvalue of the reduced interaction matrix in the case of heterogeneous carrying capacities in Fig. \ref{fig:carryingcapacities} below. This shows that, even when carrying capacities vary between species, the central conclusion of the main text remains valid. That is, one must take into account the non-Gaussian statistics of interactions between species if one is to properly predict the stability of the system. 
	
	\begin{figure}[H]
		\centering 
		\includegraphics[scale = 0.47]{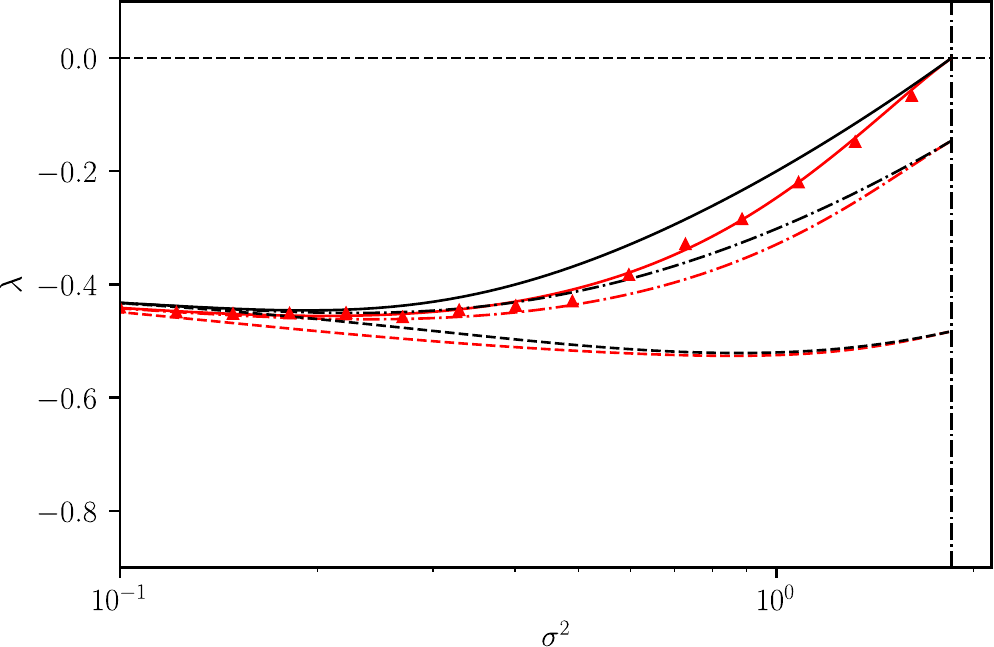}
		\includegraphics[scale = 0.47]{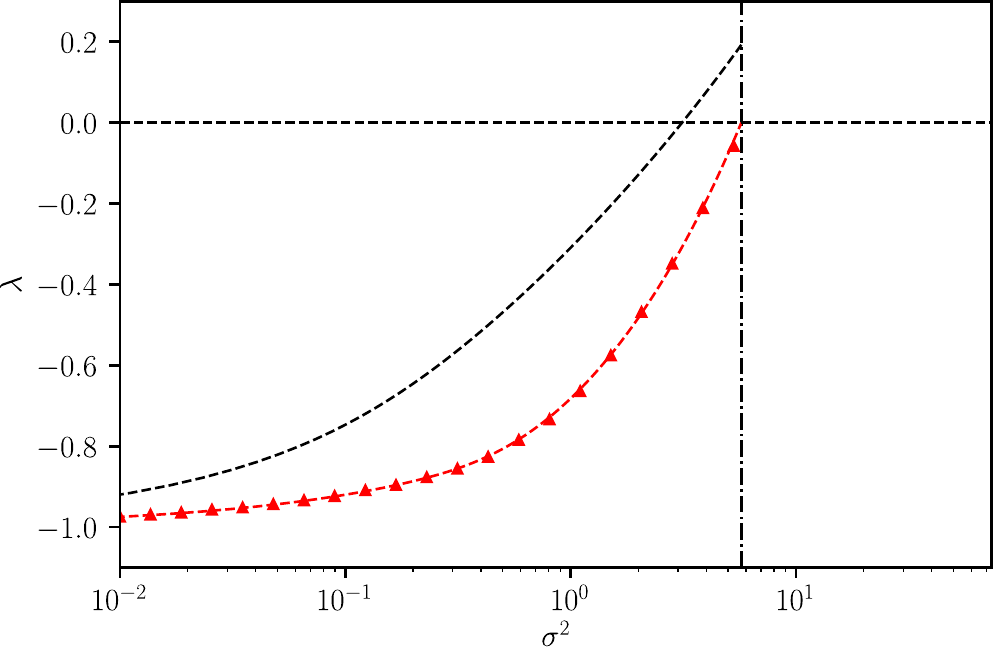}
		\caption{Panel (a): Leading eigenvalue of the reduced interaction matrix as a function of $\sigma^2$ for fixed $\mu = 0.6$, but now in a model with heterogeneous carrying capacities (analog of Fig. 4 in the main text). The red solid line is the modified theory using the values from Eq.~(\ref{orderparameterscc}) in Eq.~(8) of the main text (i.e., it is the non-Gaussian theory for the model with distributed carrying capacities). The dot dashed and dashed red lines are found by inserting the values in Eq.~(\ref{orderparameterscc}) into the expressions for $\lambda_1$ and $\lambda_0$ in the main text respectively. The black lines are the corresponding lines from the theory for homogeneous $k_i = 1$ for all species (as in the main text). The data in panel (a) thus demonstrates that one must take into account non-Gaussian statistics to correctly predict stability also in the presence of varying carrying capacities. (b) The edge of the bulk of the eigenvalue spectrum when $\mu = -5.0$. The red line is for varying carrying capacities, the black line is for $k_i = 1$ for all species. Notably, the instability point in the model with distributed carrying capacities is no longer given by $\sigma = \frac{\sqrt{2}}{1+\Gamma}$ as in the case of homogeneous $k_i=1$. In both panels, $\Gamma=0.2$. Simulations are for $N = 4000$, averaged over 10 trials. We used a dichotomous distribution of carrying capacities, $\gamma(k) = p_1 \delta_{k,k_1}+  p_2 \delta_{k,k_2}$, with $p_1 = 0.9$, $p_2 = 0.1$, $k_1 = 0.1$, $k_2 = 5.0$.}\label{fig:carryingcapacities}
	\end{figure}
	\pagebreak 
	
\section{Generation of reduced interaction matrices without elimination of extinct species in the Lotka--Volterra dynamics}
In this section, we discuss the possibility of constructing matrices with the same statistics as the reduced interaction matrices without running the dynamics of the Lotka--Volterra system and then eliminating rows and columns of extinct species from the original (full) interaction matrix. In other words, we construct the ensemble of reduced matrices directly from some prescribed distribution `from scratch' (or `bottom up').  We will refer to matrices constructed in this way as `imitation' reduced interaction matrices. By constructing the imitation ensemble, we begin to understand in more detail the origin of the non-Gaussianity of the reduced interaction matrix and thus why the universality principle fails to predict its eigenvalues.	

In Sec.~\ref{sec:idea} we first present the general idea of the bottom-up construction. In Sec.~\ref{sec:cond_stat} we calculate the statistics of the true interaction matrices conditioned on given values of the species abundances at a Lotka--Volterra fixed point. These conditional statistics are used in the construction of the imitation ensemble. Technical details of the method to produce the imitation matrices are then given in Sec.~\ref{sec:proc}, before we verify in Sec.~\ref{section:resultsimitation} that the ensemble of imitation matrices has the same properties as the ensemble of true reduced interaction matrices. We discuss and interpret these results in Section \ref{section:interpretation}.

Throughout the section we write $a_{ij}$ for elements of the actual reduced interaction matrix, and $\tilde a_{ij}$ for the elements of an imitation matrix. The size of the imitation matrices is written as $\tilde N\times \tilde N$. To construct the ensemble of imitation matrices we use, and develop further, ideas put forward in Refs. \cite{barbier2021fingerprints} and \cite{bunin2016interaction}. 
		
	\subsection{Overall idea}\label{sec:idea}

Our approach exploits the fact that an interesting structure becomes apparent in the statistics of the reduced interaction matrix  when the reduced matrix is conditioned on the abundances of the surviving species. We use this to turn the Lotka-Volterra approach on its head, so to speak. Instead of drawing a set of interaction coefficients, which then determine the equilibrium abundances and surviving species in the Lotka--Volterra system, and thus the reduced interaction matrix, we proceed in reverse. We first draw a set of mock or `imitation' abundances and then a set of reduced interaction matrix elements conditioned on these abundances.

More specifically, for a fixed set of model parameters $\mu, \sigma, \Gamma$ of the original Lotka--Volterra system, we draw a set of imitation abundances $\{\tilde x_i\}$ from the known distribution of abundances at a Lotka--Volterra fixed point. Then, we draw imitation interaction matrix elements from a carefully-constructed Gaussian distribution, whose statistics depend on the imitation abundances. The precise details are described below in Sec.~\ref{sec:proc}.

\subsection{Conditional statistics of the elements of the true reduced interaction matrix}\label{sec:cond_stat}
In this section we calculate the statistics of elements of the true reduced interaction matrices, conditioned on given values of the species abundances at a Lotka--Volterra fixed point.

\medskip

For real valued $\alpha_1, \beta_1, \alpha_2$ and $\beta_2$, we define the following quantities,
	\begin{align}
		\vartheta_1(x_i) = \begin{cases}
			1 \,\,\,\,\,\, \mathrm{if} \,\,\,\,\,\, \alpha_1 < x_i \leq \beta_1, \\
			0 \,\,\,\,\, \mathrm{otherwise},
		\end{cases}
		\end{align}
and similarly

\begin{align}
		\vartheta_2(x_i) = \begin{cases}
			1 \,\,\,\,\,\, \mathrm{if} \,\,\,\,\,\, \alpha_2 < x_i \leq \beta_2, \\
			0 \,\,\,\,\, \mathrm{otherwise}.
		\end{cases}
	\end{align}
Thus $\vartheta_1(x_i)$ indicates if a particular abundance $x_i$ is in the range $[\alpha_1,\beta_1]$, and $\vartheta_2(x_i)$ if the abundance is in the interval $[\alpha_2,\beta_2]$. We always assume that $\beta_1\geq \alpha_1>0$ and $\beta_2\geq \alpha_2>0$.
	\medskip
	
As a first step, we now find the statistics of the interaction coefficients of species whose abundances are conditioned to be within the above-stated ranges. That is, we wish to find for example
	\begin{align}
		N\,\mathrm{E}\left(a_{ij} \vert \alpha_1 < x_i \leq \beta_1, \alpha_2 < x_j \leq \beta_2 \right) = \frac{1}{N \phi_1 \phi_2}\sum_{ij} \overline{a_{ij} \vartheta_1(x_i) \vartheta_2(x_j)},
	\end{align}
	where we have defined
	\begin{align}
		\phi_1 = \frac{1}{N}\sum_{i}\overline{\vartheta_1(x_i)}
	\end{align}
	as the fraction of species with a fixed-point abundance in the interval $[\alpha_1,\beta_1]$, and similarly for $\phi_2$. We highlight that $N$ is the number of species in the initial pool, at the point when the Lotka--Volterra dynamics is started.
	
	 A similar calculation to that which was performed in Section \ref{section:modifiedstatistics} yields
	\begin{align}
	N\mathrm{E}\left(a_{ij} \vert \alpha_1 < x_i \leq \beta_1, \alpha_2 < x_j \leq \beta_2 \right) = \mu + \frac{1}{\phi_1 \phi_2} \left[ \chi_T^{(1)} M^{(2)} \sigma^2  + \Gamma \sigma^2 \chi_T^{(2)} M^{(1)}   \right] ,\label{eq:S10_aux1}
	\end{align}
	where we have
	\begin{align}
		\phi_1 &= \int_{z(\alpha_1)}^{z(\beta_1)} Dz , \nonumber \\
		M^{(1)} &= \int_{z(\alpha_1)}^{z(\beta_1)} Dz\, x(z) , \nonumber \\
		\chi_T^{(1)} &= -\frac{1}{\sqrt{2 \pi q \sigma^2}} \left[ e^{-z(\beta_1)^2/2} - e^{-z(\alpha_1)^2/2}\right]
	\label{eq:S10_aux2}
	\end{align}
	with $Dz=e^{-z^2/2}/\sqrt{2\pi}$. Analogous relations apply for the objects $\phi_2, M^{(2)},$ and $\chi_T^{(2)}$. For surviving species we also have from Eq.~(\ref{fp}),
	\begin{align}
		z(x) &=- \frac{1 + \mu M  - (1-\Gamma \chi \sigma^2) x}{\sqrt{q} \sigma} , \nonumber \\
		x(z) &= \frac{ 1 + \mu M + \sqrt{q} \sigma z}{(1-\Gamma \chi \sigma^2)}. \label{eq:S10_aux3}
	\end{align}
	\medskip
	
	Let us now examine the limit where $\beta_1 = \alpha_1 + \varepsilon$ and $\varepsilon\ll 1$. That is to say the condition $\vartheta_1(x_i)=1$ constrains the abundance $x_i$ to be in a small interval just above the value $x_i=\alpha_1$. Expanding in $\varepsilon$ to linear order we have
	\begin{align}
		\phi_1 &\approx \frac{1}{\sqrt{2\pi}} e^{-z(\alpha_1)^2/2} \frac{(1-\Gamma \chi \sigma^2)}{\sigma\sqrt{q}} \varepsilon, \nonumber \\
		M^{(1)} &\approx \alpha_1 \frac{1}{\sqrt{2\pi}} e^{-z(\alpha_1)^2/2} \frac{(1-\Gamma \chi \sigma^2)}{\sigma\sqrt{q}} \varepsilon, \nonumber \\
		\chi_T^{(1)} &\approx \frac{1}{\sqrt{2 \pi q \sigma^2}} z(\alpha_1)  e^{-z(\alpha_1)^2/2} \frac{(1-\Gamma \chi \sigma^2)}{\sigma\sqrt{q}} \varepsilon .\label{eq:S10_aux4}
	\end{align}
	Analogous expressions can be found from constraining abundance $x_j$ to be in the interval $[\alpha_2,\alpha_2+\varepsilon]$, with $\varepsilon\ll 1$.
	
	Using Eqs.~(\ref{eq:S10_aux4}) and the expression for $z(x)$ in Eq.~(\ref{eq:S10_aux3}) in Eq.~(\ref{eq:S10_aux1}) one obtains 
	\begin{align}
		 N\mathrm{E}\left(a_{ij} \vert x_i=\alpha_1, x_j=\alpha_2 \right) =  \mu -  \alpha_2\frac{1 + \mu M - (1-\Gamma \chi \sigma^2) \alpha_1 }{q }  -  \Gamma \alpha_1 \frac{1 + \mu M - (1-\Gamma \chi \sigma^2) \alpha_2}{q} .  
	\end{align}
We write this more in more compact form as 
	\begin{align}
		 N\mathrm{E}\left(a_{ij} \vert x_i, x_j \right) =  \mu -  x_j \frac{1 + \mu M - (1-\Gamma \chi \sigma^2) x_i }{q }  -  \Gamma x_i \frac{1 + \mu M - (1-\Gamma \chi \sigma^2) x_j }{q} . \label{conditionalmean}
	\end{align}
	This describes the mean of the element $a_{ij}$ in the true reduced interaction matrix, conditioned on given abundances $x_i$ and $x_j$ for species $i$ and $j$. An expression which is in agreement with this was obtained in Ref. \cite{barbier2021fingerprints} for the case $\Gamma = 0$.
	
	\medskip
	
	 In a similar fashion, we also find 
	\begin{align}
	\mathrm{Var}(a_{ij}|x_i, x_j) &= \frac{\sigma^2}{N}, \nonumber \\
	\mathrm{Cov}(a_{ij}, a_{ji}|x_i, x_j) &= \frac{\sigma^2 \Gamma}{N}, \nonumber \\
	\mathrm{Cov}(a_{ij}, a_{ik}|x_i, x_j, x_k) &= - \frac{\sigma^2}{N^2} \frac{x_jx_k}{q},    \nonumber \\
	\mathrm{Cov}(a_{ij}, a_{ki}|x_i, x_j, x_k) &= - \Gamma \frac{\sigma^2}{N^2} \frac{x_jx_k}{q},  \nonumber \\
	\mathrm{Cov}(a_{ji}, a_{ki}|x_i, x_j, x_k) &= -\Gamma^2 \frac{\sigma^2}{N^2}\frac{x_jx_k}{q}.\label{conditionalcorrelations}
	\end{align}
	We note that the third of the above expressions agrees with the result obtained for the case $\Gamma=0$ in Ref. \cite{barbier2021fingerprints}.

	\subsection{Procedure for producing `imitation' reduced interaction matrices}\label{sec:proc}
\noindent The construction proceeds in two steps.
	\medskip
	
	\noindent{\bf \underline{Step 1:}}\\
	For a given choice of $\sigma^2, \Gamma, \mu$ we draw a set of $\tilde N$ imitation equilibrium species abundances $\{\tilde x_i\}$ independently from the truncated Gaussian distribution of the true abundances [this distribution results from Eq.~(\ref{fp})],
	\begin{align}
		P(x_i) &= \Theta(x_i) \frac{1}{\phi} \frac{1}{\sqrt{2\pi \Sigma^2}} \exp\left[ \frac{\left(x_i - m\right)^2}{2 \Sigma^2}\right], \label{truncatedgaussian}
	\end{align}
	where $\Theta(\cdot)$ is the Heaviside function. We have here written $x_i$ for the argument to stress that this is the distribution of the actual fixed-point abundances of the Lotka--Volterra system. We have also defined 
	\begin{align}
		m &= \frac{1 + \mu M}{1 - \Gamma \sigma^2 \chi}, \nonumber \\
		\Sigma &= \frac{\sigma \sqrt{q}}{(1 - \Gamma \sigma^2 \chi)} .
	\end{align}
 The quantities $M, q, \chi$ and $\phi$ are available analytically for the given values of $\sigma^2, \Gamma, \mu$ (see Sec.~\ref{sec:fp_analysis}, and Refs. \cite{bunin2016interaction, galla2018dynamically}). 
 
 While step 1 relies on analytical results for the distribution of fixed-point abundances of the Lotka--Volterra system, it is important to note that the imitation abundances $\{\tilde x_i\}$ are obtained without running the Lotka--Volterra dynamics. Instead they are generated synthetically as independent samples from the distribution in Eq.~(\ref{truncatedgaussian}).
 
	\bigskip
	
	\noindent{\bf \underline{Step 2:}}\\
	For a given set of values $\tilde x_1,\dots,\tilde x_{\tilde N}$ from step 1 we then construct a random matrix of size $\tilde N \times \tilde N$, with elements $\{\tilde a_{ij}\}$ drawn from a joint Gaussian distribution $P(\{\tilde a_{ij} \} \vert \{\tilde x_i\})$ with the same conditional first and second moments as the ones in a (true) reduced interaction matrix for the given values of the model parameters $\sigma^2, \Gamma, \mu$. We calculated these conditional moments in the previous section (Sec.~\ref{sec:cond_stat}). 
	\medskip

The use of a Gaussian distribution in step 2 is motivated by the discussion in Appendix A of Ref.~\cite{barbier2021fingerprints}. The $\{a_{ij}\}$ of the original interaction matrix are Gaussian random variables. For a fixed set of abundances $\{x_i\}$, the $\{a_{ij}\}$ for surviving species at the fixed point of the GLVEs are constrained to satisfy the fixed point equations $1 -x_i +\sum_{j\neq i}a_{ij} x_j = 0$. The distribution of independent Gaussian random variables that are made to satisfy linear constraints was shown to be Gaussian, albeit with additional correlations, in Refs.~\cite{barbier2021fingerprints, Vrins2018}. It is by virtue of taking into account that $\{x_i\}$ are also random variables that the resulting distribution of reduced interaction matrices becomes non-Gaussian. 

\pagebreak

To generate the $\{\tilde a_{ij}\}$ in practice, we first draw a set of $\tilde N^2$ Gaussian i.i.d. random numbers $y_{ij}$ with mean zero and variance $\phi \sigma^2/\tilde N$. We then construct the elements of the imitation matrices as follows

\begin{align}
	\tilde a_{ij} &= \frac{\phi \mu_{ij}}{\tilde N} +A y_{ij} +B y_{ji} - \frac{A \phi}{\tilde Nq}\sum_{k\neq j} y_{ik} \tilde x_k \tilde x_j -\frac{\phi\Gamma^2}{2 \tilde N q B } \sum_{k\neq i} y_{jk} \tilde x_k \tilde x_i \nonumber \\
	&\;\;\;\;\;\;\;\;\;\;-\frac{\phi B}{ \tilde N q } \sum_{k\neq j} y_{ki} \tilde x_k \tilde x_j - \frac{\phi\Gamma^2}{2 \tilde Nq A}\sum_{k\neq i} y_{kj} \tilde x_k \tilde x_i , \label{imitationmatrix}
	\end{align}
	where
	\begin{align}
	A &= \frac{1}{2} \left[\sqrt{1 + \Gamma} +\sqrt{1- \Gamma} \right], \nonumber \\
	B &= \frac{1}{2} \left[\sqrt{1 + \Gamma} -\sqrt{1- \Gamma} \right],
\end{align}
and where we have defined 
\begin{align}
\mu_{ij} = \mu -  \tilde x_j\frac{1 + \mu M - (1-\Gamma \chi \sigma^2) \tilde x_i }{q }  -  \Gamma \tilde x_i \frac{1 + \mu M - (1-\Gamma \chi \sigma^2) \tilde x_j}{q} .
\end{align} 
We note that each element $\tilde a_{ij}$ is the sum of products of random variables (for example of the type $y_{jk} \tilde x_k \tilde x_i$), and hence is manifestly non-Gaussian. 

\medskip

The expression in Eq.~(\ref{imitationmatrix}) is designed so that the $\{\tilde a_{ij}\}$ have the following properties (which can be verified by direct calculation),
\begin{align}
	\mathrm{E}\left(\tilde a_{ij} \vert \tilde x_i, \tilde x_j \right) &=  \frac{\phi}{\tilde N}\left(\mu -  \tilde x_j \frac{1 + \mu M - (1-\Gamma \chi \sigma^2) \tilde x_i }{q }  -  \Gamma \tilde x_i \frac{1 + \mu M - (1-\Gamma \chi \sigma^2) \tilde x_j }{q} \right), \nonumber \\
	\mathrm{Var}(\tilde a_{ij}|\tilde x_i, \tilde x_j) &= \frac{\phi \sigma^2}{\tilde N} + \mathcal{O}\left(\frac{1}{\tilde N^2}\right) \nonumber \\
	\mathrm{Cov}(\tilde a_{ij}, \tilde a_{ji}|\tilde x_i,\tilde  x_j) &= \frac{\phi \sigma^2 \Gamma}{\tilde N} + \mathcal{O}\left(\frac{1}{\tilde N^2}\right) , \nonumber \\
	\mathrm{Cov}(\tilde a_{ij}, \tilde a_{ik}|\tilde x_i,\tilde x_j,\tilde x_k) &= - \frac{\phi^2 \sigma^2}{\tilde N^2} \frac{\tilde x_j\tilde x_k}{q} + \mathcal{O}\left(\frac{1}{\tilde N^3}\right) ,    \nonumber \\
	\mathrm{Cov}(\tilde a_{ij}, \tilde a_{ki}|\tilde x_i, \tilde x_j, \tilde x_k) &= - \Gamma \frac{\phi^2 \sigma^2}{\tilde N^2} \frac{\tilde x_j\tilde x_k}{q} + \mathcal{O}\left(\frac{1}{\tilde N^3}\right),  \nonumber \\
	\mathrm{Cov}(\tilde a_{ji}, \tilde a_{ki}|\tilde x_i, \tilde x_j, \tilde x_k) &= -\Gamma^2 \frac{\phi^2 \sigma^2}{\tilde N^2}\frac{\tilde x_j\tilde x_k}{q} + \mathcal{O}\left(\frac{1}{\tilde N^3}\right).\label{eq:imit_prop}
\end{align}

Comparison with Eqs.~(\ref{conditionalmean},\ref{conditionalcorrelations}) demonstrates that the imitation matrices therefore have the same {\em conditioned} statistics as the true reduced interaction matrices in the thermodynamic limit. (To make the comparison one must keep in mind that the dimension of the reduced matrix is related to that of the original Lotka--Volterra system via $\tilde N = \phi N$.)

\subsection{Properties of the imitation ensemble} \label{section:resultsimitation}	 
We now test to see whether the imitation matrices also have the same unconditioned statistics, agnostic of the species abundances, as the true reduced interaction matrices. More precisely, we compare the properties of imitation reduced matrices (generated using the procedure described in Sec.~\ref{sec:proc}) with the analytical results for the true reduced matrices from Secs.~\ref{section:outlierresolvent} and \ref{section:modifiedstatistics} of this Supplement. 

\begin{figure}[H]
		\begin{center}
		\includegraphics[scale = 0.425]{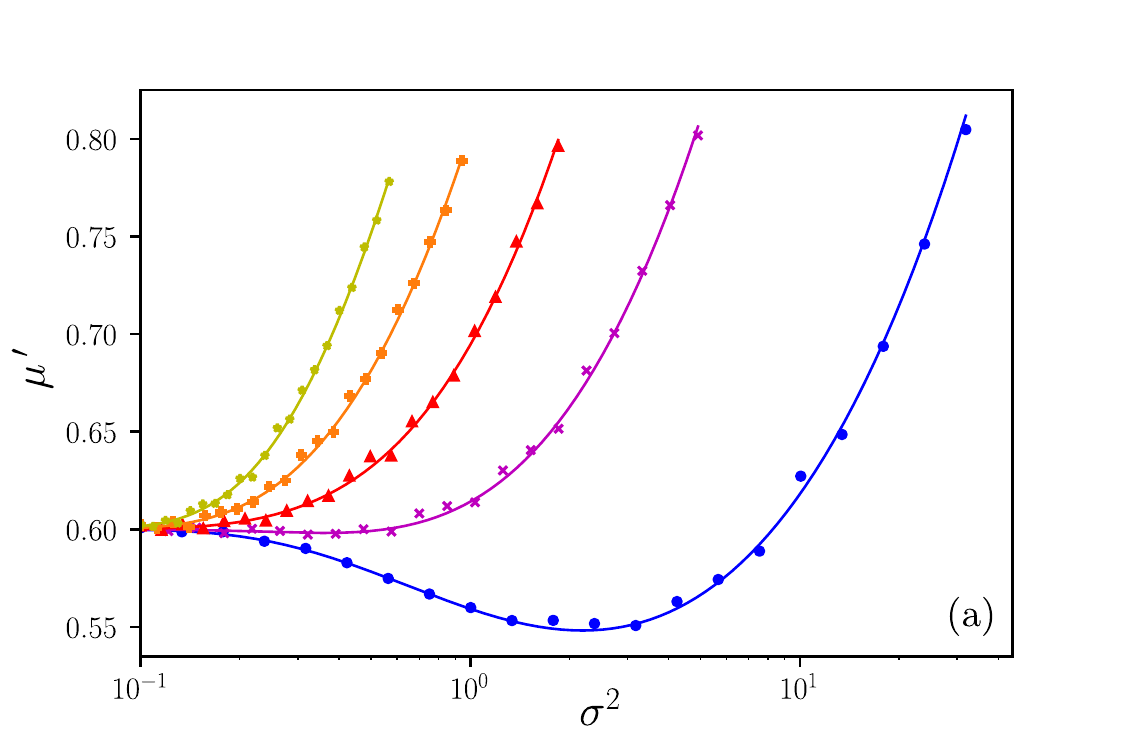}
		\includegraphics[scale = 0.425]{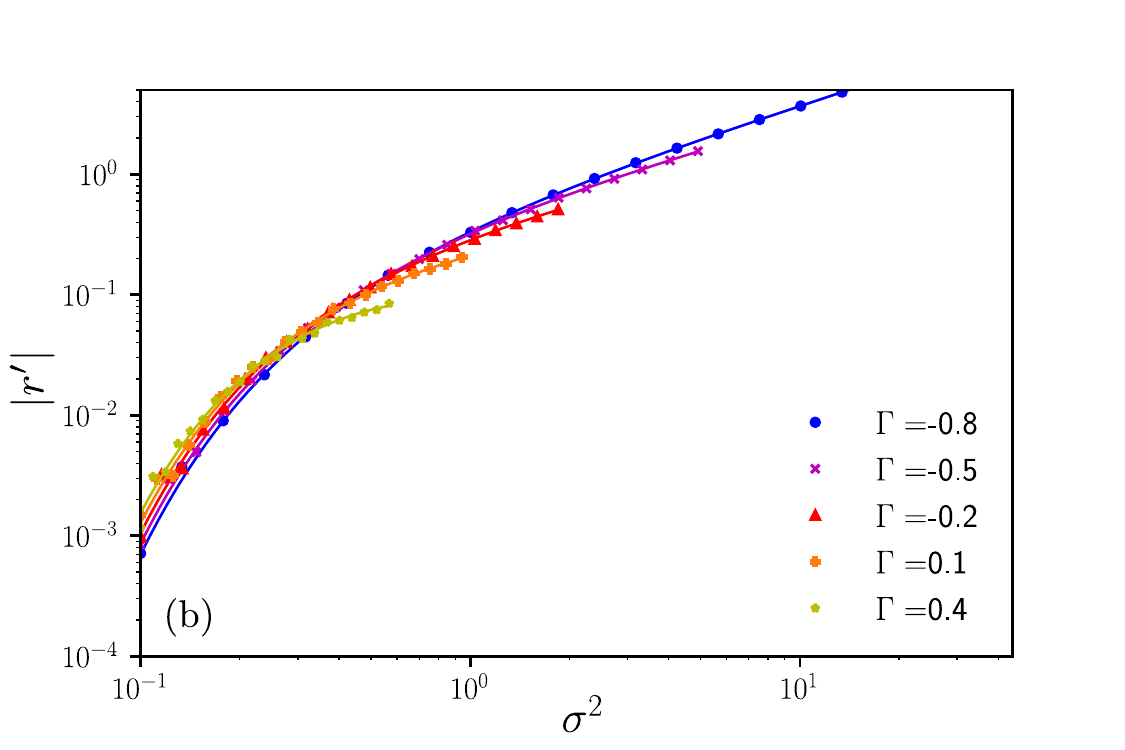}
		\includegraphics[scale = 0.425]{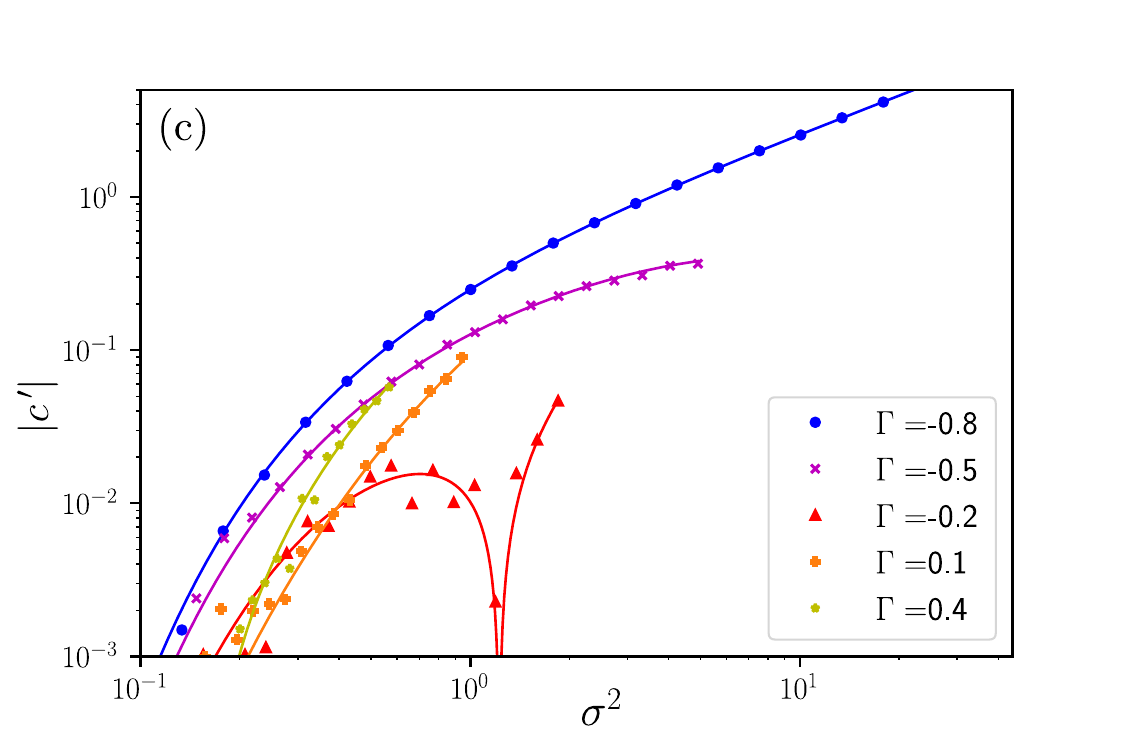}
		\includegraphics[scale = 0.425]{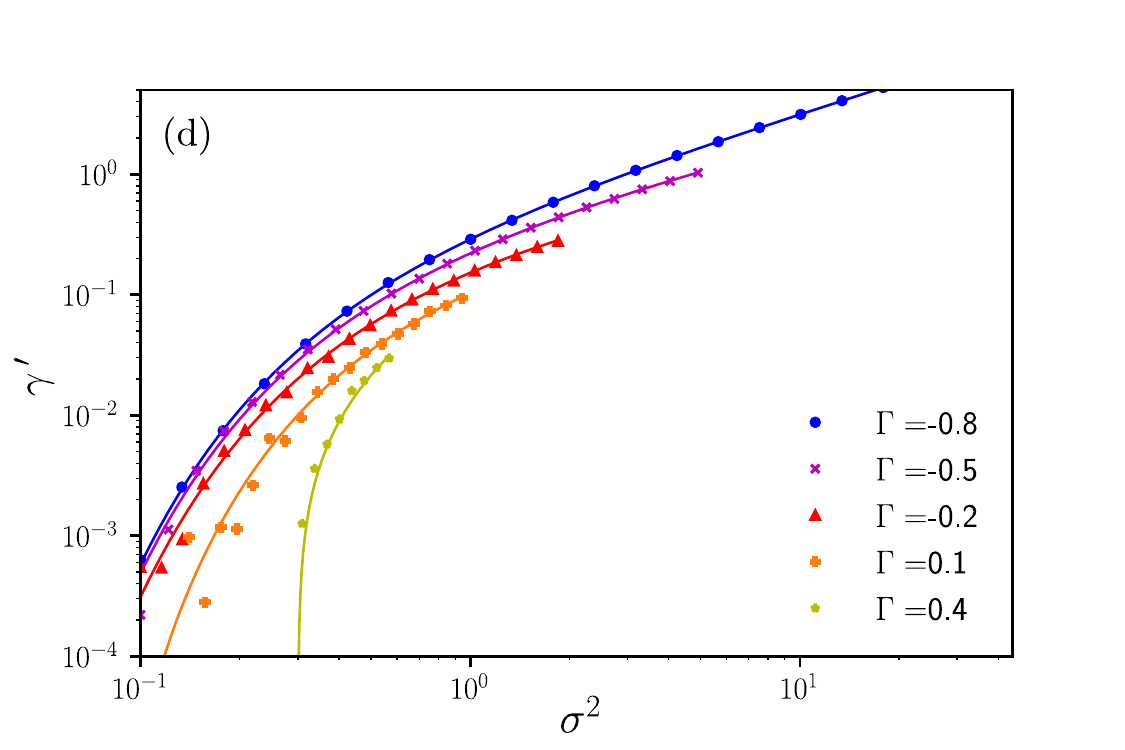}
		\end{center}
		\caption{Verifying that the imitation reduced interaction matrix has the correct `global' statistics (statistics not conditioned on the abundances). Solid lines are the analytical results for the true reduced matrices, given by Eqs.~(\ref{scaledmean}) and (\ref{correlations}). Markers are from imitation matrices generated using the procedure in Sec.~\ref{sec:proc}, in particular  Eq.~(\ref{imitationmatrix}). In all panels, $\mu=0.6$, $\tilde N = 8000$ and results were averaged over 10 trials. Panel (a): The scaled mean $\mu' = \tilde N\mathrm{E}(\tilde a_{ij})$. Panel (b): In-row correlations $r' = \tilde N^2\mathrm{Cov}(\tilde a_{ij}, \tilde a_{ik})$. Panel (c): In-column correlations $c' = \tilde N^2\mathrm{Cov}(\tilde a_{ij}, \tilde a_{ik})$. Panel (d): Correlations between elements in the $i^{th}$ row and the $i^{th}$ column $\gamma' = \tilde N^2\mathrm{Cov}(\tilde a_{ij}, \tilde a_{ki})$. To be able to use logarithmic vertical axes in panels (b) and (c) we plot the modulus of $r'$ and $c'$ respectively. }\label{fig:imitationmatrixstats}
\end{figure}

Fig.~\ref{fig:imitationmatrixstats} shows that the imitation matrix constructed according to Eq.~(\ref{imitationmatrix}) does indeed produce the correct mean and correlations of the reduced interaction matrix, given in Eqs.~(\ref{scaledmean}) and (\ref{correlations}) respectively. 

\begin{figure}[H]
	\begin{center}
		\includegraphics[scale = 0.425]{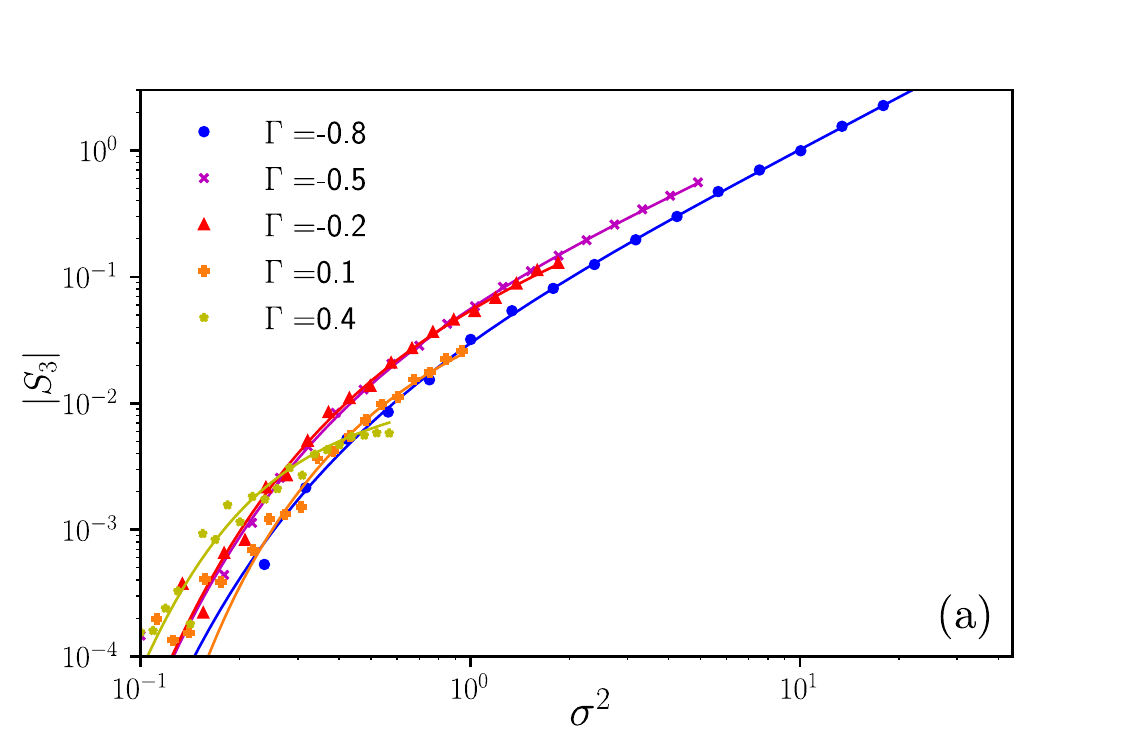}
		\includegraphics[scale = 0.425]{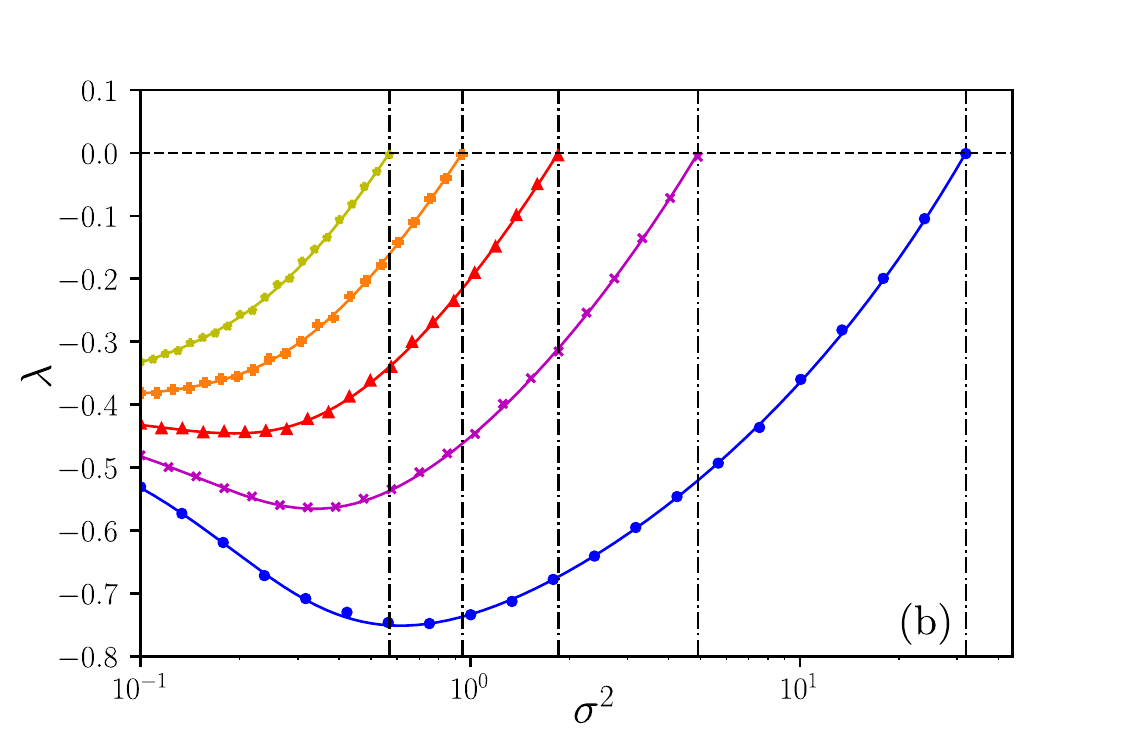}
	\end{center}
	\caption{Verifying that the imitation reduced interaction matrix has the correct higher-order statistics and leading eigenvalue. Solid lines are analytical results for the true reduced interaction matrix, given by Eq.~(\ref{s31}) and Eq.~(\ref{lambdafromg}). Markers are from imitation matrices generated as described in Sec.~\ref{sec:proc} and Eq.~(\ref{imitationmatrix}) in particular. In both panels,  $\mu=0.6$, $\tilde N = 8000$ and results were averaged over 10 trials. Panel (a): The third moment $S_3$ [defined in Eq.~(\ref{s3})]. To be able to use a logarithmic vertical axis we plot $|S_3|$. Panel (b): The leading eigenvalue. }\label{fig:higherorderimitation}
\end{figure}

Then, in Fig. \ref{fig:higherorderimitation}, we verify that the higher-order moment $S_3$ [c.f. Eq.~(\ref{s3})] of the true reduced interaction matrix is captured by the imitation matrices. This demonstrates again that the ensemble of imitation matrices is non-Gaussian ($S_3$ vanishes in a Gaussian ensemble). We also see in Fig. \ref{fig:higherorderimitation} that the leading eigenvalue, which as we saw is calculated using an infinite series of higher-order moments (Sec.~\ref{section:outlierresolvent}), is also well-replicated by the imitation matrices.

\subsection{Interpretation and further discussion}\label{section:interpretation}
The results in Figs.~\ref{fig:imitationmatrixstats} and \ref{fig:higherorderimitation} confirm that the ensemble of true reduced interaction matrices can be generated `bottom-up', without going through the elimination procedure in the Lotka--Volterra dynamics. We have verified that as well as reproducing the first and second moments of the true reduced interaction matrices, the imitation matrices also capture higher-order quantities such as $S_3$ in Eq.~(\ref{s3}), as well as the leading eigenvalue, and thus stability.
 	
Based on these observations, and the manifest non-Gaussianity of the reduced interaction matrices, we can begin to understand why the universality principle does not apply here. There is a finer structure to the interaction statistics, that becomes apparent when we condition on the abundances of surviving species. It is this structure that gives rise to the higher-order moments in the ensemble of reduced matrices. Ignoring these moments and making a simple Gaussian assumption one obtains an incorrect result for the leading eigenvalue [given in Eq.~(5) of the main text]. This is why we say that the universality principle fails in the ensemble of reduced interaction matrices.